\documentstyle[12pt,epsfig,rotating]{article}
\newcommand{\Pma}{I\!\!P}
\newcommand{\stotgp}{\mbox{$\sigma_{tot}^{\gamma{p}}$}}
\def\rap{\hat{y}}

\begin{document}
\begin{titlepage}
\begin{flushleft}
DESY 97-113 \\
June 1997
\end{flushleft}
\begin{center}
\vspace*{5cm}
\begin{Large}
{\bf High energy photoproduction}
\\
\vspace{1 cm}
G.~Abbiendi
\end{Large}
\\
\vspace{1 cm}
{\sl Deutsches Elektronen-Synchrotron DESY, \\
Notkestrasse 85, 22607 Hamburg, Germany}
\end{center}
\vspace{5 cm}
\begin{abstract}
The study of photoproduction reactions has gained a unique opportunity with the
HERA data. The high center of mass energy allows to carry out quantitative
tests of QCD, to explore the substructure of both the photon and the proton and
to shed new light on soft and diffractive processes. In this report we review
the HERA results, 
comparing them with the reach of fixed target photoproduction experiments 
and complementary measurements from $e^{+}e^{-}$ and hadron colliders.
\end{abstract}
\end{titlepage}
\section{Introduction}

High energy photoproduction processes occur with high cross sections 
at the HERA $e p$
collider ($\sqrt{s} = 300$~GeV), due to the large flux of quasi real photons 
from the electron beam. The obtained $\gamma p$ center of mass energies are up 
to ten times the values reached by fixed target experiments.
The integrated luminosity achieved in the first years of HERA running has
produced a great yield of physics output. With increasing luminosity the 
improvement of some of these measurements relies on a quite 
precise knowledge of the responses of the various detectors. 
The HERA results have also triggered a huge amount of theoretical activity.

The field covered by photoproduction at HERA has several connections with 
physics at $e^+ e^-$ and hadron colliders. The photon structure is probed in a 
complementary way with respect to the deep inelastic $e \gamma$ scattering 
at $e^+ e^-$ colliders.
Moreover most of the photoproduction processes are seen to resemble 
hadron-hadron processes and can be described in a similar framework.
It is non-trivial to say that we are testing our understanding of the high 
energy scattering with different beam particles and the universality of the 
underlying picture.

We review here the photoproduction results from HERA. 
In section \ref{introd} an introduction is given on the concept of the photon 
as a hadron,
the measurements of its hadronic structure in the $e \gamma$ deep inelastic 
scattering, the definition of parton distributions for the photon 
and the most used sets of parametrizations.
The basic characteristics of HERA, its kinematics and the experimental procedure
to select photoproduction events are described in section \ref{heragp} and 
a brief description of the ZEUS and H1 detectors is given in section 
\ref{detectors}.
The physics topics are then grouped in the following way: soft physics is dealt
with in section \ref{totsgp}, where the measurement of the total $\gamma p$ 
cross section and its decomposition in the different diffractive and non 
diffractive processes are described and in section \ref{vmsection}, where 
the elastic vector meson production is described;
hard photoproduction is then the subject of section \ref{hphp}, where the 
emphasis is on both tests of the QCD dynamics and on the determination of the 
photon and the proton structure from measurements of inclusive particle, jets, 
open charm and inelastic $J/\psi$ production;
section \ref{hdiff} is devoted to hard diffraction, where both photon 
dissociation with the appearance of jets or heavy flavours and colour singlet 
exchange are described.
Section \ref{conclu} concludes with some remarks about the future
prospects.

\section{Resolved photon and photon structure functions}
\label{introd}

In the Standard Model the photon is the gauge particle exchanged in
electromagnetic interactions. It is elementary and its fundamental
interactions occur with the charged leptons and quarks. On the other hand
hadrons are composite objects, bound states of quarks and gluons, 
and have a size of the order of $1$~fm ($10^{-13}$~cm). Despite
this classification the photon can also behave like a hadron.
The dual nature of the photon was first observed in fixed
target photoproduction experiments using real photon beams \cite{bauer78}.
These experiments
have covered center of mass energies from $1$ to about $20$~GeV.
The hadronic behaviour of the photon has been interpreted in
the Vector Dominance Model (VDM) \cite{saku60}. It describes the 
photon-nucleon interaction as
a two stage phenomenon: firstly the photon fluctuates into a vector meson
state with the same quantum numbers $J^{PC} = 1^{--}$ ($\rho^0, \omega, \phi$), 
then a normal hadronic interaction takes place between the
nucleon and the vector meson. The total cross section is written as:
\begin{equation}
\sigma(\gamma N \rightarrow X)= \sum_{V=\rho,\omega,\phi}
\frac{\pi \alpha} {\gamma_{V}^{2}} ~ \sigma (V N \rightarrow X )
\end{equation}
where $\gamma_{V}$ is the photon-meson coupling, which can be measured from
electromagnetic decay rates like $V \rightarrow e^{+}e^{-}$. Photoproduction
cross sections are suppressed by the QED coupling $\alpha$ in comparison to
purely hadronic ones, but the characteristics of the final state are found 
to be the same as e.g. $\pi N$ interactions.

The bulk of hadronic interactions are soft processes involving low momentum
transfers. However at high enough energies hard interactions can occur, their
signature being the production of high $p_{T}$ particles or jets. 
In Leading Order of QCD (LO) the strong coupling constant $\alpha_{s}$ 
has the expression:
\begin{equation}
\alpha_{s}(\mu^2) = 
\frac{12\pi}{(33-2n_f)~ln\left(\frac{\mu^2}{\Lambda^2} \right) }
\end{equation}
with $n_f$ the number of active flavours. 
Whenever the characteristic energy scale $\mu$ of the interaction, 
like the $p_{T}$ of the produced jet, is much higher than the QCD scale 
$\Lambda$ ($\approx 0.2$ GeV), $\alpha_{s}$ is small enough to apply 
perturbation theory.

In hard $\gamma N$ interactions the photon can behave as a pointlike particle 
in the so-called
{\it direct} photon processes or it can act as a source of
partons, which then scatter against partons in the nucleon, in the {\it
resolved} photon processes. 
Resolved interactions stem from the photon fluctuation to a $q \bar q$ state 
(or a more complex partonic state). Depending on the relative transverse 
momentum between $q$ and $\bar q$, this may be a bound state like in the VDM 
({\it hadronic} component) or a
perturbative calculable state ({\it anomalous} component).
Both are embedded in the definition of the photon structure functions.

\begin{figure}[p]
\centerline{\epsfig{file=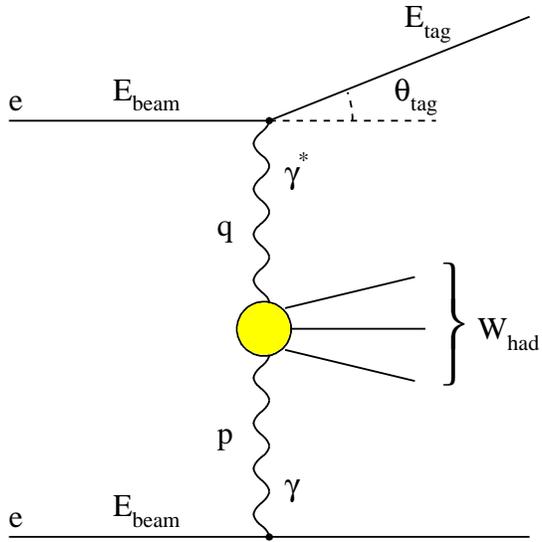,height=9.0cm}\hspace{0.5cm}}
\caption{The kinematics of a single--tag inclusive $\gamma \gamma $ event.}
\label{lep2}
\end{figure}

The apparent hadronic structure of the photon can be studied in $e \gamma$ deep
inelastic scattering at $e^{+}e^{-}$ colliders. This proceeds through the
exchange of two photons, one highly virtual (probe) and the other almost real
(target), as in Fig.~\ref{lep2}. Experimentally one electron is detected at
a large angle with respect to its initial direction, and its energy $E_{tag}$ 
and polar angle $\theta_{tag}$ are measured; the other electron is required to 
scatter at small angle, generally by an antitagging condition, 
and it is thus lost into the beam pipe (single tag). 
Without tagging the electron scattered at low angle,
the energy and small virtuality of the target photon are undetermined. It is
then necessary to measure the hadronic invariant mass $W_{had}$. The following
kinematic variables are defined, with reference to Fig.~\ref{lep2} ($k$ is
the four-momentum of the incoming electron):

\begin{eqnarray}
 Q^{2} &=& -q^{2} = 2 E_{beam} E_{tag} (1 - cos\theta_{tag})  \nonumber \\
 W^{2}_{had} &=& (q + p)^{2} \nonumber \\
 x &=& \frac {Q^{2}}{2p\cdot q} = \frac{Q^{2}}{Q^{2}+W^{2}_{had}} \\
 y &=& \frac{p\cdot q}{p\cdot k} = 
     1 - \frac {E_{tag}}{E_{beam}} cos^{2} \frac{\theta_{tag}}{2} \nonumber
\end{eqnarray}

The reaction $e^{+}e^{-} \rightarrow e^{+}e^{-} hadrons$ can be regarded as
$e ~\gamma \rightarrow e ~hadrons$ factorizing the flux of quasi-real
photons, i.e. the lower vertex in Fig.~\ref{lep2}. This is achieved in the
Weisz\"acker-Williams approximation \cite{wwa} or better in the Equivalent
Photon Approximation \cite{bud75}. Naming $z$ the energy fraction
carried by the quasi-real photon, the differential cross section versus
$z$, integrated over the photon virtuality $P^{2}$ between the limits
$P^{2}_{min}$ and $P^{2}_{max}$, is:
\begin{equation}
\frac{d\sigma}{dz} (e e \rightarrow e e X) = f_{\gamma/e}(z) ~\sigma(e \gamma
\rightarrow e X)
\end{equation}
with the photon flux $f_{\gamma/e}$ given by:
\begin{equation}
f_{\gamma/e}(z) = \frac{\alpha}{2\pi} \left[ \frac{1+(1-z)^{2}}{z} ln
\frac{P^{2}_{max}}{P^{2}_{min}} 
- 2m_{e}^{2}z \left(\frac{1}{P^{2}_{min}} - \frac{1}{P^{2}_{max}} \right)
\right]
\label{wwaflux}
\end{equation}
Here $P^{2}_{min}$ is the kinematic limit:
\begin{equation}
P^{2}_{min} = \frac{m_{e}^{2} z^{2}} {1-z}
\label{q2minwwa}
\end{equation}
while $P^{2}_{max}$ corresponds to the maximum scattering angle $\theta_{max}$
for the electron not to be detected (typically $\theta_{max} \sim 30$~mrad):
\begin{equation}
P^{2}_{max} = (1-z) E_{beam}^{2} \theta_{max}^{2}
\label{q2maxwwa}
\end{equation} 
The level of accuracy of the Equivalent Photon Approximation goes from few
per cent to about $10~\%$, depending on the $P^{2}$ range of integration.

The cross section for the deep inelastic $e \gamma$ scattering is written
in complete analogy to the electron-nucleon DIS, introducing the two photon 
structure functions $F_{2}^{\gamma}$ and $F_{L}^{\gamma}$:
\begin{equation}
\frac{d\sigma(e \gamma \rightarrow e X)}{dE_{tag} ~dcos\theta_{tag}} =
\frac{4 \pi \alpha^{2} E_{tag}}{Q^{4} y}
\left\{ \left[ 1+(1-y)^{2} \right] F_{2}^{\gamma}(x,Q^{2}) -
             y^{2}F_{L}^{\gamma}(x,Q^{2}) \right\}
\end{equation} 
In the usual experimental conditions $y^{2}$ is rather small and
$F_{L}^{\gamma}$ can be neglected. 
The cross section measurements thus correspond to determinations
of the $F_{2}^{\gamma}$ structure function. 

The most striking feature of $F_2^\gamma$
is that in the region of asymptotically high
$Q^{2}$ it is completely calculable by perturbative QCD (at large $x$). 
This has to be contrasted
with the nucleon structure functions, for which QCD predicts for example the
$Q^{2}$ evolution but not the absolute normalization at any $Q^{2}$ value.

\begin{figure}[thb]
\begin{center}
\vspace*{-0.8cm}
\epsfig{file=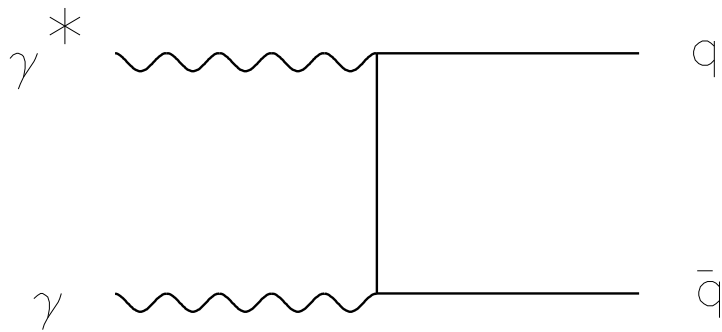,height=12cm}
\vspace*{-1.2cm}
\end{center}
\caption{Quark-Parton Model (box) contribution to the photon structure function.}
\label{box}
\end{figure}

The lowest order contribution to $F_{2}^{\gamma}$ is given by the Quark-Parton
Model diagram shown in Fig.~\ref{box} (plus its crossed one), which is purely
electromagnetic. It gives \cite{wal71}:
\begin{equation}
F_{2}^{\gamma}(x,Q^{2}) = \frac{3\alpha}{2\pi} \sum_{q=1}^{2n_f} 
{
e_q^4 ~x~
\left\{ \left[ x^2+(1-x)^2 \right] ln \frac {Q^2 ~(1-x)}{m_q^2 ~ x} + 
8x(1-x)-1 \right\}
}
\label{equf2gama}
\end{equation} 
where the sum runs over all quark and antiquark flavours $q$. 
The charge (in units of the proton charge) and mass are given by
$e_q$ and $m_q$.
This expression depends on the quark masses and roughly
describes experimental data using constituent masses of few hundred MeV.
However it shows some important qualitative features.
Comparing it to the known behaviour of the $F_2^N$
nucleon structure function two major differences appear:
\begin{itemize}
\item There is no scale invariance holding for $F_{2}^{\gamma}$ even in the
Parton Model. For nucleons the scaling violations result from gluon radiation
and have opposite signs in different $x$ ranges:
with increasing $Q^2$ $F_2^N$ increases at low $x$ while it decreases at high
$x$. Instead $F_2^\gamma$ grows at all the $x$ values with increasing $Q^2$.
\item $F_2^\gamma$ is large at high $x$, while counting rules predict for the nucleon
a vanishing $F_2^N$ for $x \rightarrow 1$. 
\end{itemize}
The different behaviour of photon and nucleon structure functions originates
from the pointlike coupling of the photon to quark-antiquark pairs $\gamma
\rightarrow q \bar q$: this coupling is what endows the photon
with a hadronic structure.
Perturbative QCD can be applied on top of the Parton Model and resummations
have been performed for $F_{2}^{\gamma}$ both in leading \cite{wit77} and
next-to-leading (NLO) order \cite{barbu79} of QCD for large $x$ and
asymptotically high $Q^2$. 
The result reads as:
\begin{equation}
F_{2}^{\gamma}(x,Q^{2}) = \alpha \left[ \frac{a(x)}{\alpha_s(Q^2)}+b(x)\right]
\end{equation}
with $a$ and $b$ calculable functions of $x$ but diverging for $x \rightarrow
0$. The first term is the LO result, the second its NLO correction. The
absolute normalization of $F_2^\gamma$
would then be a direct measurement of $\alpha_s(Q^2)$,
or equivalently of the QCD scale $\Lambda$.
This result accounts for only pointlike couplings and is valid at high $Q^2$
where the incalculable hadronic component of the photon is not important.
Unfortunately this regime is not experimentally reachable at present and 
foreseeable future accelerators.
Non-perturbative contributions due to the formation of bound states after the  
$\gamma \rightarrow q \bar q$ splitting cannot be neglected. 

The successful approach to $F_2^\gamma$, 
firstly suggested by Gl\"uck and Reya \cite{glu83}, requires the introduction of
boundary conditions, which have to take inputs from experimental data at some
arbitrary scale $Q_0^2$ already in the domain of perturbative QCD. 
At the scale $Q_0^2$ both
perturbative and non-perturbative contributions are buried together. From the
boundary conditions the evolution with $Q^2$ is then described by perturbative
QCD, most practically with generalized DGLAP equations \cite{dglap}. 
At the starting scale $Q_0^2$ parton densities are defined with a 
quark-antiquark component $q^\gamma(x,Q_0^2) = \bar q^\gamma(x,Q_0^2)$ 
and a gluon component $g^\gamma(x,Q_0^2)$.
The evolution equations for the photon parton densities differ
from those of the nucleon due to the anomalous coupling 
$\gamma \rightarrow q \bar q$. Its splitting function is given by the
coefficient of the leading logarithmic term in (\ref{equf2gama}):
\begin{equation}
k_i(x) = 3 e_i^2 \left[ x^2+(1-x)^2 \right]
\end{equation}
The anomalous coupling introduces an inhomogeneity in the evolution equations, 
which are written in the form:
\\
\begin{displaymath}
\frac{dq_{i}^\gamma(x,t)}{dt} = \frac{\alpha_{em}}{2\pi} k_i(x) +
\frac{\alpha_{s}(t)}{2\pi}
\int_{x}^{1}\frac{dx'}{x'}\left[ P_{qq}(\frac{x}{x'})\,q_{i}^\gamma(x',t)+
 P_{qg}(\frac{x}{x'})\,g^\gamma(x',t) \right]
\end{displaymath}
\begin{equation}
\label{apequ}
\end{equation}
\begin{displaymath}
\frac{dg^\gamma(x,t)}{dt} =
\frac{\alpha_{s}(t)}{2\pi}
\int_{x}^{1}\frac{dx'}{x'}\left[\sum_{j=1}^{2n_f} P_{gq}(\frac{x}{x'})\,
  q_{j}^\gamma(x',t)+P_{gg}(\frac{x}{x'})\,g^\gamma(x',t) \right]
\end{displaymath}
\\
Here $t = ln \left( \frac{Q^2}{\Lambda^2} \right)$, $q_{i}^\gamma$ and
$g^\gamma$
are the quark (antiquark) and gluon density in the photon and
$P_{ij}~(i,j=q,g)$ the ordinary splitting functions. 
Equations (\ref{apequ}) hold in this form at both LO \cite{wit77,dewit79} and 
NLO \cite{rossi84,fon92,grv92}, 
provided the right one or two loop splitting functions and
coupling constant $\alpha_{s}$ are used. 
Recently a full
NLO calculation of the heavy quark contribution to the photon structure
functions has been carried out \cite{lae94}. 

A peculiar characteristic of the photon structure is that there is no momentum 
sum rule constraining the photon parton distributions. In contrast
in the nucleon case the early DIS data showed that gluons should carry about
half of the nucleon momentum, thus fixing the size of $g^N(x,Q_0^2)$.
The reason is that in the quantum decomposition of
the real photon the hadronic states and thus the parton densities are of order
$\alpha_{em}$, while the bare photon state is of zeroth order. Thus any
variation, even large, in the parton densities can be reabsorbed in a
renormalization of the bare photon wavefunction \cite{vog94,sto94}. 

Quark and antiquark distributions are however constrained by the measurements of
$F_2^\gamma$, since in LO:
\begin{equation}
F_2^\gamma(x,Q^2) = \sum_{i=1}^{n_f} e_i^2 \left[ x q_i^\gamma(x,Q^2) + x {\bar
q}_i^\gamma(x,Q^2) \right]
\end{equation}

\begin{figure}[thb]
\begin{center}
\vspace*{-1.cm}
\epsfig{file=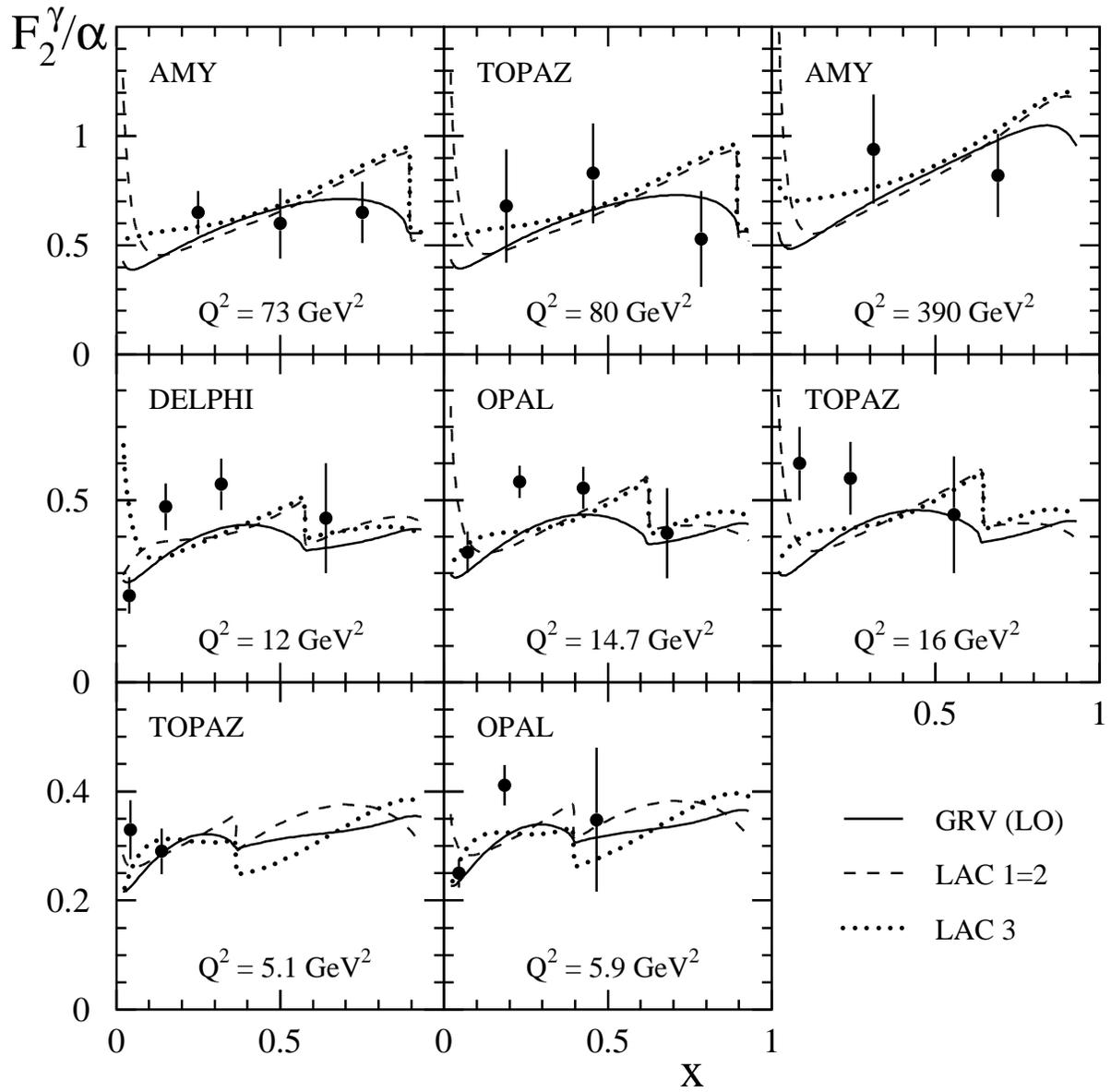,height=17cm}
\vspace*{-1.2cm}
\end{center}
\caption{Recent data on $F_{2}^{\gamma}(x,Q^2)$ from TRISTAN 
\protect\cite{f2topa94,f2amy95} and LEP \protect\cite{f2opa94,f2del95} 
compared to the LO fits to all previous data of LAC \protect\cite{lac} 
and GRV \protect\cite{grv}. Compilation from \protect\cite{ssv96}.}
\label{f2new}
\end{figure}

The first data on $F_2^\gamma$ came from experiments at the $e^+e^-$
colliders PETRA and PEP (for a review see \cite{mor94}). Most of the
available photon parton distributions have been parametrized from those data.
Later results have come from the TOPAZ
\cite{f2topa94} and AMY \cite{f2amy95} experiments at TRISTAN and from the
OPAL \cite{f2opa94,f2opa96} and DELPHI \cite{f2del95}
experiments at LEP. They are shown in Fig.~\ref{f2new}
compared to LO parametrizations obtained from previous data.
Most of the data points suffer from quite large errors. One of the limitations
are the low statistics in comparison with similar measurements of the nucleon 
structure functions. 
The reason is that the two-photon exchange implies a cross-section of order 
$\alpha_{em}^4$ instead of $\alpha_{em}^2$ as for lepton-nucleon DIS.
Another important contribution to the experimental errors is of a systematic 
nature. As we said, the measurement of $x$ requires that of $W_{had}$. 
Due to the limited acceptance of the detectors, particularly around the beam
pipe, one is not sensitive to the whole hadronic final state.
To correct for detector effects a model
of the hadronic final state is necessary and this is usually done with the help
of Monte Carlo generators. Unfortunately the unfolding of the true $W_{had}$ 
from the measured one is strongly dependent on the different models,
especially at low $x$. The
expected accuracy of the measurements which can be done at LEP2 in an enlarged
kinematic domain is certainly very desirable \cite{lep2proc}. 
The knowledge of the photon structure is of utmost importance at LEP2  for the
research of new physics, given that the $\gamma\gamma$ processes are an 
important source of background.

\begin{figure}[p]
\begin{center}
\epsfig{file=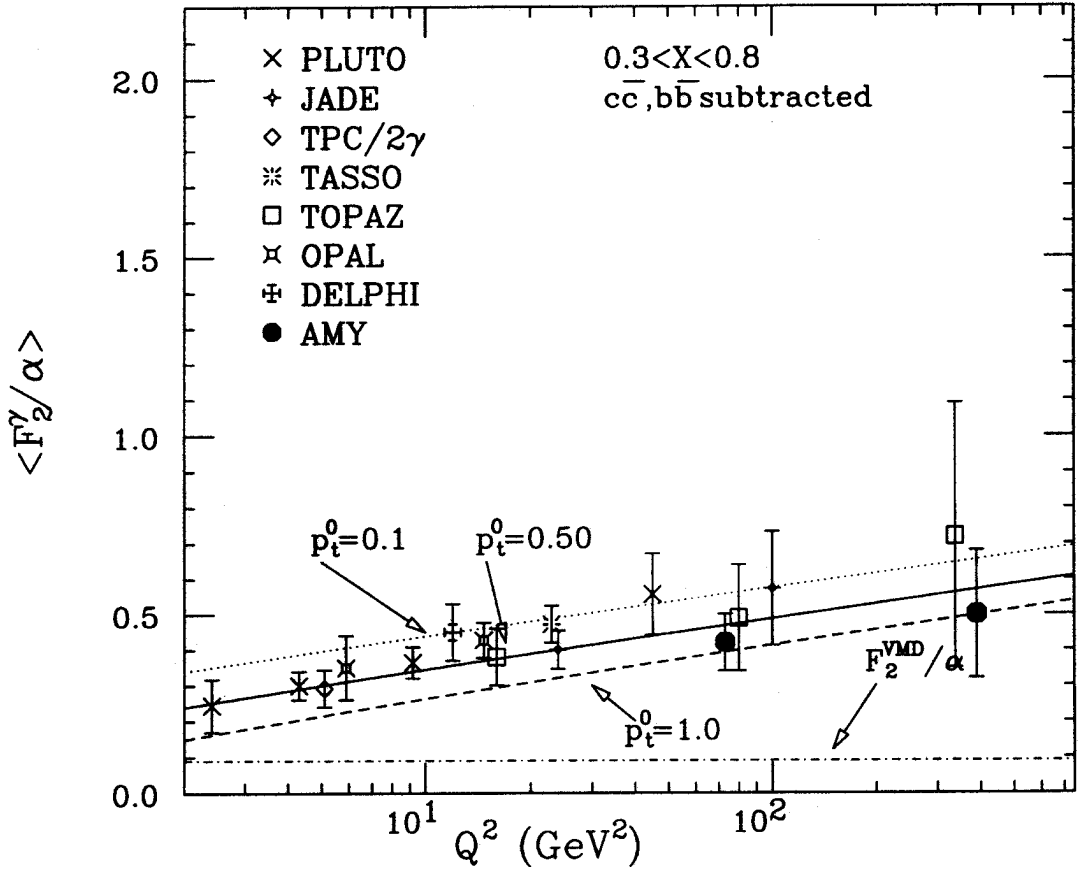}
\end{center}
\caption
{$Q^2$ evolution of $F_2^\gamma$ averaged over $0.3 < x < 0.8$ 
(from \protect\cite{nozaki95}). 
The lines are predictions of the FKP model \protect\cite{fkp86} 
with different phenomenological values of the cutoff parameter $p_t^0$: 
0.1 (dotted), 0.5 (solid) and 1.0 GeV (dashed). The VDM contribution is
represented by the dot-dashed line.}
\label{f2q2}
\end{figure}

The logarithmic increase of $F_2^\gamma$ with $Q^2$, predicted by
(\ref{equf2gama}), is clearly established from experimental data
in Fig.~\ref{f2q2}, 
where $F_2^\gamma$ is averaged in the large-$x$ region $0.3 < x <
0.8$. This behaviour is a proof of the anomalous component of the photon. To
account for the normalization a non-perturbative VDM component is also needed,
like in the FKP model \cite{fkp86} shown by the curves in the figure.

Most of the current sets of parton distributions of the photon follow the
approach of evolution equations. A deeper description can be
found in \cite{godb95}. 
The Drees and Grassie parametrization (DG) \cite{dg85} was obtained by LO
evolution equations starting from input parton distributions at $Q_0^2=$ 
1~GeV$^2$, to fit only one data point at that time available at 
$Q^2=$ 5.9~GeV$^2$. Its input consists of three quark flavours with densities 
assumed proportional to
the squared quark charges and flavour thresholds for charm and beauty crossed
through the $Q^2$ evolution. The input gluon is generated radiatively and does
not enter in the fit. 

A similar strategy was used in the parametrizations by Levy, Abramowicz
and Charchula (LAC) \cite{lac}, which were
based on the world data set available in 1991. However they included in the fit
the gluon distribution at the starting scale, demonstrating that data on
$F_2^\gamma$ do not constrain it significantly. This is connected with the lack
of a momentum sum rule for the parton distributions in the photon. Among the
LAC sets, LAC1 and LAC2 have a very soft gluon distribution, rising steeply at
low $x$, while LAC3 has a very hard distribution, 
with a maximum at $x \sim 0.9$. The
LAC parametrizations are LO and use four massless quark flavours. LAC3 has
already been excluded by data on jet production in $\gamma p$ and $\gamma
\gamma$ scattering.

The WHIT parametrizations \cite{whit} (after Watanabe, Hagiwara, Izubuchi and
Tanaka) follow a similar approach as
LAC, but treating correctly the charm threshold. This is more important
than in the nucleon case, because the photon can easily develop a charm content
through the anomalous splitting $\gamma \rightarrow c \bar c$. Six sets have been
provided, each one with a different assumption on the shape of the gluon
distribution. The total momentum carried by gluons is however fixed at about
half the momentum carried by quarks (at $Q_0^2 = 4$~GeV$^2$) in WHIT 1,2,3,
while gluons and quarks carry about the same momentum fraction in WHIT 4,5,6.

The Gordon-Storrow parametrizations (GS) \cite{gs} exist at both LO and NLO. 
Their first version starts
from a scale $Q_0^2 = 5.3$~GeV$^2$ with a VDM ansatz modified by the addition
of a pointlike term:
\begin{equation}
(q^\gamma,g^\gamma)(Q_0^2) = k \frac{4\pi\alpha}{f_\rho^2}(q^{\pi^0},g^{\pi^0})
(Q_0^2) + (q^{QPM},g^{QPM})(Q_0^2)
\end{equation} 
Here the structure of the vector meson is assumed equal to the pion structure.
The free parameters are the momentum fractions carried by gluons and sea quarks
in the pion, the constant $k$ and the light quark masses from the QPM term. In
the GS2 distribution $g^\gamma(Q_0^2)$ comes entirely from the first term, 
while in GS1 the second term also contributes through radiation from the 
pointlike quark component. 
Recently these distributions have been updated \cite{gsnew} 
by lowering their starting scale to 3~GeV$^2$,
including all available data on $F_2^\gamma$ and constraining the gluon from jet
production data at TRISTAN.

The GRV parametrizations (after Gl\"uck, Reya and Vogt) \cite{grv} 
exist both in LO and NLO and are built at a very low
starting scale: $Q_0^2 = $0.25~GeV$^2$ (LO), 0.3~GeV$^2$ (NLO). Such a low
value is considered the minimum for the applicability of perturbative QCD. 
The authors apply the idea of {\em dynamically} generated parton distributions,
which they had previously used to describe the nucleon and pion structure
\cite{grvhadold}.
The ansatz for
the nucleon was originally to have only valence quarks at the initial $Q_0^2$
scale. It was then modified introducing valence-like distributions of gluons
and sea quarks. For the pion no sea quarks are needed at the starting scale
$Q_0^2$ but a valence-like gluon density, 
proportional to the valence quark density. The GRV
approach has attained considerable success with HERA data on the proton $F_2$
structure function \cite{h1f2,zeusf2}, 
although it is not obvious that perturbative QCD implemented in
DGLAP equations works already at $Q^2 < $ 1~GeV$^2$. 
The GRV parametrizations for the photon start from a VDM input:
\begin{equation}
(q^\gamma,g^\gamma)(Q_0^2) = k \frac{4\pi\alpha}{f_\rho^2}(q^{\pi^0},g^{\pi^0})
(Q_0^2)
\end{equation} 
where only a normalization factor $k$ ($1\leq k \leq 2$) remains free,
accounting for $\rho/\omega/\phi$ mixing. 
In this way the gluon distribution at the scale $Q_0^2$ is constrained
to be equal to that of the pion. 

The AFG parametrization (after Aurenche, Fontannaz and Guillet) \cite{afg} 
is built in NLO and, like GRV, starts from a low scale 
$Q_0^2 =$ 0.25~GeV$^2$ assuming a VDM ansatz. However a coherent sum of 
$\rho,~\omega,~\phi$ is supposed. Pion distributions are taken from
\cite{aur89}. The authors choose the $\overline{MS}$
regularization scheme pointing out that the $DIS_\gamma$ scheme adopted by GRV
is more stable but includes process dependent terms in the parton distributions.

Schuler-Sj\"ostrand parametrizations \cite{sas} 
are quite recent and follow a similar approach as GRV and AFG.
They exist only in LO. Different sets are provided, including parton
distributions for the virtual photon.

\begin{figure}[p]
\begin{center}
\vspace*{-0.5cm}
\epsfig{file=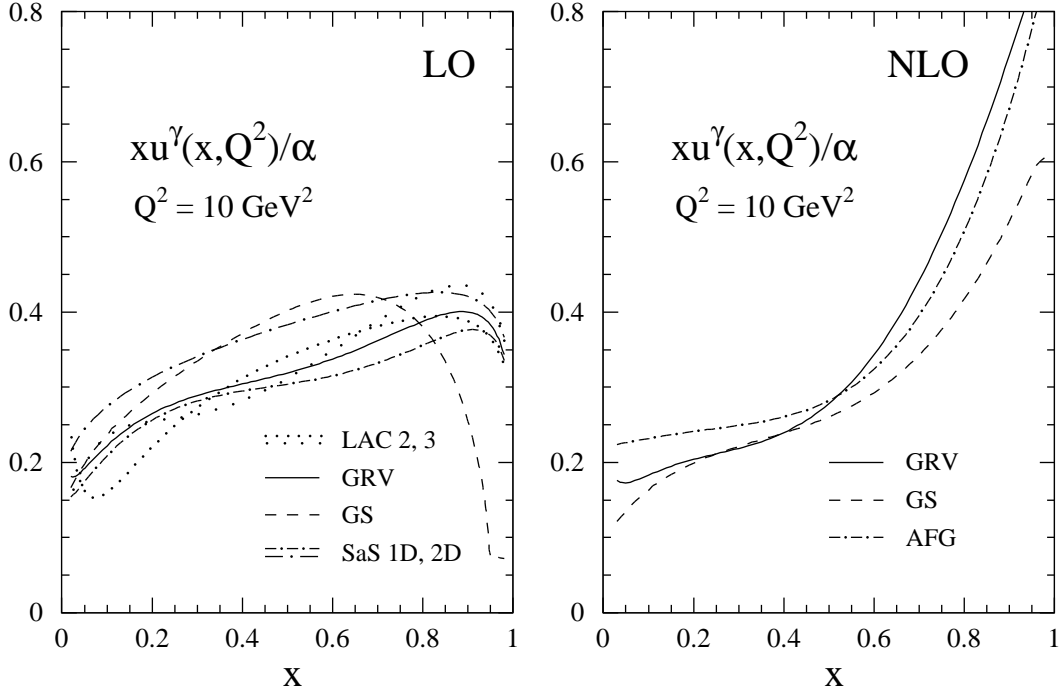,height=9.8cm}
\vspace*{-1.2cm}
\end{center}
\caption
{Parametrizations of the $u$-quark distribution in the photon at LO
\protect\cite{lac,grv,gs,sas} and NLO \protect\cite{grv,gs,afg}, from
\protect\cite{ssv96}.
The NLO results are given in the $\overline{\mbox{MS}}$ scheme.}
\label{qpara}
\end{figure}

Summarizing, the VDM ansatz allows one to constrain the gluon distribution in 
the photon, otherwise free due to the lack of a momentum sum rule. Unfortunately
with typical starting scales $Q_0^2 > $1~GeV$^2$ a pure VDM input is not able
to reproduce data at higher $Q^2$. The various parametrizations solve the
problem by adopting one of these two approaches:
\begin{itemize}
\item starting from a very low scale $Q_0^2$ with a pure VDM input, in a region
where perturbative QCD may be doubted;
\item keeping the scale at $Q_0^2 >$1~GeV$^2$ and using a more complex input 
with the addition of a pointlike component given by the Quark-Parton Model or 
simply guessed.
\end{itemize}
A comparison between the quark distributions from different sets is shown in
Fig.~\ref{qpara}.
There is reasonable agreement between them in the region $0.05 < x < 0.8$, 
where $F_2^\gamma$ data exist. 
Instead large differences appear at high $x$, for example comparing GS and GRV. 
The striking difference between the quark distribution in LO and NLO is
characteristic of the $\overline{MS}$ scheme, and this has 
been discussed in \cite{vog94}. It does not affect the final 
predictions on observable quantities like $F_2^\gamma$. 
It can however be avoided using the $DIS_\gamma$ scheme which is adopted 
by the GRV parametrizations \cite{grv92}.

\begin{figure}[p]
\begin{center}
\vspace*{-0.5cm}
\epsfig{file=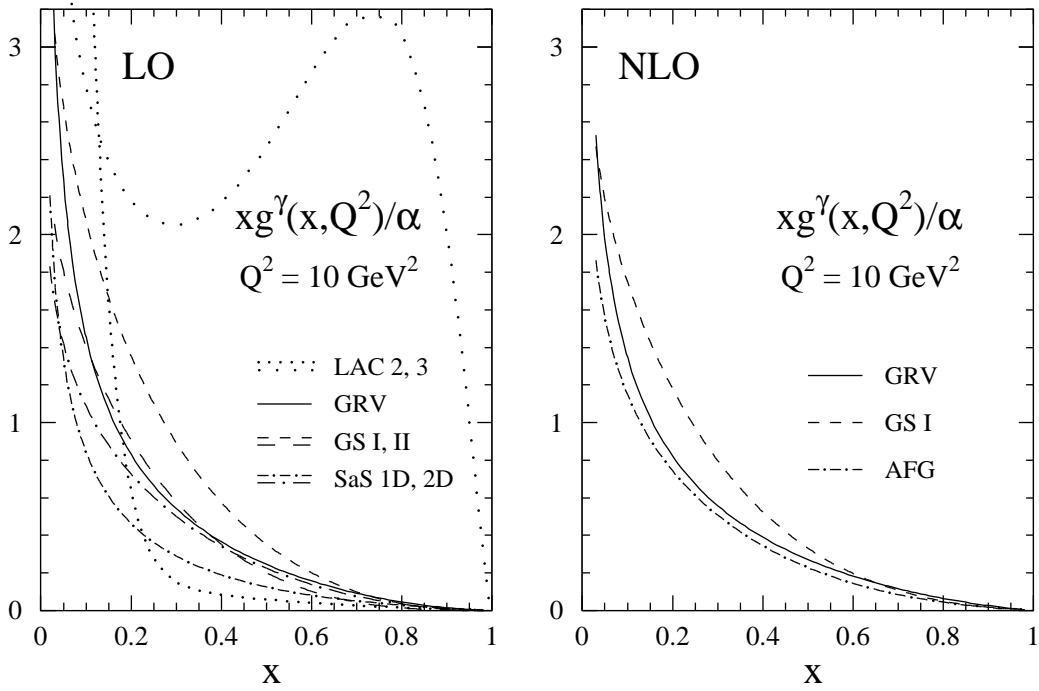,height=9.8cm}
\vspace*{-1.2cm}
\end{center}
\caption
{Parametrizations of the gluon distribution in the photon 
at LO \protect\cite{lac,grv,gs,sas} and NLO \protect\cite{grv,gs,afg}, from
\protect\cite{ssv96}.}
\label{gpara}
\end{figure}

Large differences between the parametrizations also occur at low $x$.
This is clearly visible in Fig.~\ref{gpara}, where different LO gluon 
distributions are compared. The LAC sets are much steeper at low $x$
than the others, demonstrating that the $e\gamma$ DIS data do not constrain 
the gluonic component of the photon.
At NLO the similarity of the curves is due to the
similarity of the models, it is not borne out by data.

\begin{figure}[thb]
\vspace*{-0.8cm}
\hspace*{-3.cm}
\epsfig{file=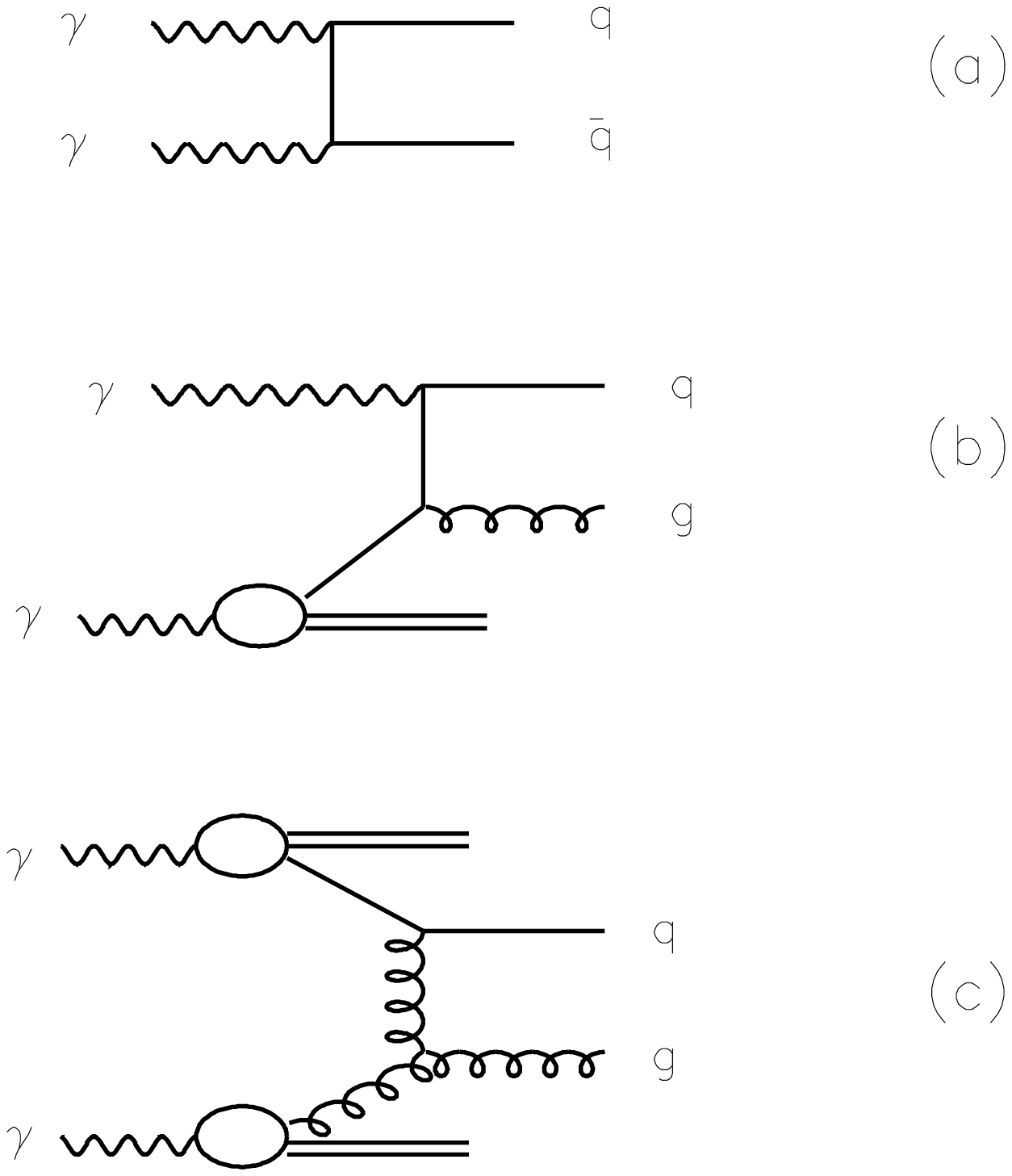,height=20.cm}
\vspace*{-1.2cm}
\caption
{Typical leading order diagrams contributing to jet production 
in two-photon exchange processes at $e^+ e^-$ colliders:
(a) direct photon; (b) single-resolved photon; (c) double-resolved photon}.
\label{qpmgg}
\end{figure}

\begin{figure}[p]
\begin{center}
\epsfig{file=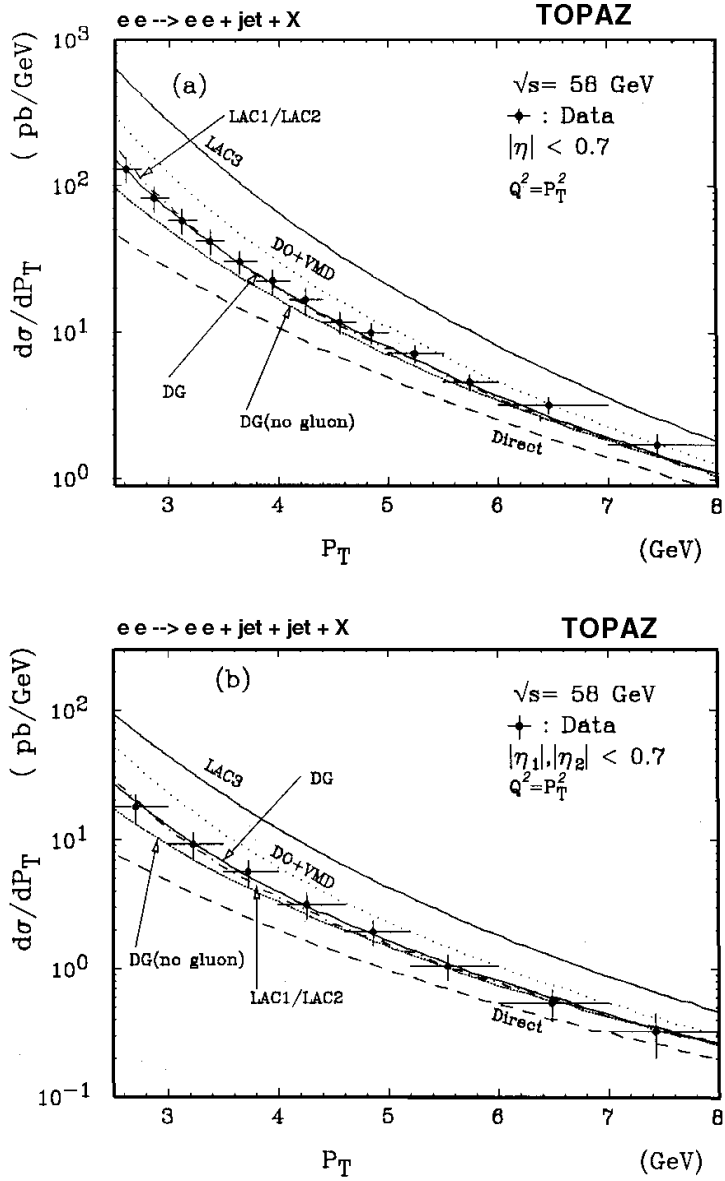,height=16cm}
\end{center}
\caption{Inclusive (a) jet and (b)  two-jet cross sections as a function 
of the jet transverse
momentum for the central pseudorapidity region $|\eta^{jet}| < 0.7$ from
\protect\cite{topajet}.
The contribution of the direct process only is represented by the dashed line.
The other curves are predictions including direct and resolved processes,
from different LO parametrizations \protect\cite{dg85,lac,dukow}. 
The expectation from the DG parametrization
excluding the contribution of gluons from the photon is also shown.}
\label{topaz2}
\end{figure}

The first experimental constraints on the gluon density in the photon have come
from $e^+e^-$ experiments at TRISTAN ($\sqrt{s}=58$~GeV) from the
study of $\gamma\gamma$ interactions, with both photons almost real. Here jet
production occurs in LO through the Quark-Parton Model process as in
Fig.~\ref{qpmgg}/a, 
or through {\em single-} or {\em double-resolved} processes 
as in Fig.~\ref{qpmgg}/b-c. The three kinds of processes
are all of the same order in the coupling constants since the photon parton 
distributions are \cal{O}$\left(\alpha_{em}/\alpha_s \right)$.
The first experimental evidence for a non zero gluon
content came from AMY \cite{amy92}. 
Single jet and dijet inclusive cross sections have then
been measured by both TOPAZ \cite{topajet} and AMY \cite{amyjet}. 
The results of TOPAZ compared to theoretical predictions are
shown in Fig.~\ref{topaz2}. 
The need for the hadronic component of the photon is evident when comparing
data to the expected cross section for direct photon only. Moreover it is clear
that there is a need for a gluon component in the photon. 
These data also rule out the LAC3 parametrization which would predict a far 
too high cross section.

\section{HERA as a $\gamma p$ collider}
\label{heragp}

The HERA storage ring at DESY (Hamburg, Germany) is the first electron-proton
collider ever built. Two different magnetic systems, respectively
superconducting for protons and conventional for electrons, drive the beams in
separate rings along a circumference of 6.3 Km. The design energies are $E_e=
30$~GeV, $E_p = 820$~GeV, giving a center of mass energy $\sqrt{s}=314$~GeV
(currently $E_e = 27.5$~GeV, so that $\sqrt{s}=300$~GeV).
This corresponds to an electron beam energy of 52 TeV against a fixed target.
The beams cross each other at zero angle in two interaction regions, 
occupied by the experiments H1 and ZEUS. 
The bunch crossing occurs every 96~ns. The
designed number of bunches in each beam is 210. To check beam related
backgrounds about 10 bunches are left unpaired in each beam. 

\begin{figure}[p]
\begin{center}
\epsfig{file=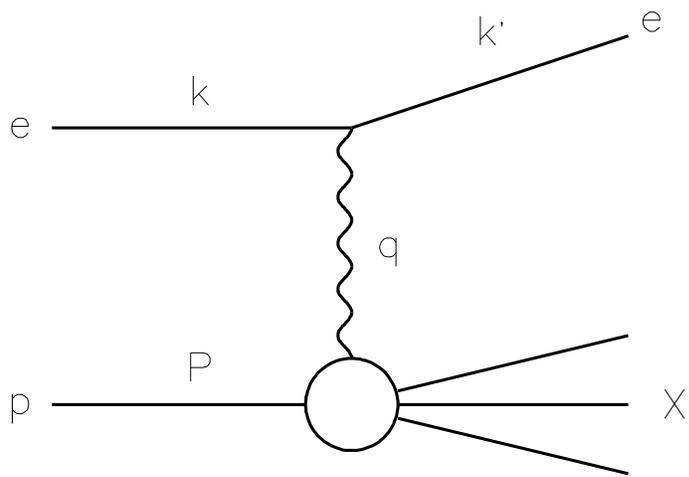,height=10cm,%
bbllx=80pt,bblly=140pt,bburx=490pt,bbury=510pt,clip=}
\end{center}
\caption{Kinematics of the semi-inclusive electron-proton scattering 
{\em $e p \rightarrow e X$}.}
\label{dis}
\end{figure}

The HERA kinematics is represented in Fig.~\ref{dis}.
The invariant kinematic variables of the semi-inclusive reaction 
$e p \rightarrow e X$ are:
\begin{eqnarray}
 Q^{2} &=& -q^{2} = -(k-k')^2  \\
 x &=& \frac {Q^{2}}{2 P\cdot q}  \\
 y &=& \frac{P \cdot q} {P \cdot k}  \\
 W^{2} &=& (P + q)^{2}
\end{eqnarray}
Two of them can be chosen as independent variables.
$Q^2$ is the virtuality of the exchanged photon. The photoproduction regime is 
defined by low $Q^2$, such that the photon is almost real. 
In this case we may regard HERA as a photon-proton collider, where the photons
are emitted almost collinearly with the electron beam.
The photon spectrum can be again evaluated with the Equivalent Photon
Approximation, factorizing the vertex $e \rightarrow e \gamma$ in the flux
factor of formula (\ref{wwaflux}) with the kinematic limits given by 
(\ref{q2minwwa}) and (\ref{q2maxwwa}), in which the variable $z$ is substituted
by $y$.
Due to the asymmetric energies of the HERA
beams the $\gamma p$ center of mass is boosted typically by two units of
pseudorapidity $\eta$, defined as:
\begin{equation}
\eta = \frac{1}{2} ln \frac{p+p_z}{p-p_z} = -ln \left( tan \frac{\theta}{2}
\right)
\end{equation}
This means that a particle going at $90^o$ in the $\gamma p$ center of mass
appears at about 15$^o$ in the HERA laboratory frame. In the following sections
we will refer to forward or backward (rear) directions intending respectively 
the proton and the photon directions. 

The photoproduction regime is experimentally defined in two ways:
\begin{itemize}
\item tagged photoproduction, when the scattered electron is detected at a very
small angle. This can be done in both ZEUS and H1 by dedicated calorimeters
placed near the beam line at several meters from the interaction point. 
The $Q^2$ acceptance extends in this case from $10^{-8}$~GeV$^2$ to 
0.02~GeV$^2$;
\item untagged photoproduction, defined by the antitagging condition that the
scattered electron is not detected in the main calorimeter. This definition
has been used by both the experiments. It corresponds to $Q^2 \leq $ 4~GeV$^2$,
with median $Q^2$ about 10$^{-4}$~GeV$^2$.
\end{itemize}

In contrast the region of $Q^2 \geq$ 4~GeV$^2$ is the traditional Deep
Inelastic Scattering (DIS) region, where the process in Fig.~\ref{dis} is
understood as a pointlike interaction between the electron and a quark
constituent of the proton. In this regime the structure of the proton can be
studied by exploiting the almost pointlike nature of the exchanged virtual 
photon. 
This region is experimentally defined by tagging the scattered electron
in the main calorimeter. The $x$ variable represents in this context the
momentum fraction of the struck quark relative to its parent proton. 

The meaning of $y$ is that of scaled photon energy: in the proton rest frame 
$0<y = E_\gamma/E_e <1$ and $W$ is the photon-proton center of mass energy
or equivalently the invariant mass of the hadronic final state. 
In the photoproduction regime these are related by $W \simeq \sqrt{s y}$. 
In comparison with previous photoproduction experiments,
carried out with fixed targets, HERA enlarges by one order of magnitude the
range of center of mass energies. Depending on the specific process the HERA 
experiments can span the whole range of $W$ between $30$~GeV and $300$~GeV
running at the nominal beam energies. 
Lower values of $W$ could be accessed by reducing the energy of the proton beam.

In the tagged mode $y$ (and hence $W$) is easily determined by measuring the
energy of the scattered electron $E_e'$ in the low-angle electron tagger by:
\begin{equation}
 y = 1 - \frac{E_e'}{E_e}
\end{equation}
In the untagged mode $y$ can be determined from the measurement of the hadronic
final state by the Jacquet-Blondel variable \cite{jbmeth}, defined as:
\begin{equation}
y_{JB} = \frac{\sum_{i=1}^{N_{cell}} (E - p_z)_i }  {(E - p_z)_{tot}} = 
\frac {\sum_i E_i(1-\cos \theta_i)}{2E_e}
\end{equation}
where the sum runs over all the calorimeter cells, $E_i$ being the energy
contained in the i-th cell at polar angle $\theta_i$
with respect to the proton direction.
$y_{JB}$ gives a systematically
lower value than the true $y$, mostly due to particles escaping detection
inside the beam pipe in the photon direction or inactive material 
in front of the calorimeter, and is corrected by Monte Carlo detector
simulation.

The final states produced in photoproduction events can be very diverse,
from hadron-like minimum bias $\gamma p$ interactions to (quasi)elastic
vector meson production, to the production of large $p_T$ jets and heavy
flavours. Detection of all those processes demands multipurpose detectors 
such as ZEUS and H1 with many capabilities and good overall performances.

\section{The experiments ZEUS and H1}
\label{detectors}

Both ZEUS and H1 have multipurpose detectors. Here only a brief description is
given, more details can be found in \cite{zeus,h1}. 

\begin{figure}[p]
\begin{center}
\epsfig{file=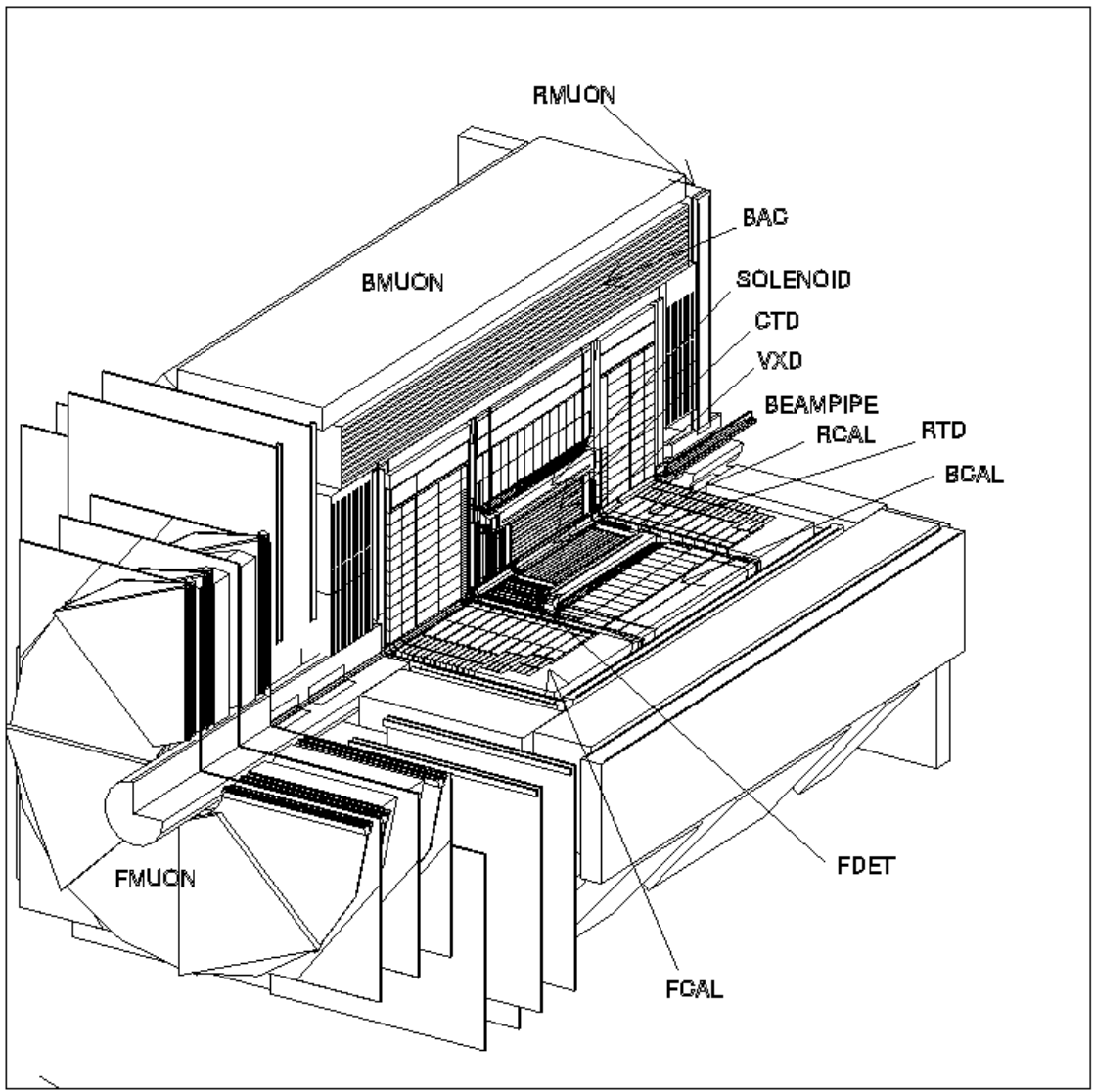,width=15.cm}
\end{center}
\caption{Schematic view of the central part of the ZEUS detector.}
\label{zeusdet}
\end{figure}

A longitudinal section of the central part of the ZEUS detector is shown in
Fig.~\ref{zeusdet}. Charged particles are
measured by the inner tracking detectors: the vertex detector (VXD), the
central tracking detector (CTD) and forward and rear tracking detectors
(FDET,~RTD). They operate in a magnetic field of $1.43$~T produced by 
a thin superconducting solenoid. 
The VXD and the CTD are concentric cylindrical drift chambers covering the 
angular region $15^o < \theta < 164^o$ 
(where $\theta$ is the polar angle with respect to the proton direction).
The measured resolution for tracks going through all the CTD layers is: 
$\sigma_{p_T}/p_T = 0.005 ~p_T$(GeV)$ \bigoplus 0.016$.

Surrounding the coil is the high
resolution uranium-scintillator calorimeter (CAL), mechanically divided into 
three parts, the
forward (FCAL), barrel (BCAL) and rear (RCAL) calorimeter. 
The calorimeter is compensating, the thickness of the 
uranium and scintillator tiles were optimized to achieve the best possible 
energy resolution for hadrons.
It covers the angular region: $2.6^o < \theta
< 176.1^o$, corresponding to $99.7 \%$ of the solid angle, with holes of $20
\times 20$~cm$^2$ in the center of the forward and rear calorimeters to
accomodate the HERA beam pipe. Each part of the calorimeter is subdivided into
towers of transverse size $20 \times 20$~cm$^2$
which are segmented longitudinally into one electromagnetic (EMC) and 
one or two hadronic (HAC) sections.
In the electromagnetic section the readout cells have finer granularity of
$5 \times 20$~cm$^2$ (in FCAL and BCAL) or $10 \times 20$~cm$^2$ (in RCAL). 
Each cell is read out by two photomultipliers.
Under test beam conditions the energy resolution is 
$\sigma_E / E = 0.18 / \sqrt{E}$ for electrons and 
$\sigma_E / E = 0.35 / \sqrt{E}$ for hadrons, with
$E$ in GeV. The time resolution is better than $1$~ns for energy deposits
greater than $4.5$~GeV.

Outside the uranium calorimeter a moderate resolution calorimeter (BAC), made
of iron layers interleaved with proportional tubes,
measures the tails of high energy jets and acts as a filter for muons. 
The muon detectors are also divided into three sections, covering the forward,
the barrel and the rear regions. In the barrel and rear regions large chambers
made of limited streamer tubes measure the position and the direction of muons
in front and behind the iron yoke (BMUON, RMUON). 
The yoke is magnetized to $1.6$~T to allow the momentum measurement for
penetrating muons. In the forward direction a spectrometer (FMUON), made of two 
iron toroids, drift chambers and planes of limited streamer tubes, 
identifies muons and measures their momenta.

Charged particles scattered at very small forward angles and carrying a
substantial fraction of the proton beam momentum are detected 
in the leading proton spectrometer (LPS). This is a set of six stations placed
along the beam line at distances from $20$ to $90$ meters from the interaction
point, using planes of silicon microstrip detectors very close to the proton 
beam. The track deflection induced by the magnets in the proton beam line is
used to measure the momentum. The LPS measures a leading proton for fractional
momentum $x_L \geq 0.4$ and $p_T \leq 1$~GeV.

The luminosity is measured from the rate of the Bethe-Heitler process $e p
\rightarrow e \gamma p$, whose cross section is large and calculable to
high precision. Two lead-scintillator calorimeters placed downstream the
electron beam pipe detect respectively the outgoing electron and the photon.
The electron calorimeter lies at $35$~m from the interaction point and accepts
electrons with energy between $7$ and $20$~GeV scattered at angles up to about
$5$~mrad.
The photon calorimeter is placed at $107$~m from the interaction point and
accepts photons produced at angles below $0.5$~mrad.
Bremsstrahlung events are tagged by the coincidence of the two calorimeters 
where the energies add up to the energy of the electron beam. 
The luminosity is calculated from the photon tagger alone.

\begin{figure}[p]
\begin{center}
\epsfig{file=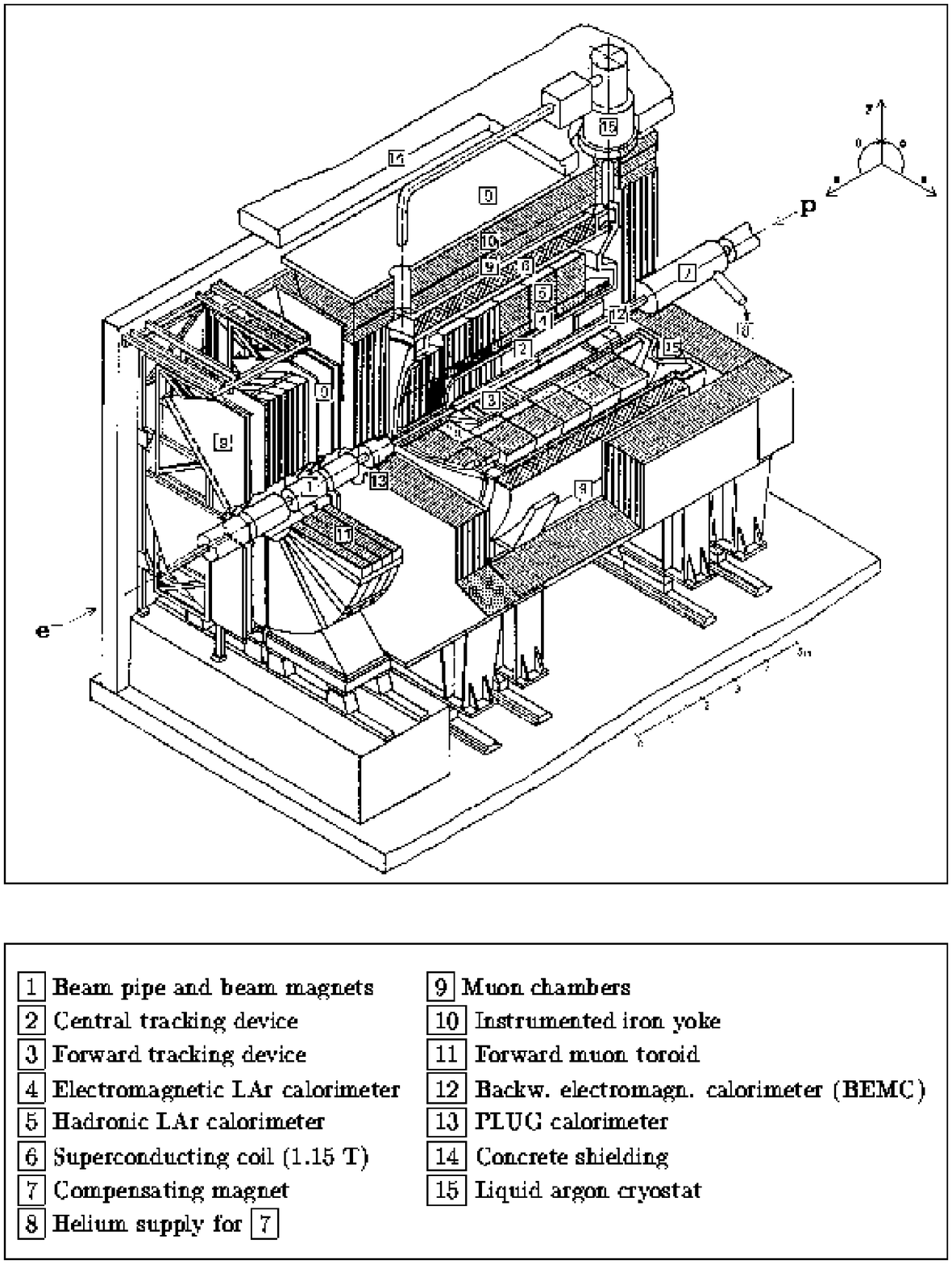,height=20cm}
\end{center}
\caption{Schematic view of the central part of the H1 detector.}
\label{h1det}
\end{figure}

Figure~\ref{h1det} shows a schematic view of the H1 detector. 
Charged particles are tracked by the central and forward tracking systems,
both consisting of drift and multiwire proportional chambers. The momentum
resolution achieved is $\sigma_{p_T}/p_T \approx 0.009 \cdot p_T
$(GeV)$~\bigoplus 0.015$ and the angular coverage is the same as for the ZEUS
detector.

The tracking region is surrounded by a fine grained liquid argon calorimeter
(LAr) with an electromagnetic section made of lead absorber plates and an
outer hadronic section with steel absorber plates. Its angular coverage is $4^o
< \theta < 153^o$. The energy resolution under test beam conditions
is $\sigma_E/E = 0.12/\sqrt{E}$ for electrons and
$\sigma_E/E = 0.50/\sqrt{E}$ for hadrons, with $E$ in GeV.
The calorimeter is not compensating and offline
corrections are applied according to the shower shape and energy. Its
segmentation varies as a function of the polar angle and is finer in the 
proton direction where, due to the asymmetry of the HERA beam energies, the
particle density is higher. In this region the transverse cell size is 
as small as $3 \times 3$~cm$^2$.
The backward region $155^o < \theta <  176^o$ is covered by another
calorimeter,
which was primarily intended for the detection of the scattered electron in deep
inelastic scattering. This consisted of a lead-scintillator sandwich with depth
of $22.5 X_0$ and energy resolution $\sigma_E/E = 0.10/\sqrt{E}$. 
From the 1995 data taking it has been replaced by a scintillating fiber
calorimeter with photomultiplier readout.

The calorimeters are contained inside a superconducting solenoid providing a 
uniform magnetic field of $1.15$~T for the inner tracking detectors.
Outside the coil there is the iron yoke, which is instrumented with LST planes 
for the muon identification. 
Further muon chambers in front and behind the iron complete the muon detectors
in the central region, while in the forward region muons are measured in a 
toroidal spectrometer similar to the one in ZEUS.
The luminosity is measured in a similar way as in ZEUS, from the coincidence 
rate of two TlCl/TlBr crystal calorimeters.

\begin{figure}[p]
\begin{center}
\epsfig{file=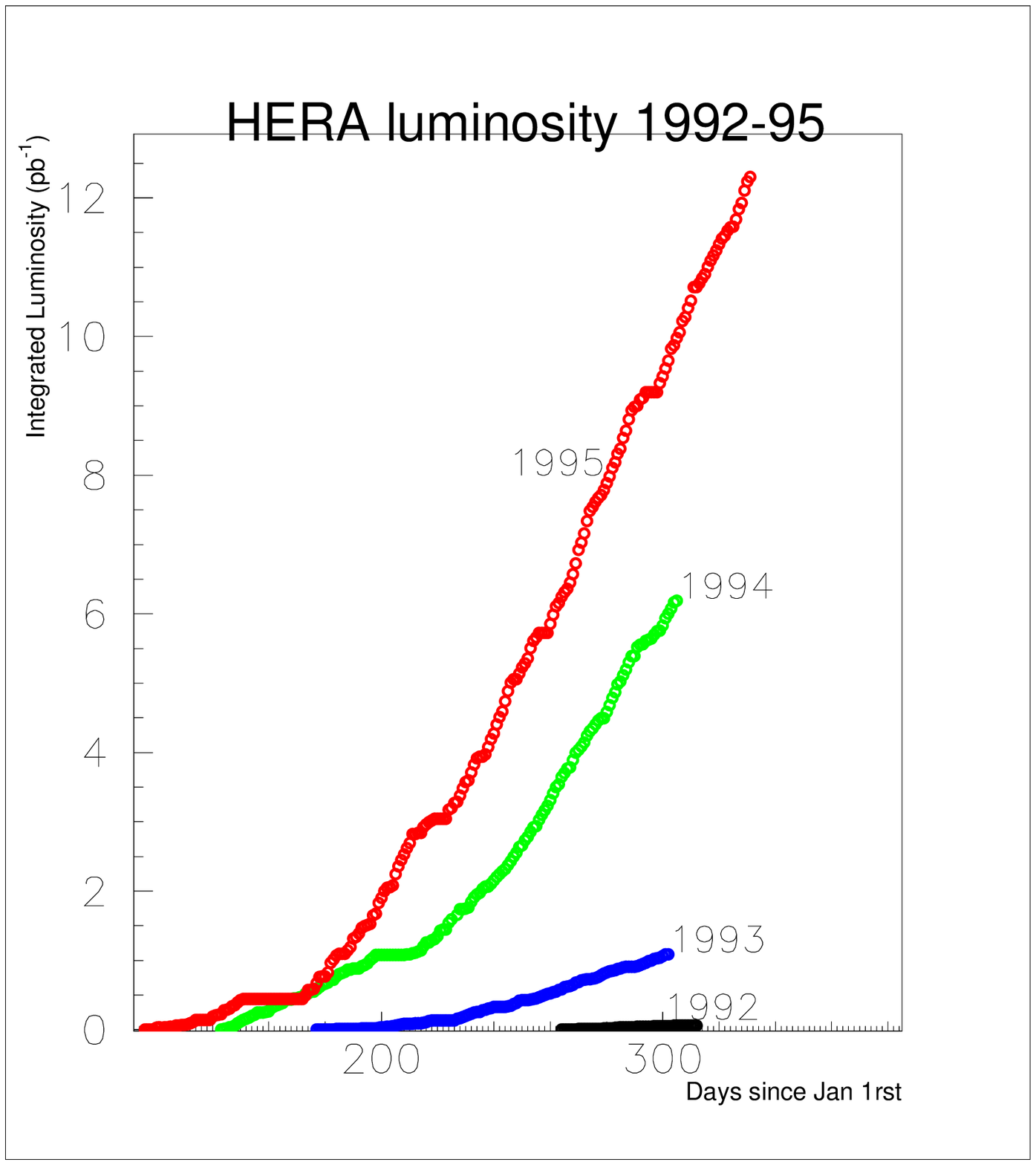,%
bbllx=74pt,bblly=170pt,bburx=515pt,bbury=670pt,clip=}
\end{center}
\caption{Integrated luminosity delivered by HERA as a function of the date of 
the year in the period 1992-1995.}
\label{lumi}
\end{figure}

HERA operation began in
summer 1992 with electron beam energy of $26.7$~GeV. Since 1994 positrons 
are run at energy of $27.5$~GeV. 
The luminosity has continuously increased from year to year. In Fig.~\ref{lumi}
the integrated luminosity delivered by HERA is shown as a function of time for
the period 1992-1995.
We review here the final results available at present, which have been obtained
analyzing data taken up to the 1994 run.

\section{Total $\gamma$p cross section and its decomposition.}
\label{totsgp}

The high energy behaviour of total hadronic cross sections is related to 
fundamental properties of particle interactions. 
$p \bar p$ colliders have provided data up to a center of mass energy
$\sqrt s =$~1.8 TeV, demonstrating the logarithmic rise of the cross section
\cite{ua1tot,e710,cdftot}.
Asymptotically this rise
cannot violate the Froissart-Martin bound \cite{froiss,mart},
which states that total cross sections
cannot grow faster than $ln^2(s)$, as a consequence of unitarity.
At HERA the $\gamma p$ cross section has been measured up to a center of mass 
energy of $200$~GeV, 
an order of magnitude higher than previous photoproduction
measurements from fixed target experiments \cite{baldini}.

The most popular approach to hadronic cross sections employs Regge theory ideas
\cite{collins}.
In this framework the simplest description is due to Donnachie and
Landshoff \cite{donnala}.
They have been able to fit all the available cross section data 
(for $\sqrt s > 5$~GeV) using a
parametrization inspired by Regge theory, of the form:
\begin{equation}
\sigma_{tot} = X s^\epsilon + Y s^{- \delta}
\label{dlform}
\end{equation}
The first term arises from pomeron exchange and the second from reggeon exchange
($\rho$, $\omega$, $f$, $a$). The exponents $\epsilon$ and $\delta$ are
effective powers, summing effects of multiple pomeron or reggeon exchanges.
Such effects are supposed to be small, therefore $\epsilon$ and $\delta$
are related to the intercepts of the pomeron and reggeon trajectories:
\begin{equation}
\alpha_{\Pma}(t) = \alpha_{\Pma}(0) + \alpha_{\Pma}^{'} \cdot t
\label{pomtraj}
\end{equation}
\begin{equation}
\alpha_{\cal R}(t) = \alpha_{\cal R}(0) + \alpha_{\cal R}^{'} \cdot t
\end{equation}
that is:
\begin{equation}
\epsilon = \alpha_{\Pma}(0) - 1 = 0.0808 
\label{dlpow}
\end{equation}
\begin{equation}
\delta = 1 - \alpha_{\cal R}(0) = 0.4525
\end{equation}
The pomeron has the quantum numbers of the vacuum, so that its coupling to a
particle $a$ and its antiparticle $\bar a$ are equal. Thus the coefficient X is
set equal for $\sigma (ab)$ and $\sigma (\bar a b)$. This is a way of enforcing
the Pomeranchuk theorem \cite{pome} in the fit. 
The universal rise of the cross sections with energy is interpreted by 
Donnachie and Landshoff as sign of the universality of something that 
is exchanged, the pomeron,
which is sometimes also referred to as the ``soft'' pomeron. This simple
parametrization becomes inconsistent at very high energies, violating the
Froissart-Martin bound. 
However this takes place only at $\sqrt s > 10^{24}$~GeV,
very far from the presently conceivable energies.

On the other hand several models have been based on perturbative QCD,
attributing the rise of the cross section to the onset of ``minijets''
\cite{minij,godb95}. 
In this context the total
inelastic $\gamma$p cross section is assumed to have a energy independent part,
coming from soft non-perturbative interactions, and a QCD part rising with
energy, due to the increase of gluons in the proton and in the photon at low
partonic energy fractions. Predictions based on this approach depend on the
parton distributions of the proton and the photon, the minimum
transverse momentum down to which the perturbative calculation is assumed to be
valid ($p_T^{min}$), and the probability for a photon to go into a 
hadronic state. These
models use the eikonal formalism to ensure unitarity \cite{forsh91} and
in general have a large number of unknown parameters.

\begin{figure}[p]
\begin{center}
\epsfig{file=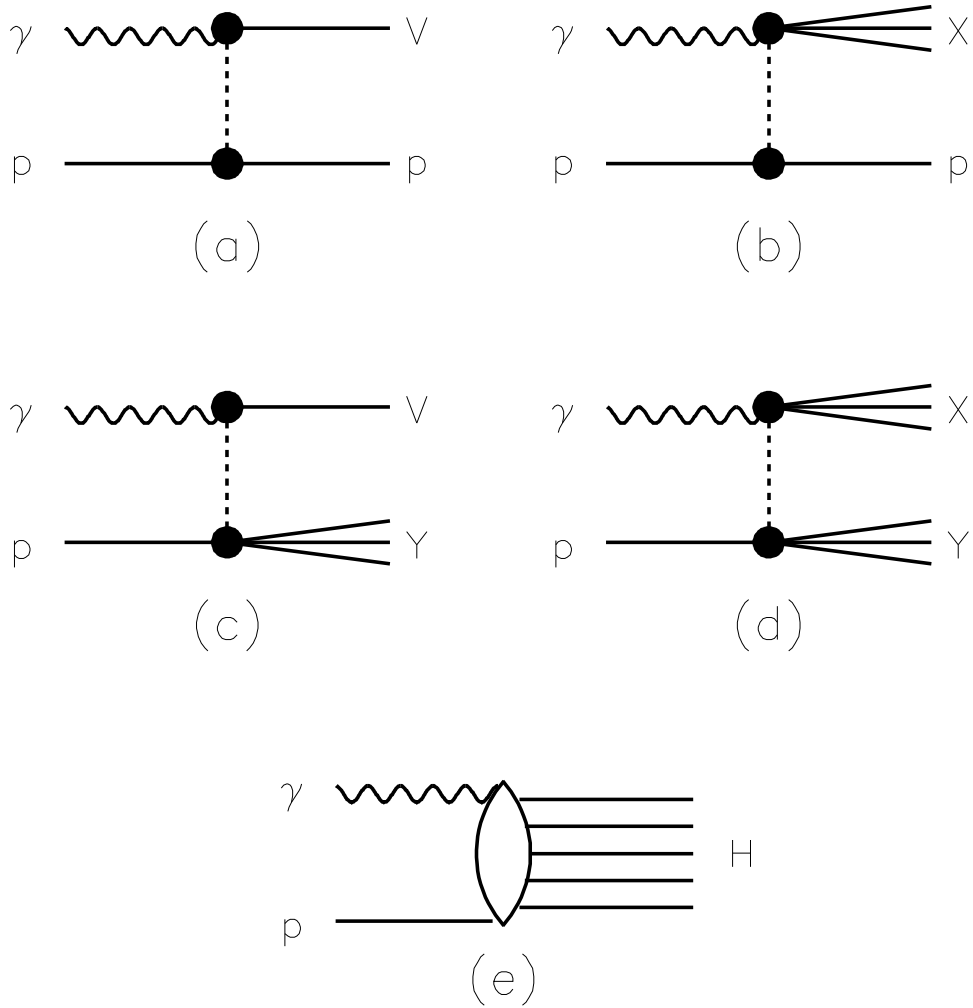,height=18cm}
\end{center}
\caption{Representative diagrams for the different classes of photoproduction
processes: (a) elastic vector meson production $\gamma p \rightarrow V p$; (b)
diffractive photon dissociation $\gamma p \rightarrow X p$; (c) diffractive
vector meson production with proton dissociation $\gamma p \rightarrow V Y$;
(d) double diffractive dissociation $\gamma p \rightarrow X Y$; (e)
non-diffractive inelastic scattering $\gamma p \rightarrow H$.}
\label{zoo}
\end{figure}

The total $\gamma p$ cross section is the sum of quite different processes,
represented in Fig.~\ref{zoo}. Let us first divide diffractive (\ref{zoo}/a-d)
from non-diffractive processes (\ref{zoo}/e). 
In diffractive processes there is no colour exchange between the proton and the 
hadronic state of the photon. In the figure they are represented to proceed 
through pomeron exchange. Assuming the validity of the VDM \cite{saku60} 
for the transition $\gamma \rightarrow V$ ($V = \rho$, $\omega$, $\phi$) 
the following four classes are identified:
\begin{itemize}
\item elastic vector meson production $\gamma p \rightarrow V p$
(Fig.~\ref{zoo}/a). The truly elastic Compton scattering $\gamma p \rightarrow
\gamma p$ has a cross section about two orders of magnitude smaller and can be
neglected.
\item photon diffractive dissociation $\gamma p \rightarrow X p$
(Fig.~\ref{zoo}/b). In this case the photon gives rise to a low mass hadronic
system X with the same quantum numbers, while the proton stays almost untouched
and is the leading particle in the final state.
\item proton diffractive dissociation $\gamma p \rightarrow V Y$
(Fig.~\ref{zoo}/c). The photon converts to a vector meson and the proton breaks
up into a low mass hadronic system Y.
\item double diffractive dissociation $\gamma p \rightarrow X Y$
(Fig.~\ref{zoo}/d). Here both the photon and the proton dissociate to low mass 
hadronic systems.
\end{itemize}
In all cases the transverse momentum exchanged is
generally low and the topology of the final state is characterized by large
rapidity intervals (or {\em gaps}) without particles.
This is expected if there is no colour flow between the interacting particles
and is exploited as an experimental signature.

Non-diffractive processes (Fig.~\ref{zoo}/e) are mostly soft ones,
similarly to minimum bias events at hadron colliders. In this case colour
exchange occurs and thus final state particles are distributed throughout the
available phase space without significant gaps. 

It should be clear that the different processes listed above are qualitatively
quite different from one another. 
Experimentally, to trigger and select inclusive photoproduction events at HERA 
one needs a coincidence between tagging of the
scattered electron in the low-angle electron taggers and activity in the main
apparatus, at least a charged track or a minimum energy deposition. 
The detection
efficiency for each of the photoproduction processes is quite different, thus
complicating the measurement of the total cross section. Introducing the
partial cross sections $\sigma_i$ (i=1,5) for the five considered classes of
reactions, with $\sigma (\gamma p \rightarrow X) = \sum_{i=1}^{5} \sigma_i$,
the observed number
of events $N_{ev}$ is given by:
\begin{equation}
N_{ev} = {\cal L}~\Phi_{\gamma /e}~A_e~\sum_{i=1}^{5} \epsilon_i \sigma_i
\label{sumsigmas}
\end{equation}
with $\cal{L}$ the integrated luminosity, $\Phi_{\gamma /e}$ the photon flux, 
$A_e$ the efficiency to tag the low angle electron and $\epsilon_i$ 
the efficiency of the
main detector for the {\it i}-th process. 
The photon flux is obtained by integrating expression (\ref{wwaflux}) 
in the kinematic region corresponding to the low-angle electron tagger.
The $\epsilon_i$ are calculated by Monte Carlo simulations of each different 
process. To reduce the systematic error it is important to determine 
separately each of the $\sigma_i$, otherwise one would have
to rely on model dependent assumptions in the Monte Carlo, such as the
relative magnitude of each contribution and its characteristics. 

The H1 experiment has measured the total $\gamma$p cross section 
\cite{h1stot} by short dedicated runs with special triggers in 1994.
They took $24~nb^{-1}$ data with nominal vertex position and the same amount of
data with vertex displaced by about 70~cm towards the forward part of the
detector.  In fact the acceptance for all the diffractive processes increases 
if the position of the interaction vertex is shifted in this way. 
The modest integrated luminosity is already enough due to the large 
photoproduction cross sections. 
The partial cross sections for the diffractive reactions are determined from 
fits to selected distributions, related to the rapidity gaps in the pattern of
the final state particles. This is done on the shifted vertex data.
The double dissociation cross section ($\sigma_{DD}$) is assumed to be in the 
reasonable range $0-40~\mu$b.
The non-diffractive cross section is then determined from
nominal vertex data, giving as input in (\ref{sumsigmas}) 
the diffractive partial cross sections previously determined.
The measured cross sections are summarized in table 1.
The elastic cross section 
$\sigma_{EL}$ is almost independent of the assumed value of $\sigma_{DD}$. 
Instead the single photon dissociation ($\sigma_{GD}$) and the single proton
dissociation ($\sigma_{PD}$) cross sections depend markedly on $\sigma_{DD}$, 
but for any chosen value of $\sigma_{DD}$ the photon dissociation cross section
is found to be larger than the proton dissociation one. The dominant systematic
error ($\sim 5~\%$) in the result is on the acceptance of the electron tagger. 
This depends strongly on the electron beam optics, particularly on the offset 
and the tilt of the beam in the horizontal plane.

\begin{figure}[p] \centering
\setlength{\unitlength}{1mm}
\begin{picture}(140,110)

\put( 50,88){\small \bf H1  }
\put( 50,80){\small \bf ZEUS }
\put( 50,72){\small \bf low-energy data }
\put(  73, -3){$W_{\gamma{p}}\,$ ({\bf GeV})}
\put( -1,+72){\begin{sideways}$\stotgp\,$ ($\mu${\bf b})\end{sideways}}

\epsfig{file=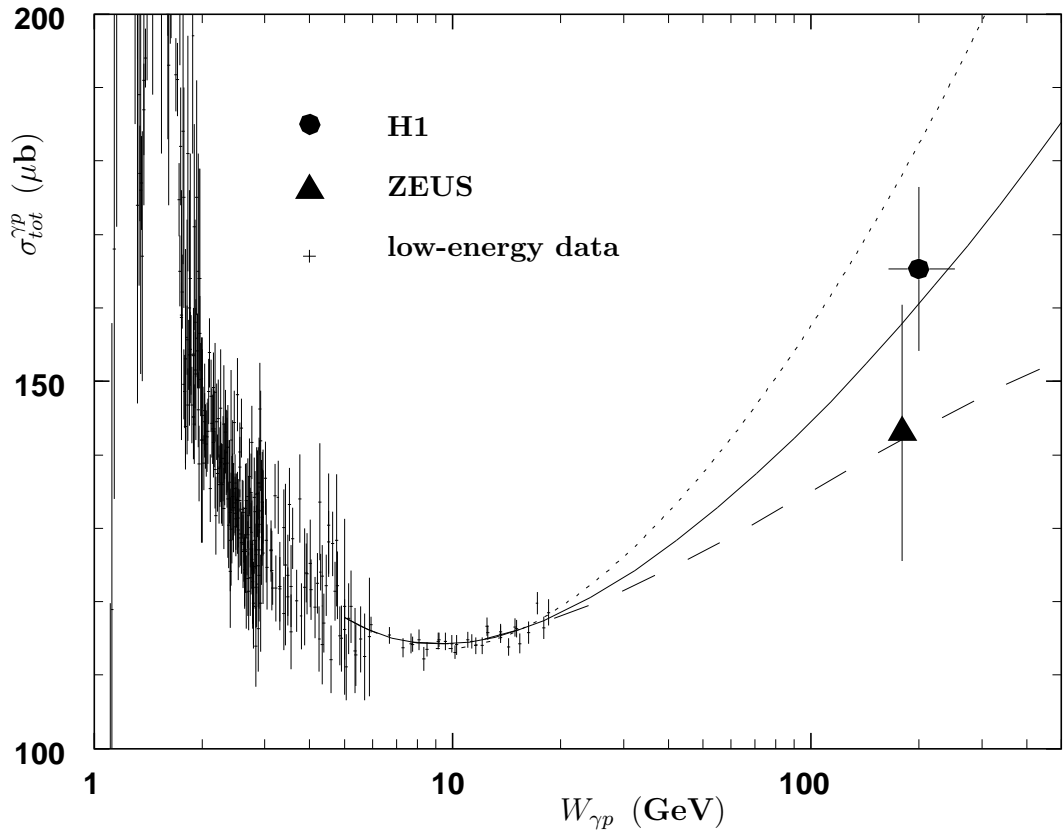,angle=90,width=14cm}

\end{picture}
\vspace{1cm}
\caption
{ Total photoproduction cross section measurements as function of the 
  $\gamma{p}$ centre of mass energy $W_{\gamma{p}}$. 
  Data are shown together with parametrizations which did not include 
  HERA measurements:
  the solid curve is DL \protect\cite{donnala} with default pomeron 
  intercept $\alpha_{\Pma}(0) = 1.08$, while the dotted one is DL with
  $\alpha_{\Pma}(0) = 1.11$ (which is obtained including in the fit the CDF 
  result \protect\cite{cdftot}); the dashed curve is ALLM \protect\cite{allm}.}
\label{stot}
\end{figure}

The total cross section as measured by H1 at an average c.m.s. energy $W \simeq
200$~GeV is shown in Fig.~\ref{stot}, together with previous data at lower
energies \cite{baldini} and a measurement from ZEUS \cite{zeustot}.
The curves are predictions of Donnachie-Landshoff (DL) \cite{donnala} and 
Abramowicz-Levin-Levy-Maor (ALLM) \cite{allm}
parametrizations, both of which did not include HERA data. The ALLM is based
on both $\gamma$p and $\gamma^\star$p data at fixed target experiments. It
allows a unified representation of photon-proton interactions with $Q^2$ from
$0$ up to the DIS regime. A DL fit
taking into account the CDF measurement \cite{cdftot} gives the exponent
$\epsilon = $~0.11 in (\ref{dlform}).
The HERA data are well accomodated in the soft pomeron phenomenology expressed
by these parametrizations, thus supporting the VDM and the universality of the
rise in the cross sections at high energy.

Minijet models give predictions spread over a larger range due to uncertainties
in the different parameters involved \cite{terr92}. 
The present HERA results can be brought into accordance with
calculations having a not too small $p_T^{min}$ ($\geq 2$~GeV) and a moderately
rising gluon density in the photon, but strong conclusions cannot be drawn.

In table 2
the diffractive partial cross sections measured by H1 are compared to
predictions of several theoretical models, based on different assumptions about
the structure and dissociation of the photon and the proton. CKMT \cite{ckmt95}
is a calculation using Regge theory plus absorptive corrections for all the
different processes. The measured cross sections are derived fixing
$\sigma_{DD} = 15~\mu$b, in agreement with the models. 
The elastic cross section is
correctly predicted by all the models. The photon dissociation cross
section is measured to be three times larger than the proton dissociation one, 
in agreement with predictions of CKMT. Instead SaS \cite{sas93} and GLM 
\cite{glm95} predict $\sigma_{PD} > \sigma_{GD}$.

\section{Elastic vector meson production}
\label{vmsection}

The elastic reaction $\gamma p \rightarrow V p$ (V = $\rho$, $\omega$, $\phi$,
$J/\psi$, \ldots) represented in Fig.~\ref{zoo}/a is equivalent to a hadronic
elastic interaction $V p \rightarrow V p$, assuming the validity of the VDM.
The optical theorem relates the total $V p$ cross section to the imaginary part
of the forward elastic scattering amplitude:
\begin{equation}
\sigma_{Vp}(s) = \frac{Im~T(s,t=0)}{s}
\end{equation}
where $t$ is the square of the momentum transfer.
Soft pomeron phenomenology thus gives straightforward predictions on the
elastic cross sections \cite{dolavm}. 
The elastic cross section is predicted to rise slowly
with the c.m.s. energy and to be dominated by forward scattering, with a sharp
peak at $t=0$ of roughly exponential behaviour:
\begin{equation}
\frac{d\sigma}{d|t|} \simeq A e^{- b |t|}
\label{eldsdt}
\end{equation}
The $b$-slope is related to the radius of the hadronic interaction and it can be
understood in the context of a simple geometrical model. 
In a fixed target configuration, for
small angle of scattering $t \simeq -(p \theta)^2$, where $p$ is the incident 
beam momentum and:
\begin{equation}
\frac{d\sigma/dt}{(d\sigma/dt)_{t=0}} = e^{- b |t|} \simeq 1 - b (p\theta)^2 
\end{equation}
In the classical optical diffraction of light from a circular aperture of
radius R the intensity of light is given by:
\begin{equation}
\frac{I}{I_0} \simeq 1 - \frac{R^2}{4} (k\theta)^2
\end{equation}
where $k$ is the wave number of the light. Comparing these two expressions one 
has a relation between the slope parameter and the radius of the interaction:
\begin{equation}
b = \frac{R^2}{4}
\end{equation}
For a typical radius of the strong interaction $R = 1/m_\pi$ (with $m_\pi$ the
pion mass) the last expression gives $b \simeq 12.5$~GeV$^{-2}$. 
This is in the measured range of values for
elastic scattering of hadrons at high energy. The $t$ spectrum is
predicted to exhibit a dip followed by a secondary diffraction maximum, as for
light diffraction. This is in fact
observed in high energy $p p$ and $p \bar p$ elastic scattering
\cite{goul83}. Regge theory predicts a weak dependence of the $b$-slope on the
center of mass energy, referred to as shrinkage, of the kind:
\begin{equation}
b = b_0 + 2 \alpha'_{\Pma} ~log\frac{s}{s_0}
\label{shrink}
\end{equation}
This effect actually allows the determination of the slope of the pomeron
trajectory in (\ref{pomtraj}): $\alpha'_{\Pma} \simeq 0.25$~GeV$^{-2}$. 

Fixed target experiments have studied the elastic vector meson photoproduction
up to $W \approx 20$~GeV. 
Light ($\rho$, $\omega$, $\phi$) vector meson data \cite{bauer78,fixvm} 
have been found consistent with soft pomeron phenomenology and the VDM.
Elastic $J/\psi$ production \cite{fixelpsi} cross sections were overestimated 
by VDM by more than an order of magnitude, 
but other predictions like the energy dependence agreed with data
within the errors.

HERA experiments have measured the elastic reaction $\gamma p \rightarrow V p$
for V = $\rho$ \cite{zeusrho,h1rho,zeuslps},
 $\omega$ \cite{zeusomega}, 
 $\phi$ \cite{zeusphi},
 $J/\psi$ \cite{h1psi93,zeuspsi93,h1psi94,zeuspsi94}.
These measurements are complemented by similar ones carried out in the DIS
regime \cite{zeusvmdis,h1vmdis}.
The reconstruction of the vector meson decays:
\begin{displaymath}
\rho^0 \rightarrow \pi^+ \pi^-
\end{displaymath}
\begin{displaymath}
\omega \rightarrow \pi^+ \pi^- \pi^0
\end{displaymath}
\begin{displaymath}
\phi \rightarrow K^+ K^-
\end{displaymath}
\begin{displaymath}
J/\psi \rightarrow e^+ e^-, \mu^+ \mu^-
\end{displaymath}
has allowed the measurements at W from 40 to 100 GeV for light mesons, 
from 30 to 150 GeV for $J/\psi$. 
One of the hardest experimental tasks is to trigger on such elastic events. 
In fact both the
outgoing electron and the proton are generally not tagged. Then only the decay
products of the meson decay can be observed in the detector. The typical
topology of these events is of two charged tracks 
in the photon hemisphere (plus an electromagnetic calorimeter cluster due to the
$\pi^0$ in $\omega$ decays).
The acceptance in W is determined by the detector geometry and trigger.

In tagged photoproduction events the $\rho$ production has also been measured 
without reconstruction of the decay at $\langle W \rangle \simeq 200$~GeV, 
from fits to distributions of the hadronic final state
by comparison to Monte Carlo distributions for $\rho$ production 
\cite{zeustot,h1rho}. At these high W values the decay
pions are boosted in the photon direction outside the acceptance of the
tracking detectors and their reconstruction is carried out by the calorimeter. 
In this case the worse spatial and energy resolutions have prevented
measurements of differential distributions.

The kinematic variables $W$ and $t$ can be easily determined 
from the vector meson, reconstructed from its decay products, in the limit of
$Q^2 \approx 0$. In fact in this case $(p_z)_\gamma \simeq -E_\gamma$ and:
\begin{equation}
W^2 \simeq 4 E_\gamma E_p \simeq 2 (E-p_z)_V E_p
\end{equation} 
\begin{equation}
t \simeq -(p_T^2)_V
\end{equation} 
where the subscript $V$ indicates the vector meson.
The main background to the elastic reactions comes from events where the proton 
dissociates (fig.~\ref{zoo}/c). This can be 
reduced by requiring a very low energy deposit in the forward part of the 
calorimeter ($E < 1$~GeV). The residual background is
estimated studying the effect of changing this cut in comparison to Monte Carlo
simulations of the relevant diffractive process. In the final samples it is
generally $10-30~\%$ and is statistically subtracted.

\begin{figure}
\vspace{-1.0cm}
\begin{center}
\epsfig{file=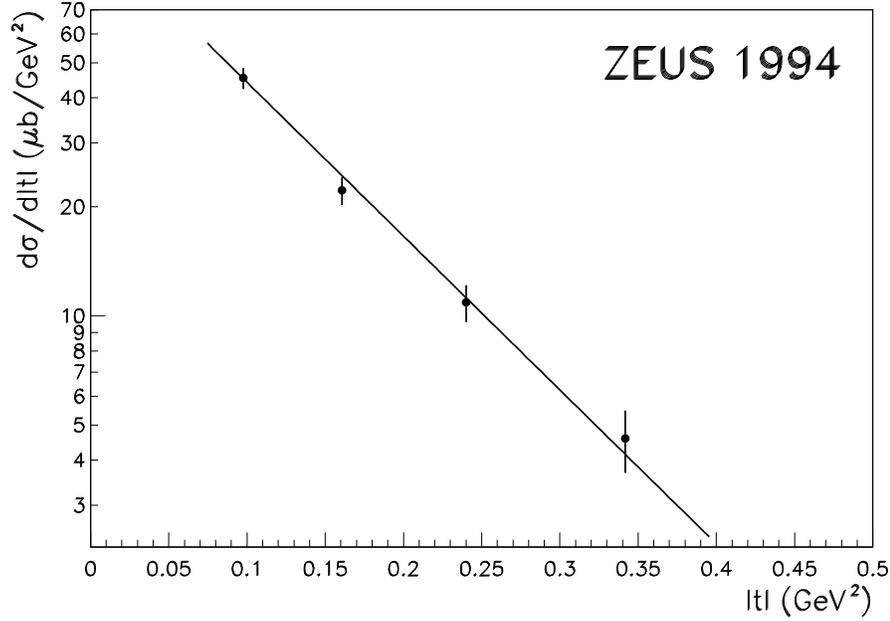,width=13cm}
\end{center}
\vspace{-1cm}
\caption{Differential cross section $d\sigma/d|t|$ 
for elastic $\rho^0$ photoproduction, $\gamma p \rightarrow \rho^0 p$,
at $\langle W \rangle=73$~GeV, as obtained by ZEUS through direct 
measurement of the scattered proton.   
The line is the exponential fit described in the text.}
\vspace{2cm}
\label{lpsslope}
\end{figure}

Proton dissociation can be completely avoided by detecting the leading proton
scattered at a very small angle. This has been done by ZEUS using the leading
proton spectrometer (LPS), 
which directly measures the momentum of the leading proton with
high accuracy (with the configuration the detector had in 1994 run:
$0.4\%$ on the longitudinal component, $\approx 5$~MeV on $p_T$). For truly
elastic events the momentum of the proton is reduced by less
than $0.2\%$ and thus events are selected with $0.98 < x_L < 1.02$, 
where $x_L = p'/p$ is the ratio between outgoing and incoming proton momentum
 ($x_L$ can be measured greater than unity due to the experimental resolution).
One is thus guaranteed that the proton scatters elastically and $t$ is 
directly measured. Owing to the energy-momentum
balance the $Q^2$ can also be determined in this case. 
In fact in the elastic reaction the knowledge of the $p_T$ of the vector meson 
and of the scattered proton allows a determination of the $p_T$ of the untagged 
scattered electron, and from this the $Q^2$:
\begin{equation}
Q^2 = \frac {p_T^e}{1 - y}
\end{equation}
where $y \simeq W^2/s$.
The measured $d\sigma/dt$ for elastic $\rho^0$ production is shown in 
Fig.~\ref{lpsslope}.
A fit of the form (\ref{eldsdt}) gives $b = 9.8
\pm 0.8$~(stat.)~$\pm 1.1$~(syst.)~GeV$^{-2}$, with dominant systematic
uncertainties on the LPS acceptance and the beam spread in the transverse
momentum.

\begin{figure}[htbp]
\vspace*{-3.0cm}
\begin{center}
\epsfig{file=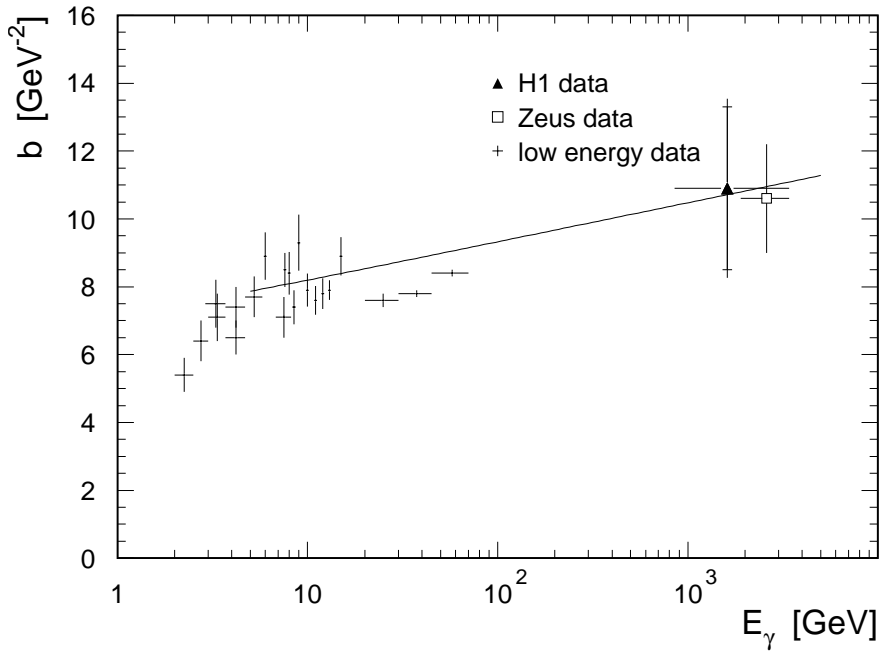,width=13cm}
\end{center}
\vspace{-1cm}
\caption {
Exponential slope parameter $b$ as a
function of $E_{\gamma}$, the photon energy in the rest frame of the
proton. The curve is a prediction based on pomeron exchange 
\protect\cite{sas93}.}
\label{h1rhoslope}
\end{figure}

HERA measurements of the slope for $\rho$ elastic photoproduction are
shown in Fig.~\ref{h1rhoslope} in comparison with fixed target results.
They agree with a slow shrinkage of the diffractive peak as predicted by the 
soft pomeron formula (\ref{shrink}). This is also true for the measurements of
the other light vector mesons, summarized in table \ref{slopes}.
Recalling that $b$ is related to the radius of the interaction, the values in
table~\ref{slopes} indicate a smaller radius for $\phi p$ as compared to $\rho
p$ interaction.

\begin{figure}
\begin{center}
\epsfig{file=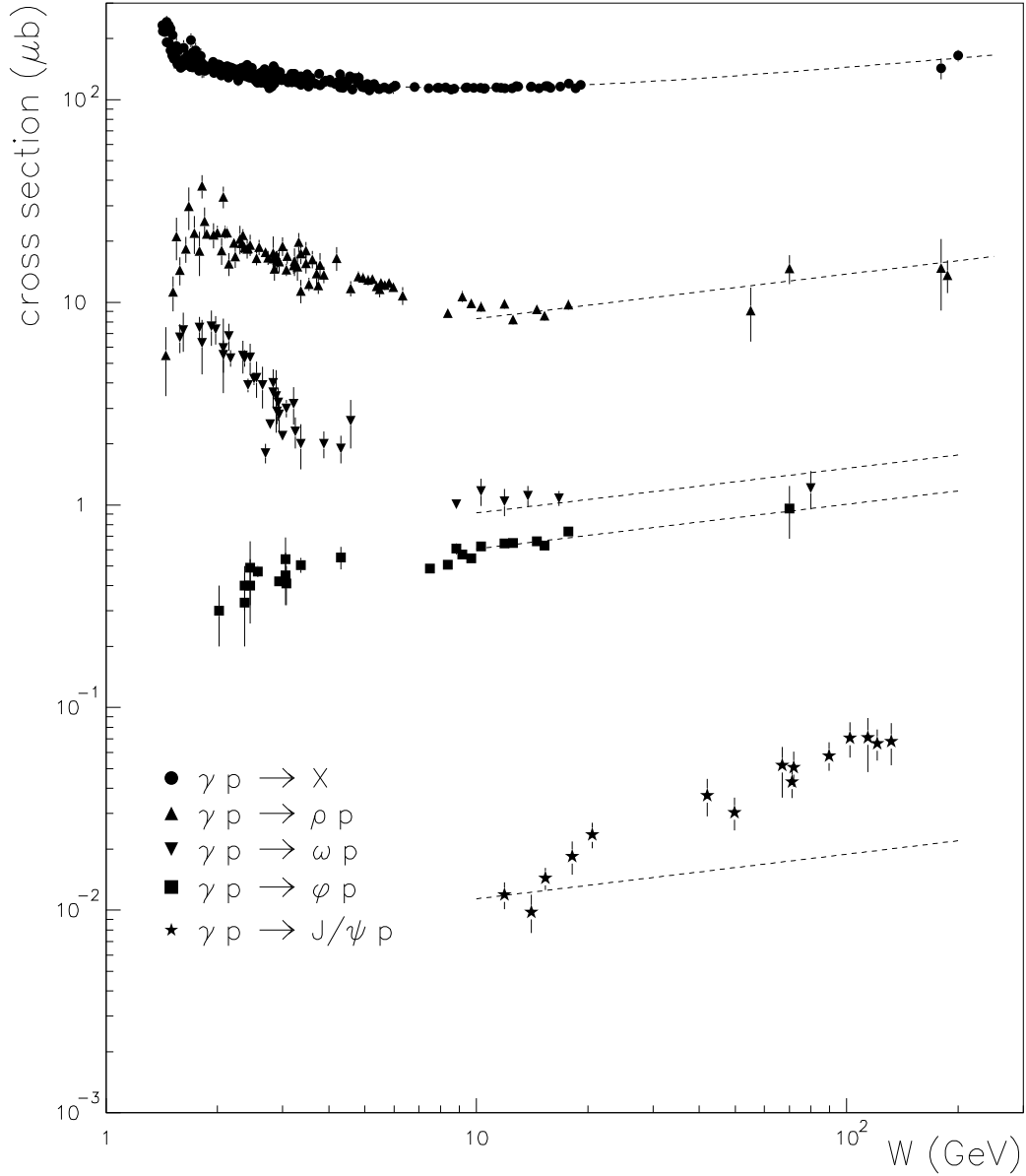,width=15cm}
\end{center}
\caption {
Elastic vector meson photoproduction cross sections for $\rho^0$, $\omega$,
$\phi$, $J/\psi$ as function of the $\gamma p$ center of mass energy.
For comparison the total
photoproduction cross section measurements are shown in the upper part of the
plot.
The data points at $W \geq 40$~GeV are HERA measurements, the lower energy
points come from fixed target experiments.
The dashed curves are expectations from the soft pomeron model
\protect\cite{dolavm}, arbitrarily normalized at fixed target points in the
elastic cross sections.}
\label{mesonitutti}
\end{figure}

The integrated elastic cross sections for $\rho$, $\omega$, $\phi$, $J/\psi$ 
photoproduction measured at HERA as functions of $W$ are shown in 
Fig.~\ref{mesonitutti} together with
data from fixed target experiments. In the soft pomeron model \cite{dolavm}
one expects:
\begin{equation}
\sigma_{\gamma p \rightarrow V p}(W) \propto \frac {W^{4 \epsilon}}{b(W)}
\simeq W^{0.22}
\end{equation}
where $\epsilon$ is given by (\ref{dlpow}) and $b(W)$ is given by 
(\ref{shrink}) with $s=W^2$. 
This is in good agreement with data on light vector mesons.
The same model predicts for the total photoproduction cross
section a dependence $W^{2 \epsilon} \simeq W^{0.16}$, which also agrees with 
data, as shown. 

Another characteristic of the elastic vector meson production at fixed target
energies is the so-called s-channel helicity conservation (SCHC). For real or
quasi-real photons it means that the transverse polarization of the photon is
transferred to the vector meson. The polarization of the vector meson can be
experimentally determined from the angular distributions of the decay products, 
which depend on the spin-density matrix elements \cite{schil70}. 
The analyses reveal in all cases that SCHC continues to hold at the 
HERA energies. 

\begin{figure}[p]
\hspace{2cm}
\epsfig{file=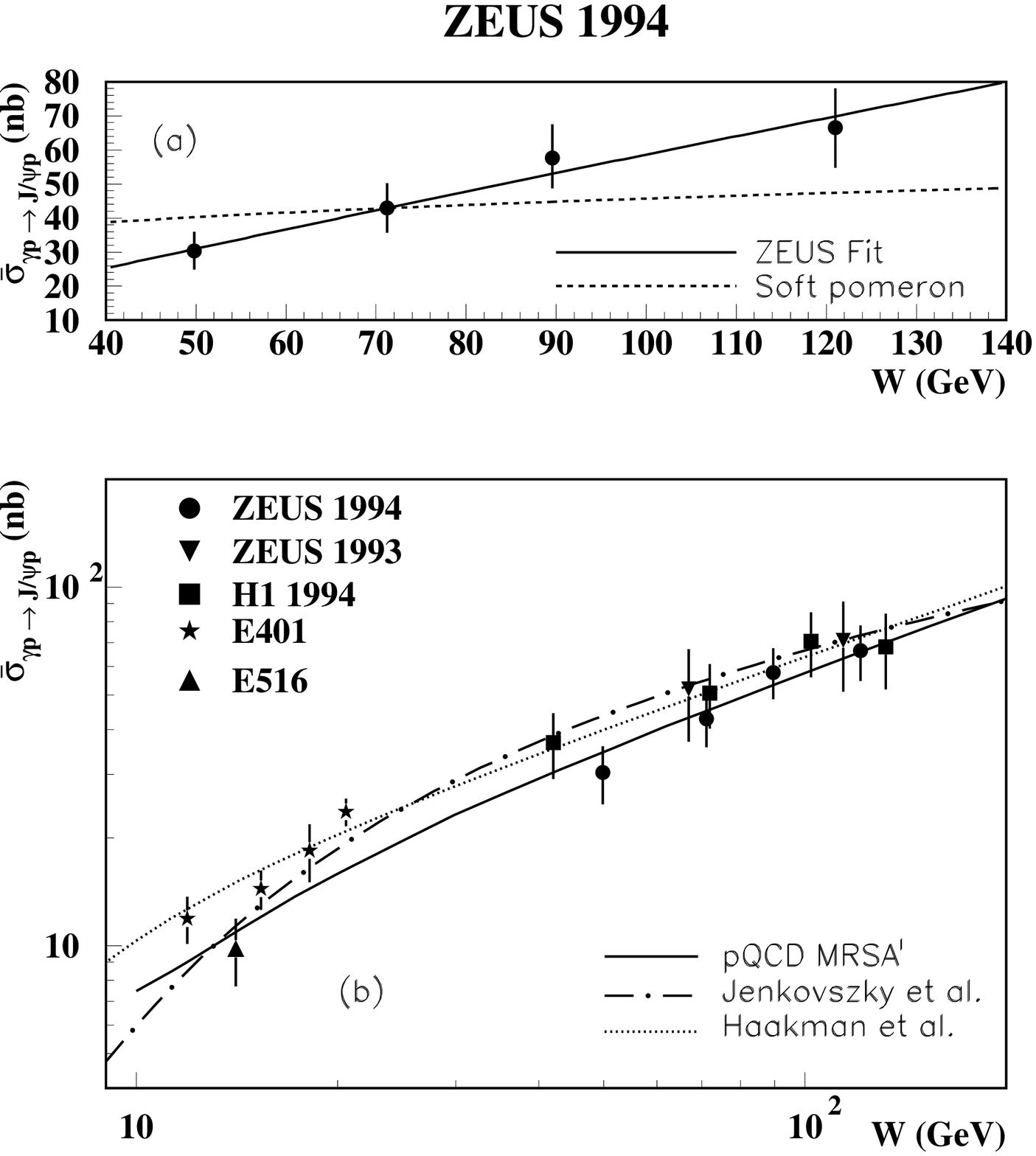,width=15cm}
\caption{
The elastic $J/\psi$ photoproduction cross section as a function of the 
$\gamma p$ center of mass energy $W$. In (a) the ZEUS data alone are fitted to
a dependence $W^\delta$ with $\delta=0.92$; the dashed line shows the
prediction of a soft pomeron model \protect\cite{dolavm} in which $\delta
\approx 0.22$.
In (b) HERA data are shown together with results from fixed target experiments.
The full line is the result of a perturbative QCD calculation
\protect\cite{rysk} using MRSA'
proton parton distributions, the dotted and the dash-dotted lines are results 
of two pomeron models \protect\cite{haa96,jen96}.}
\label{zeuspsicros}
\end{figure}

\begin{figure}[p]
\begin{center}
\epsfig{file=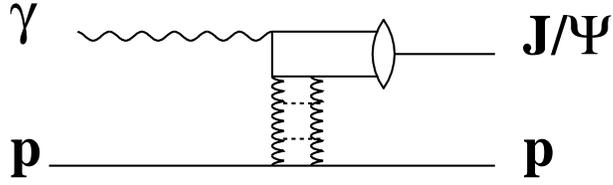,width=8cm,
bbllx=87pt,bblly=460pt,bburx=523pt,bbury=600pt,clip=}
\end{center}
\caption{Diagram representing elastic $J/\psi$ production
according to QCD-inspired models with
the exchange of a gluon ladder.}
\vspace{2cm}
\label{ladder}
\end{figure}

This homogeneous situation changes dramatically when one turns to the elastic
production of $J/\psi$. Coming back to Fig.~\ref{mesonitutti}, 
the $J/\psi$ elastic
cross section is steeply rising with $W$ and the Donnachie-Landshoff model 
fails, predicting a slow rise as for the light vector mesons. 
In Fig.~\ref{zeuspsicros}/a only the ZEUS data
are shown. The fit with a function $W^\delta$ gives 
$\delta = 0.92 \pm 0.14 \pm 0.10$. Thus even in this restricted
range the soft pomeron model is excluded by data.
Instead other two pomeron models are found to give a good description of the 
total cross section \cite{haa96,jen96} (Fig.~\ref{zeuspsicros}/b). 
They involve either a scale dependent pomeron intercept or scale dependent 
pomeron couplings and mass threshold effects.
The data can also be reproduced by models
based on perturbative QCD \cite{rysk,brod94,nnzak94}, 
where the pomeron is considered as a system
of two interacting gluons, or gluon ladder, as in Fig.~\ref{ladder}. 
In these models 
the cross section is proportional to the square of the gluon density 
in the proton. For a singular gluon density at small $x$, rising like:
\begin{equation}
 x^{-\lambda}  ~~~as~  x\rightarrow 0
\label{xpower}
\end{equation}
as implied by HERA $F_2$ measurements \cite{h1f2,zeusf2}, 
a steep rise is predicted with increasing $W$. 
HERA $F_2$ data are well described by the MRS(A') parametrization \cite{mrsap}, 
which corresponds to a power $\lambda \approx 0.2$ in (\ref{xpower}), 
and by the GRV parametrization \cite{grv95} corresponding to 
$\lambda \approx 0.3-0.4$.
In the Ryskin model \cite{rysk} 
the effective energy scale $\bar q$ and momentum fraction $\bar x$ at which
the gluon distribution is tested are given by:
\\
\begin{equation}
{\bar q}^2 = \frac{Q^2 + m_{J/\psi}^2 - t}{4}
\label{ryskinq}
\end{equation}
\begin{equation}
\bar{x} = \frac{Q^2 + m_{J/\psi}^2 - t} {W^2}
\label{ryskinx}
\end{equation}
\\
For $Q^2 \approx 0$, $t \approx 0$ the only hard scale is represented by the 
$J/\psi$ mass and the applicability of perturbative QCD relies on it.
From (\ref{ryskinq},\ref{ryskinx}) one has:
${\bar q}^2 \simeq 2.5$~GeV$^2$
and $5 \cdot 10^{-4} \leq \bar{x} \leq 5 \cdot 10^{-3}$.
Due to the quadratic dependence on the gluon density the shape of the 
cross section as a function of $W$ is very 
sensitive to different proton parton density parametrizations. However
the normalization of the theoretical result suffers presently from quite large
uncertainties.
In Fig.~\ref{zeuspsicros}/b the elastic $J/\psi$ data are compared to the 
prediction of the Ryskin model with input gluon density given by 
MRS(A'). A good agreement is found, while the steeper
behaviour of GRV has been reported to overshoot HERA data \cite{h1psi94}.

\begin{figure}[p]
\vspace{-4cm}
\hspace{-2cm}
\epsfig{file=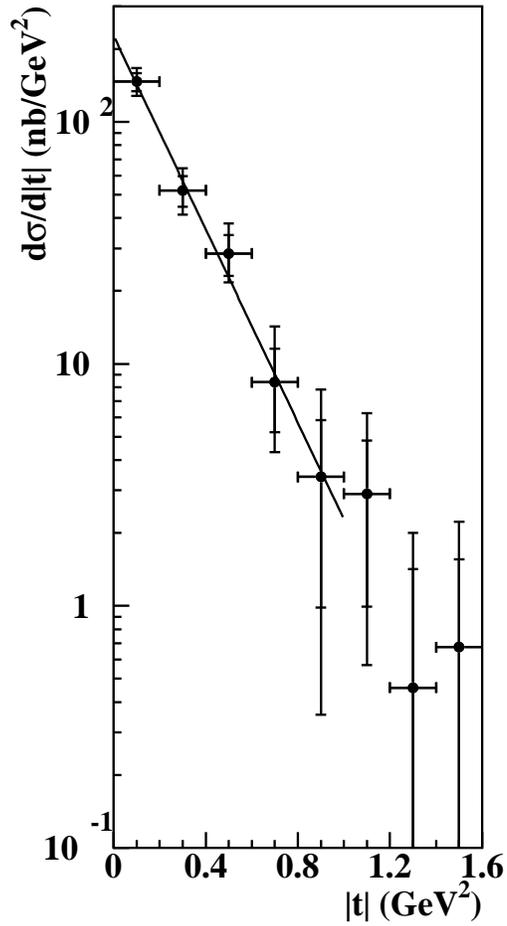,width=14cm}
\caption
{The differential cross section ${d\sigma}/{d|t|}$ for the elastic process
$\gamma p \rightarrow J/\psi ~p$ as measured by ZEUS in the kinematic range 
$40 < W < 140$~GeV.
The result of the exponential fit in the range $|t|<1$~GeV$^2$ is shown as 
the solid line.}
\label{zeusdsdtpsi}
\end{figure}

H1 has determined the slope of the $J/\psi$ ~$p_T^2$ distribution by fitting 
data with $p_T^2 \leq$~1 GeV$^2$ to an exponential $e^{-b' p_T^2}$ in the range 
$30 < W < 150$~GeV, achieving $b' = 4.0 \pm 0.2 \pm 0.2$~GeV$^{-2}$. 
The error in the slope due
to using $p_T^2$ instead of $t$ is evaluated from Monte Carlo to be $-10\%$. No
clear evidence for shrinkage is obtained dividing the sample in W bins, mainly
due to the still limited statistics. 
A consistent result comes from the ZEUS analysis, which also corrects
explicitly the $p_T^2$ differential cross section to obtain $d\sigma/d|t|$, 
shown in Fig.~\ref{zeusdsdtpsi}. 
Fitting over the range $|t| < 1$~GeV$^2$ with an exponential shape 
$e^{-b |t|}$ gives $b = 4.6 \pm 0.4 ^{+0.4}_{-0.6}$~GeV$^{-2}$, which
is consistent with the slope values from fixed target experiments
\cite{fixelpsi}.  

H1 also measured the cross section for
$J/\psi$ production with proton dissociation, exploiting their forward
detectors. This turns out to be of the same order of the elastic cross
section, in a similar range of $W$. The $p_T^2$ distribution gives a slope: 
$b' = 1.6 \pm 0.3 \pm 0.1$~GeV$^{-2}$, a factor of $2.5$ smaller than for pure
elastic scattering. 

The study of the decay angular distributions of the
$J/\psi$ (both elastic and with proton dissociation) shows consistency with
s-channel helicity conservation, i.e. the $J/\psi$, like lighter mesons, 
is mainly transversely polarized.

Diffractive production of $\psi'$ has been observed  by H1 \cite{h1psi94} in the
decay chain $\psi' \rightarrow J/\psi ~\pi^+ \pi^-$, $J/\psi \rightarrow \mu^+
\mu^-$. The rate of its production compared to $J/\psi$ is found to be $0.20
\pm 0.09$, in agreement with previous measurements at fixed target
experiments.

Elastic vector meson production has also been measured at high $Q^2$ for   
$\rho^0$, $\phi$ and $J/\psi$ \cite{zeusvmdis,h1vmdis}. 
Preliminary results exist for low and high-$t$ $\rho^0$ photoproduction
with proton dissociation \cite{gallo}. 
These data, compared to fixed target data, suggested a hard production
mechanism when a high energy scale is present in the interaction, be it a high
mass (as the $J/\psi$ mass) or high $Q^2$ or $t$. 
However recent results from the E665 Collaboration \cite{e665} appear at 
variance with the previous fixed target data. 
The experimental situation for light vector mesons is still to be clarified.
Thus vector meson production at
HERA allows to study the transition from a non-perturbative description of the
pomeron (the soft pomeron), to the hard perturbative pomeron.

\section{Hard photoproduction and QCD}
\label{hphp}

\begin{figure}[p]
\vspace{-4cm}
\hspace{0.cm}
\setlength{\unitlength}{1mm}
\epsfysize=200pt
\epsfbox[200 400 500 800]{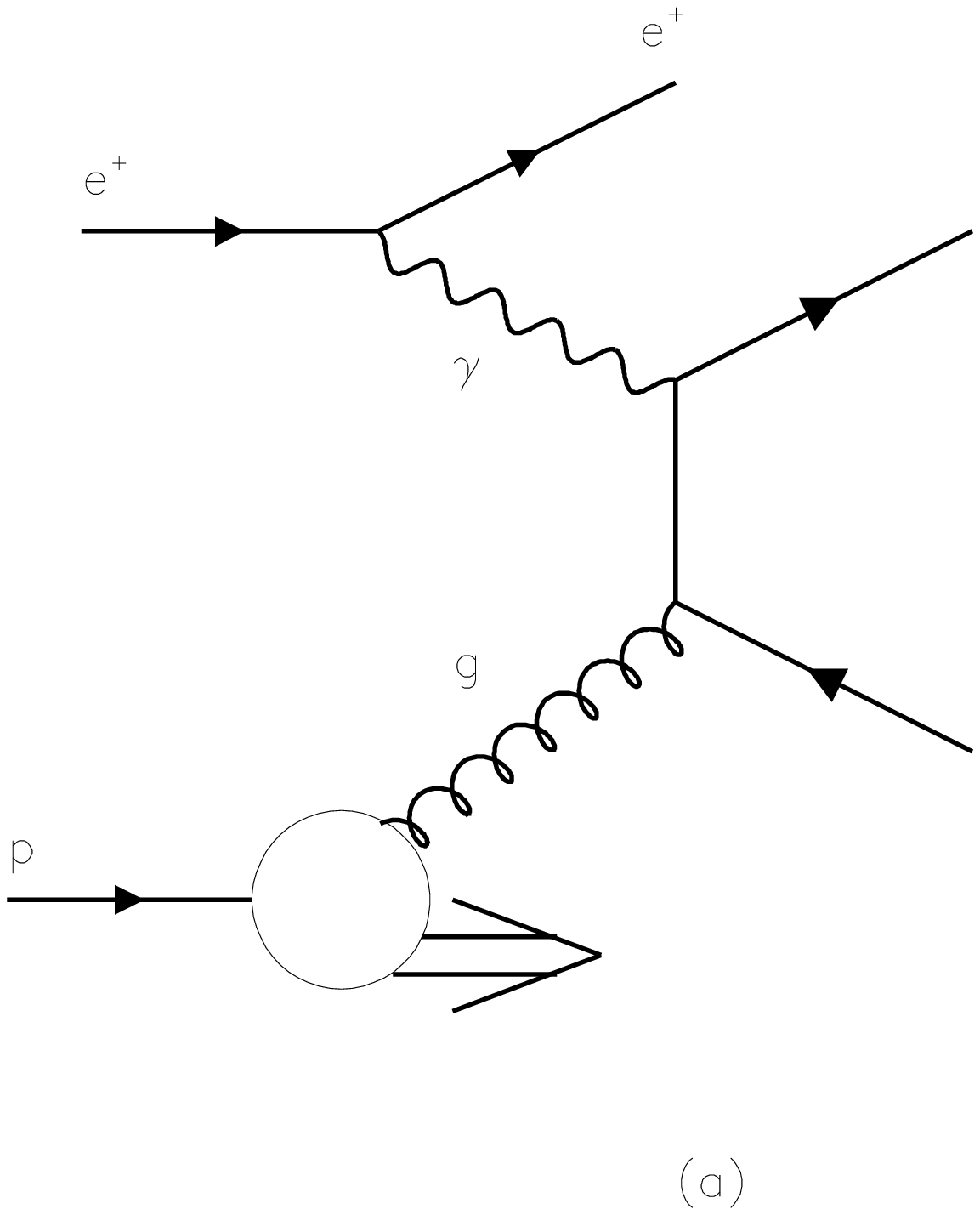}
\epsfysize=200pt
\epsfbox[0 400 300 800]{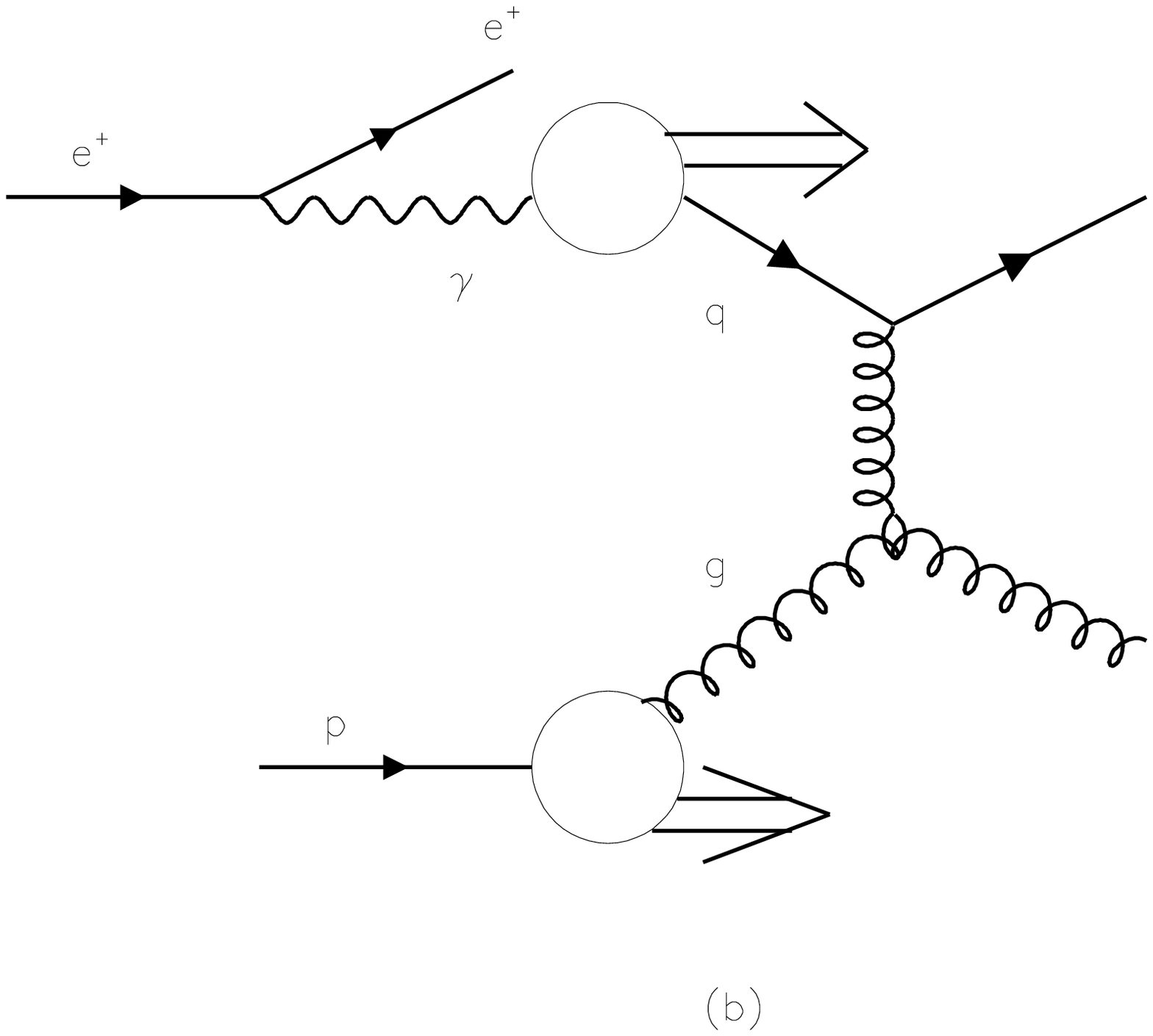}
\vspace{3.5cm}
\caption{Examples of LO QCD (a) direct and (b) resolved photon processes.}
\label{dires}
\end{figure}

Hard photoproduction
processes at HERA are of particular interest. The large cross sections and the
range of high center of mass energies allow precise studies of the photon
structure in a way complementary to the $e \gamma$ deep inelastic scattering.
Moreover the QCD dynamics and the peculiar features of photon induced reactions 
can be tested.
The hard scale is provided by the highest
transverse momentum of the produced particles or jets in the final state.
Application of perturbative QCD requires that this scale is at minimum a few GeV. 
As mentioned earlier, in leading order of QCD two classes of
processes contribute to the production of high $p_T$ particles or jets. The
photon may interact directly with a parton in the proton, as in
Fig.~\ref{dires}/a, or it may first
fluctuate into a hadronic state as in Fig.~\ref{dires}/b and then act as source
of partons, which scatter off partons in the proton. In the first case,
known as {\em direct} photon, the whole photon momentum enters in the hard
subprocess and the final state consists of two jets balancing each other
in $p_T$ plus the
proton remnant and the scattered electron, emerging with small $p_T$. In the
second case, known as {\em resolved} photon, only a fraction of the photon
momentum participates in the hard process and the final state has in addition a
photon remnant, keeping approximately the original photon direction. The two
diagrams represented in Fig.~\ref{dires} are of the same order in the coupling
constants, $\alpha_{em}^2 \alpha_s$. In fact the resolved diagram involves the
photon parton densities which are of order $\alpha_{em} / \alpha_s$, as
we discussed in section \ref{introd}, bringing the power counting to the same
result. 

The kinematics of final state particles or jets are generally specified by
their transverse momentum $p_T$ with respect to the beam axis (or equivalently 
the transverse energy $E_T$) and by the pseudorapidity $\eta$. 
The definition of a jet requires a jet algorithm at both the
experimental and the theoretical level. We will describe in some detail 
the relevant definitions. 
At the Born level the knowledge of $E_T$ 
(${E_T}_1 = {E_T}_2$), $\eta_1$ and $\eta_2$ for the two final state partons,
identified with the final jets, allows one to obtain the energy fractions
$x_\gamma$ and $x_p$ of the initial partons from the photon and the proton
respectively. Let us write the four-momentum balance 
in the hard $2 \rightarrow 2$ subprocess:
\begin{equation}
y x_\gamma \mathbf{k} + x_p \mathbf{P} = \mathbf{j_1} + \mathbf{j_2}
\end{equation}
where $\mathbf{k}$, $\mathbf{P}$, $\mathbf{j_1}$, $\mathbf{j_2}$ 
are respectively the four-momenta of the incoming electron and proton and 
of the two final partons. From this one obtains:
\begin{equation}
x_\gamma = \frac{E_T}{2 E_\gamma} (e^{-\eta_1} + e^{-\eta_2})
\end{equation}
\begin{equation}
x_p = \frac{E_T}{2 E_p} (e^{\eta_1} + e^{\eta_2})
\end{equation}
This holds only at LO. At this level direct photon events have $x_\gamma = 1$
and resolved ones $x_\gamma < 1$. 

The distinction between direct and resolved photon diagrams becomes
theoretically ambiguous in NLO of QCD. Both components are related to each other
through the factorization scale $M_\gamma$ at the photon leg. The $M_\gamma$
dependence of the NLO direct cross section cancels against the $M_\gamma$
dependence in the resolved cross section via the photon structure function. In
a consistent NLO calculation both the components have to be summed using the
same $M_\gamma$ scale. 
In general one defines observables
$x_\gamma^{OBS}$ and $x_p^{OBS}$ which are determined from the two highest 
$E_T$ jets in the event:
\begin{equation}
x_\gamma^{OBS} = \frac {{E_T}_1 e^{-\eta_1} + {E_T}_2 e^{-\eta_2}} {2 E_\gamma}
\label{xgobs}
\end{equation}
\begin{equation}
x_p^{OBS} = \frac {{E_T}_1 e^{\eta_1} + {E_T}_2 e^{\eta_2}} {2 E_p}
\end{equation}
Direct and resolved photon events can be experimentally defined by a cut on 
$x_\gamma^{OBS}$ when two or more jets are measured in the final state 
\cite{zeusdir1,zeusdir2}. 

The cross section for n-jet production is expressed formally as a
convolution of the partonic cross section with the photon and proton parton 
densities:
\begin{eqnarray}
\label{factor1}
d\sigma (e~p \rightarrow e ~jets~ X) = \int_{0}^{1} dy~
\Phi_{\gamma/e} (y) ~ \times 
~~~~~~~~~~~~~~~~~~~~~~~~~~~~~~~~~~~~~~~~~~~~~~~~~~~~~~~~
\\
\times~ \sum_{a,b} \int_{0}^{1} dx_\gamma \int_{0}^{1} dx_p ~f_{a/\gamma}
(x_\gamma,M_\gamma^2) ~f_{b/p}(x_p,M_p^2) 
~d\hat{\sigma} (a~b \rightarrow jets) \nonumber
\end{eqnarray}
where $a$ and $b$ are the parton types from $\gamma$ and $p$, and 
$M_\gamma^2$, $M_p^2$ the factorization scales for the photon and 
proton parton densities.
For a direct photon $a=\gamma$ and formally $f_{\gamma/\gamma} (x_\gamma) =
\delta (1 - x_\gamma)$.

For the production of a particle $h$ the inclusive cross section is expressed
by introducing the fragmentation function $D_{h/c}(z)$, which gives the 
probability to obtain $h$ from a final state parton $c$, with momentum fraction 
$z = p_h/p_c$. It can be written:
\begin{eqnarray}
\label{factor2}
d\sigma (e~p \rightarrow e~ h~ X) = \int_{0}^{1} dy~ \Phi_{\gamma/e} (y) ~\times 
~~~~~~~~~~~~~~~~~~~~~~~~~~~~~~~~~~~~~~~~~~~~~~~~~~~~~~~
\\
\times ~\sum_{a,b,c} 
\int_{0}^{1} dx_\gamma \int_{0}^{1} dx_p \int_{0}^{1} dz 
~f_{a/\gamma} (x_\gamma,M_\gamma^2) ~f_{b/p}(x_p,M_p^2) ~D_{h/c}(z,M_f^2)
~~d\hat{\sigma} (a ~b \rightarrow c ~X) \nonumber
\end{eqnarray}
where $M_f^2$ is the factorization scale of the fragmentation function, which is
usually set to the $p_T^2$ of the produced particle.
The fragmentation functions need to be parametrized from experimental results
\cite{bin95}.

The validity of equations (\ref{factor1}) and (\ref{factor2}) rests on the QCD
factorization theorems (for a review see \cite{thfact}), 
which state that universal parton distribution functions
can be defined for a given hadron independently from the process. Thus 
comparing QCD calculations with experimental data may serve to constrain 
the parton distributions or, when the latter have negligible uncertainties,
to test the parton dynamics.

\subsection{Inclusive charged particle distributions}

\begin{figure}[p]
\begin{center}
\epsfig{file=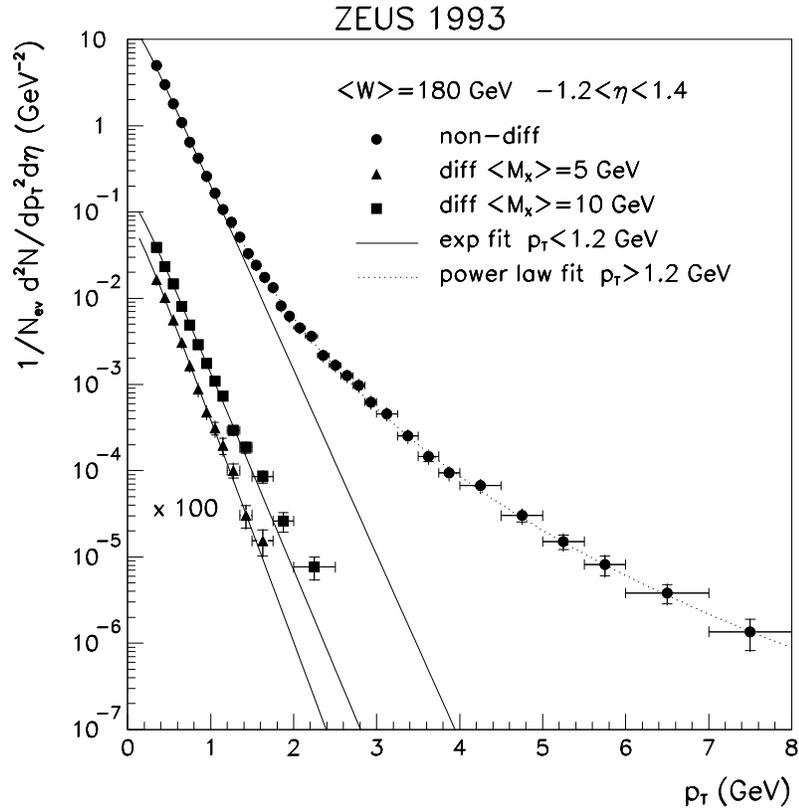,width=12cm}
\end{center}
\caption
{Inclusive transverse momentum distributions of charged particles in
photoproduction events at $\langle W \rangle = 180$~GeV
averaged over the pseudorapidity interval $-1.2<\eta<1.4$.
Solid lines indicate fits of equation (\protect\ref{termo}) to the
data in the region of $p_T < 1.2$~GeV.  The dotted line shows a
power law formula (\protect\ref{powlaw}) fitted to the non-diffractive data 
for $p_T > 1.2$~GeV. The diffractive data are scaled
down by two orders of magnitude.}
\label{pt1}
\end{figure}

\begin{figure}[p]
\begin{center}
\epsfig{file=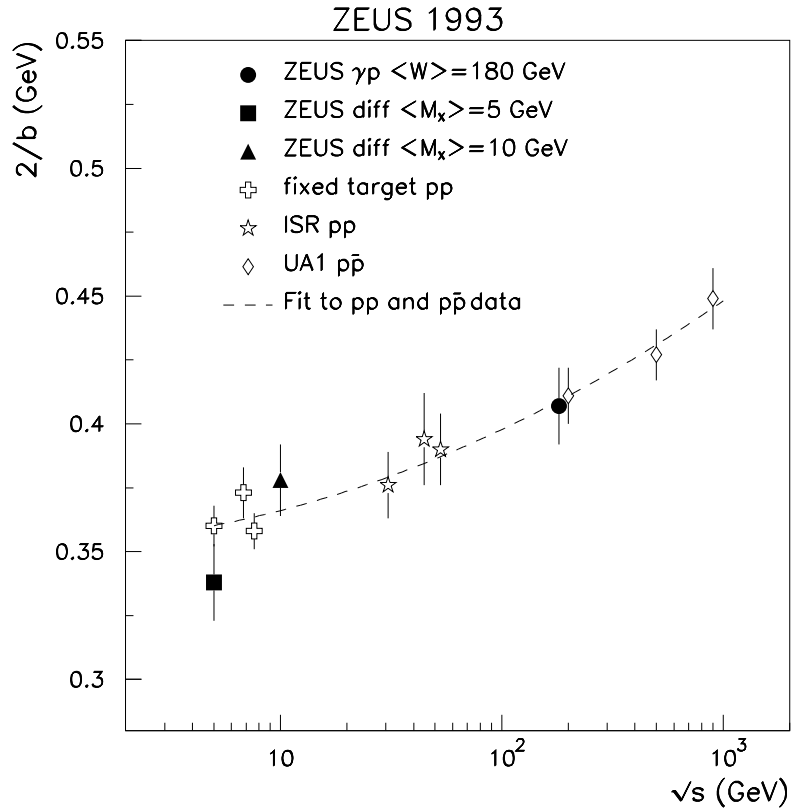,width=12cm}
\end{center}
\caption
{Inverse slope of the exponential fit (\protect\ref{termo}) 
as a function of
the center of mass energy $\protect\sqrt s$ for ZEUS $\gamma p$ data 
and hadron-hadron data from fixed target \protect\cite{fixpppp}, ISR
\protect\cite{capi74} and UA1 \protect\cite{ua190}.
The ZEUS non-diffractive point is given at the $\gamma p$ c.m.s. energy,
while the diffractive points are plotted at the energies 
corresponding to the mean value of the invariant mass $\langle M_X \rangle$
of the dissociated photon system.
The dashed line is a parabola in $log(s)$ and was fitted to
all the hadron-hadron points to indicate the trend of the data.}
\label{pt2}
\end{figure}

Inclusive distributions of charged particles in photoproduction have been
measured by both H1 \cite{h1incpa} and ZEUS \cite{zeusincpa}.
In Fig.~\ref{pt1} the transverse momentum distributions are shown averaged over
the pseudorapidity interval $-1.2 < \eta < 1.4$ (corresponding to the region 
$-3.2 < \eta^\ast < -0.6$ in the $\gamma p$ center of mass frame, 
in the photon hemisphere)
as measured by ZEUS at $\langle W \rangle = 180$~GeV for tagged photoproduction
events. The
events have been separated in a non diffractive sample and two diffractive
samples with different average values of the diffractive system mass $M_X$ of 5
and 10 GeV. The separation is done on an event by event basis, labelling as
diffractive the events with a rapidity gap around the proton beam direction
larger than a fixed cut. 
Soft interactions of hadrons can be described by thermodynamic models
\cite{hage83}, which
predict an approximately exponential $p_T$ spectrum of the kind:
\begin{equation}
\frac{1}{N_{ev}} \frac{d^2N}{dp_T^2 d\eta} = exp \left( a - b \sqrt{p_T^2 +
m_\pi^2} \right)
\label{termo}
\end{equation}
where $m_\pi$ is the pion mass. This expression is fitted to the ZEUS data in
the range $0.3 < p_T < 1.2$~GeV resulting in the solid lines. 
The fitted values of the slope $b$ are compared to results
from $p p$ and $p \bar p$ data as a function of the center of mass
energy in Fig.~\ref{pt2}. The slope of the ZEUS non-diffractive spectrum agrees
with data from hadron-hadron scattering at similar energies. The
diffractive slopes agree better with hadronic data taken at a lower energy:
here they are plotted at 5 and 10 GeV, the values of the invariant mass of the
dissociated photon, in agreement with data on proton diffractive dissociation
at hadron colliders \cite{goul83,giac90}. 

\begin{figure}[p]
\begin{center}
\epsfig{file=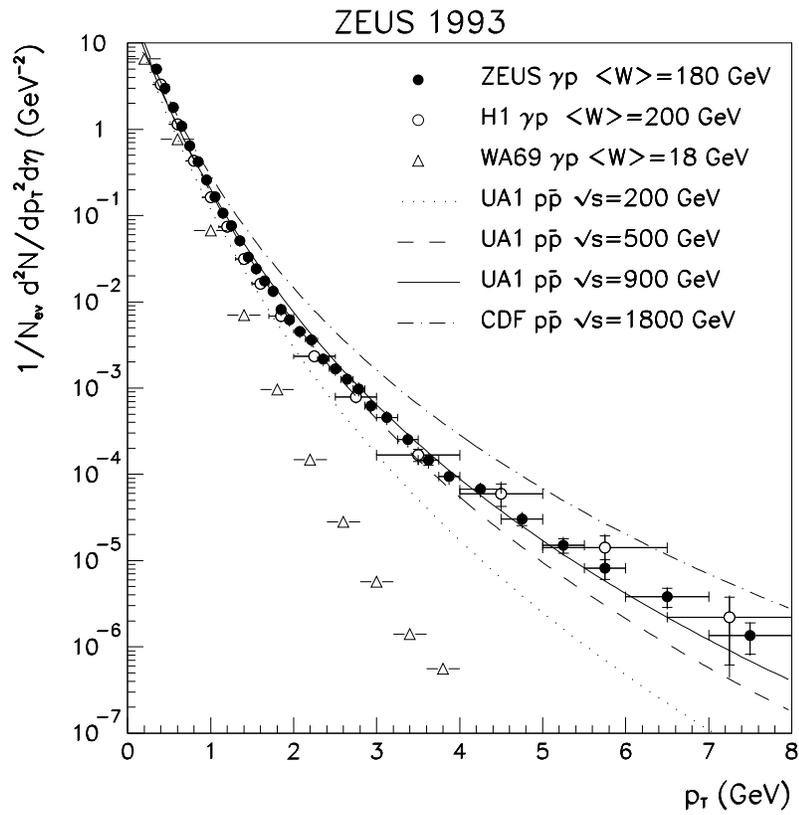,width=12cm}
\end{center}
\caption{Transverse momentum spectra of charged particles in different 
experiments: ZEUS and H1 photoproduction data are compared to data from
OMEGA \protect\cite{omega89}, UA1 \protect\cite{ua190}, and CDF 
\protect\cite{cdf88}.}
\label{pt3}
\end{figure}

The $p_T$ spectrum for non-diffractive events in Fig.~\ref{pt1} 
deviates from an exponential at high $p_T$ values. 
This is expected from QCD as a result of hard scattering between 
constituent partons. The high $p_T$ behaviour can be approximated by a power 
law:
\begin{equation}
\frac{1}{N_{ev}} \frac{d^2N}{dp_T^2 d\eta} = A \left( 1 + \frac{p_T}{{p_T}_0}
\label{powlaw}
\right)^{-n}
\end{equation}
with ${p_T}_0 = 0.54$~GeV and $n = 7.25$.
In Fig.~\ref{pt3} HERA data are compared to
data from the WA69 \cite{omega89} fixed target photoproduction experiment 
at $\langle W \rangle= 18$~GeV. This shows that the $p_T$ spectrum becomes
harder as the center of mass energy increases.
The figure also shows fits of the form (\ref{powlaw}) to $p \bar p$ data from
UA1 and CDF at several different energies \cite{ua190,cdf88}. The $p_T$ 
spectrum for 
photoproduction at HERA is clearly harder than the one from $p \bar p$
interactions at a similar energy, and is in fact similar to $p \bar p$ at
$\sqrt{s} = 900$~GeV.

High energy $\gamma p$ and $p \bar p$ interactions show
a strong similarity for the production of particles at low $p_T$, while the
rate of high $p_T$ particles is quite different, revealing the two-fold nature
of the photon.
In fact the most appealing reason for the observed difference is that at
high $p_T$ the pointlike coupling of the photon becomes important: thus $\gamma
p$ interactions are not just like $\pi p$, instead direct and anomalous 
processes may occur. 
Other possible explanations could be the different pseudorapidity ranges 
used by the various experiments ( $| \eta | < 2.5$ for UA1, 
$| \eta | < 1$ for CDF, $-3.2 < \eta < -0.6$ for ZEUS) or harder 
parton distributions in mesons (to which the photon converts according to the
VDM) than in baryons. 

HERA data have also been compared to a NLO QCD calculation \cite{borzu93} and
are in good agreement in the region of high $p_T$. The high $p_T$ inclusive
particle cross section has recently been considered an important
tool to extract precise parton distributions in the photon in the poorly known
regions of $x_\gamma > 0.8$ and $x_\gamma < 0.05$ \cite{binproc}.
It has the advantage of being insensitive to possible multiple interactions or
soft underlying event effects, discussed in section \ref{sue}, 
as opposed to jet measurements.

\subsection{Jets}

Apart from the long tail in the $p_T$ distribution of particles, the
observation of jets is a direct means to establish the occurence of hard 
scattering between constituent partons. A jet is a bunch of collimated
particles coming from the fragmentation of a high energy quark or gluon. Jet
measurements need a jet definition criterion or jet algorithm. 
Until now cone algorithms \cite{conejet,snowmass} have
been used more widely in photoproduction at HERA, given the framework similar
to hadronic collisions. 

Experimentally one starts from the calorimeter cells
with energies above a threshold. For each cell the
pseudorapidity $\eta$ and azimuth $\phi$ are determined from the segment 
joining the interaction vertex to the geometric center of the cell. 
Each cell with transverse energy above a few hundred MeV is considered as a 
possible seed for the jet finding. 
The seeds are combined if their distance in the $\eta-\phi$ space, $R=
\sqrt{\Delta\phi^2 + \Delta\eta^2}$, is less than some cut (usually $1$ or
sometimes $0.7$). Then
a cone of radius $R$ is drawn around each seed and the energy of all the
included cells is summed up in a cluster. The cluster axis is defined
along the Snowmass convention \cite{snowmass} from the $E_T$-weighted mean
of the cells inside the cone. A new cone of radius $R$ is then drawn around the
calculated axis of the cluster and the axis is recalculated. 
The procedure is
iterated until stability is reached. At the end it may happen that two clusters
overlap. In this case if the shared transverse energy is more than $75\%$ of
the total $E_T$ of one of the clusters they are merged, otherwise
the common cells are assigned to the nearest one.
In the end a cluster is called a jet if its transverse energy exceeds
a minimum value. At HERA this is set typically in the range of $6 - 8$~GeV.

\begin{figure}[p]
\begin{center}
\epsfig{file=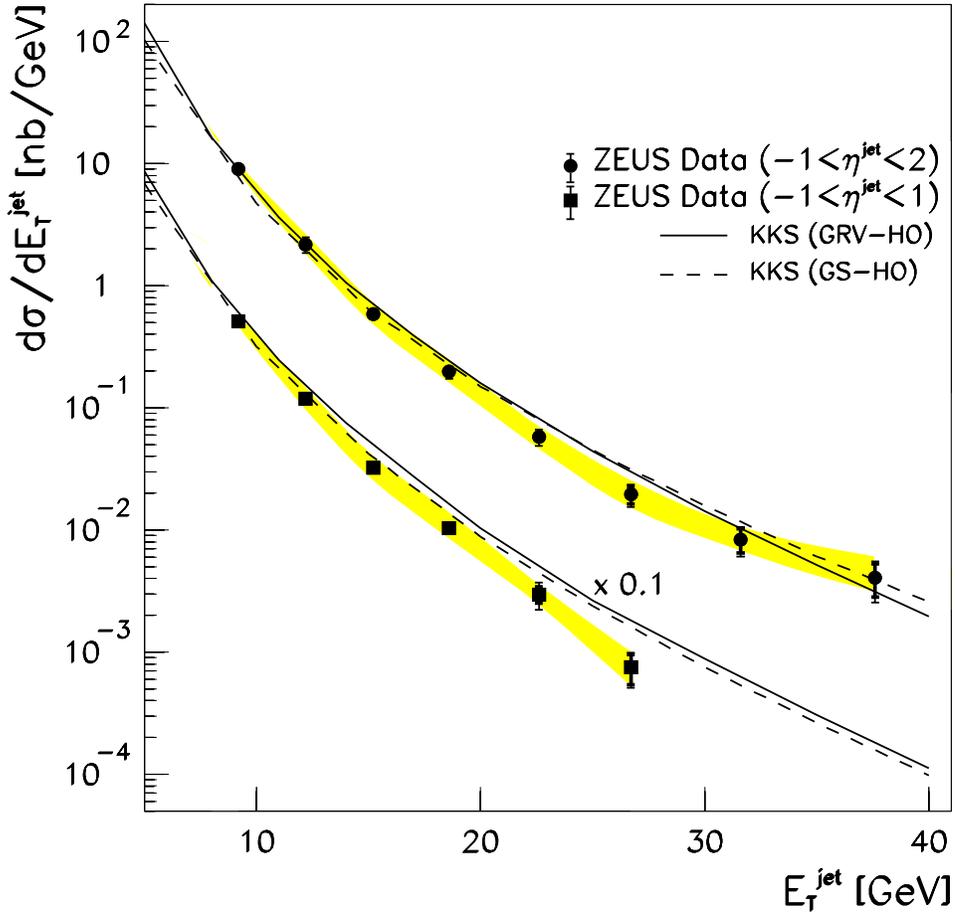,width=14cm,
bbllx=65pt,bblly=200pt,bburx=500pt,bbury=610pt,clip=}
\end{center}
\caption{Differential $e p$ cross section $d\sigma/dE_T^{jet}$ for inclusive
jet production integrated over two $\eta^{jet}$ ranges: $-1<\eta^{jet}<2$ and
$-1<\eta^{jet}<1$ (scaled down by an order of magnitude). The kinematic range
is $Q^2 < 4$~GeV$^2$ and $0.2<y<0.85$. The shaded bands give the uncertainty due
to the energy scale of the jets. 
The superimposed lines are results of a NLO QCD calculation \protect\cite{kks}
using MRS(D-) \protect\cite{mrsd} proton parton distributions and two 
different photon parton distributions: GRV-HO \protect\cite{grv} (solid line) 
and GS-HO \protect\cite{gs} (dashed line).}
\label{etjet}
\end{figure}

Inclusive jet production has been measured by ZEUS \cite{zeusincjet}
for $E_T^{jet} > 8$~GeV in
the laboratory pseudorapidity range $-1 < \eta^{jet} < 2$, from
$0.55$~pb$^{-1}$ of untagged data ($Q^2 < 4$~GeV and $0.2 < y < 0.85$).
Jets have been defined with a cone radius $R = 1$.
Similar measurements have been reported by H1 \cite{h1incjet} from a lower
statistics of tagged data ($290$~nb$^{-1}$) with $Q^2 < 0.01$~GeV$^2$ and
$0.25 < y < 0.7$.
Fig.~\ref{etjet} shows the differential cross section 
$d\sigma/dE_T^{jet}$ integrated
over two $\eta$ ranges: $-1 < \eta^{jet} < 2$ and $-1 < \eta^{jet} < 1$.
Data are corrected to the hadron level for
detector effects by a complete detector simulation. 

NLO QCD calculations of
inclusive jet cross sections have been carried out by several groups
\cite{gsincjet,greco,kks,aur94,owe96}. 
These calculations involve at most three partons in the final
state, from which jets are defined by the same cone algorithm as used for data.
NLO corrections can be very large, up to a factor of two. This is reflected in
a strong dependence of the LO results upon variations in the energy scale
at which the strong coupling constant is evaluated or in the factorization
scales $M_\gamma^2$ and $M_p^2$. In LO these scales are often set 
at about the $p_T^2$ or $(p_T/2)^2$ of the hard jets. For those values the 
size of the NLO corrections is about its minimum. 
The stability of the NLO results against variations of the scales is much 
improved. The NLO curves plotted in Fig.~\ref{etjet} \cite{kks}
are parton level results, 
not corrected for hadronization. They are consistent with the data.

\begin{figure}[p]
\begin{center}
\epsfig{file=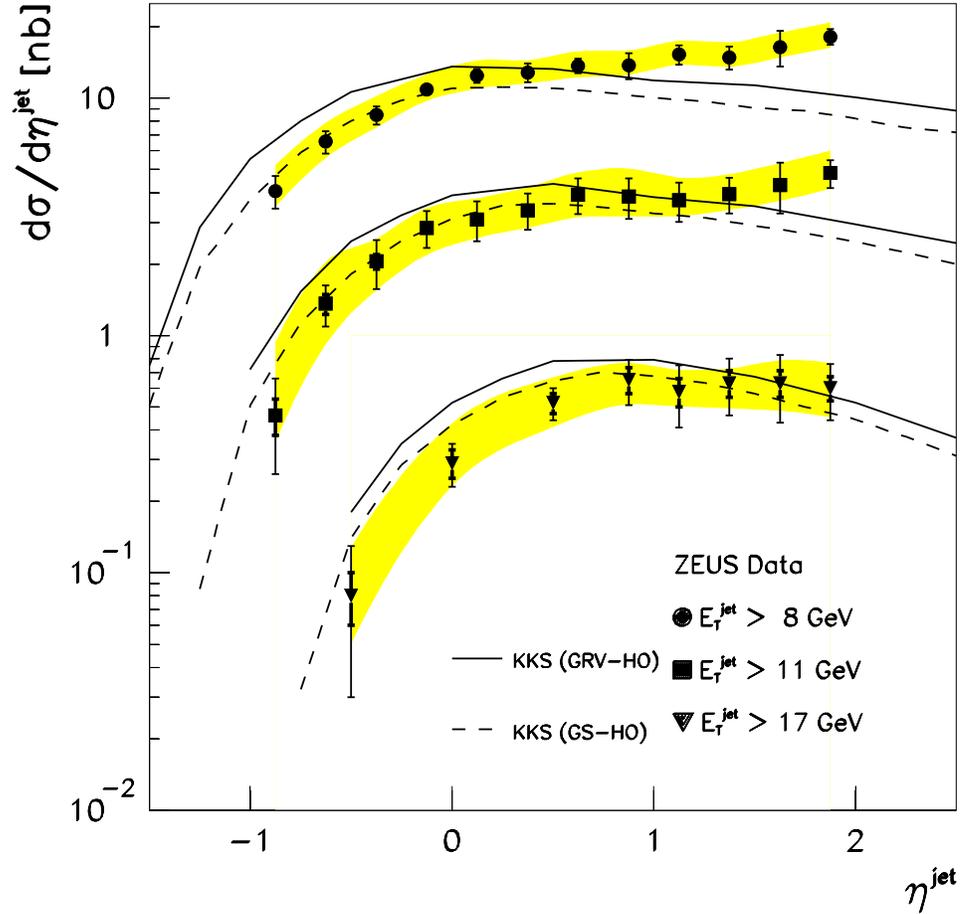,width=14cm,
bbllx=65pt,bblly=200pt,bburx=500pt,bbury=610pt,clip=}
\end{center}
\caption{Differential $e p$ cross section $d\sigma/d\eta^{jet}$ for inclusive
jet production integrated over $E_T^{jet}$ from three different thresholds
($E_T > 8$, $11$ and $17$~GeV) in the kinematic range 
$Q^2 < 4$~GeV$^2$ and $0.2<y<0.85$. 
The shaded bands give the uncertainty due to the energy scale of the jets. 
The superimposed lines are NLO QCD results as in the previous figure.}
\label{etajet}
\end{figure}

Fig.~\ref{etajet} presents $d\sigma/d\eta^{jet}$ for jets with $E_T^{jet}$ 
greater than $8$, $11$ or $17$ GeV. 
Comparison with the theoretical NLO results \cite{kks}
shows a preference for the GS parametrization over GRV in the region
$\eta^{jet} \leq 0$, where the contribution from high $x_\gamma$ partons 
is important. 
This is the photon fragmentation region, where the calculation is most reliable,
thus giving a handle to test the photon structure. 
In the region $\eta^{jet} \geq 1$ the data, particularly at the 
lowest $E_T^{jet}$, are systematically above the predictions. 
Here the fragmentation
corrections are sizeable and moreover the data are biased by a pedestal energy
which has not been subtracted. This is thought to arise from a combination of 
soft and semi-hard interactions between the beam remnants in resolved
photoproduction and it will be subject of a further discussion in the
next section. A thorough understanding of this region is essential to carry out
quantitative QCD tests in the low $x_\gamma$ region accessible to HERA
measurements ($x_\gamma \geq 0.01$). 

The largest experimental errors are due to the absolute energy scale of the
calorimeters, whose uncertainty is within $\pm 5\%$ for the measurement of
hadronic jets in both ZEUS and H1. 
This has been determined comparing real
data and Monte Carlo simulated events. H1 studied the $E_T$ imbalance between
the scattered electron and the hadronic jet in high $Q^2$ events.
ZEUS studied the ratio between $E_T^{jet}$ (measured by the calorimeter)
and the summed $p_T$ of the charged tracks associated with the jet (measured 
by the central tracking detector), for jets with $|\eta^{jet}| < 1$; 
for jets with $1 < \eta^{jet} < 2$ the $E_T$ imbalance has been studied 
in dijet events with the second jet in the central region. 
Due to the exponential fall of the jet cross section with increasing 
$E_T^{jet}$, the $5\%$ energy scale error
translates into a dominant systematic uncertainty of about $20\%$ on the
normalization of the cross sections. This uncertainty is almost completely
correlated from point to point in figures \ref{etjet} and \ref{etajet} and is 
indicated as a shaded band. It has to be interpreted as the maximum
systematic shift that might be applied overall. The cross section measurements 
could then move all up or down of the indicated amount.  

\begin{figure} [p]
\begin{center}
\epsfig{file=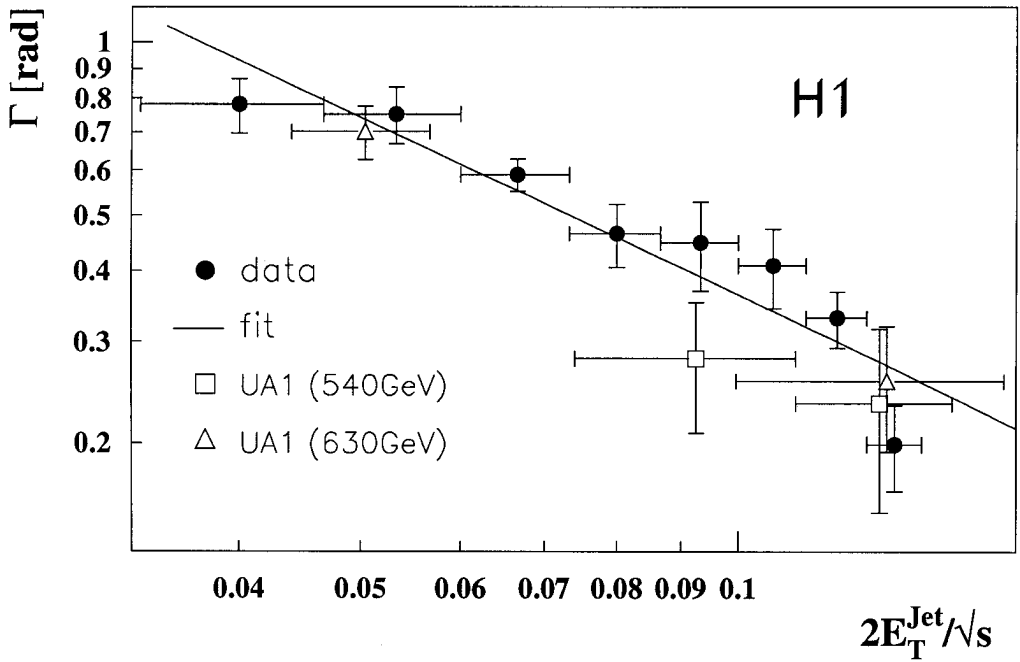}
\end{center}
\caption{
Jet width measurements as function of the scaled jet transverse energy 
$2 E_T^{jet}/\protect\sqrt{s}$ from H1 \protect\cite{h1incjet} and UA1
\protect\cite{ua1shape}. 
H1 data are normalized to the $e p$ center of mass energy.
$\Gamma$ is defined as the full width at half maximum above the pedestal
energy determined from the $E_T$ azimuthal profile distribution 
around the jet axis in H1 data and from the pseudorapidity profile 
in UA1 data.
The line represents a $1/E_t^{jet}$ fit to the H1 data.}
\label{jshape}
\end  {figure}

It is expected from QCD that jets become narrower with increasing energy.
The jet shape, that is the transverse energy flow
versus $\Delta\phi$ or $\Delta\eta$, is approximately gaussian around the
position of the jet axis. The full width $\Gamma$ at half maximum of a jet
is expected to decrease as $1/E_T^{jet}$ \cite{sale93}.
H1 has measured the dependence of $\Gamma$ on the scaled $E_T$ of the jet, 
shown in Fig.~\ref{jshape}. The height of the
gaussian is determined after subtraction of the pedestal energy. 
The jet transverse energy is normalized to the
$e p$ center of mass energy according to \cite{sale93}.
H1 data are compared to similar
measurements from jet profiles in $p \bar p$ interactions \cite{ua1shape}.
The decrease of $\Gamma$ with increasing $E_T^{jet}$ is apparent and, within the
errors, $\gamma p$ and $p \bar p$ jets show the same behaviour.

\begin{figure}
\begin{center}
\epsfig{bbllx=0pt,bblly=410pt,bburx=280pt,bbury=650pt,clip=,%
file=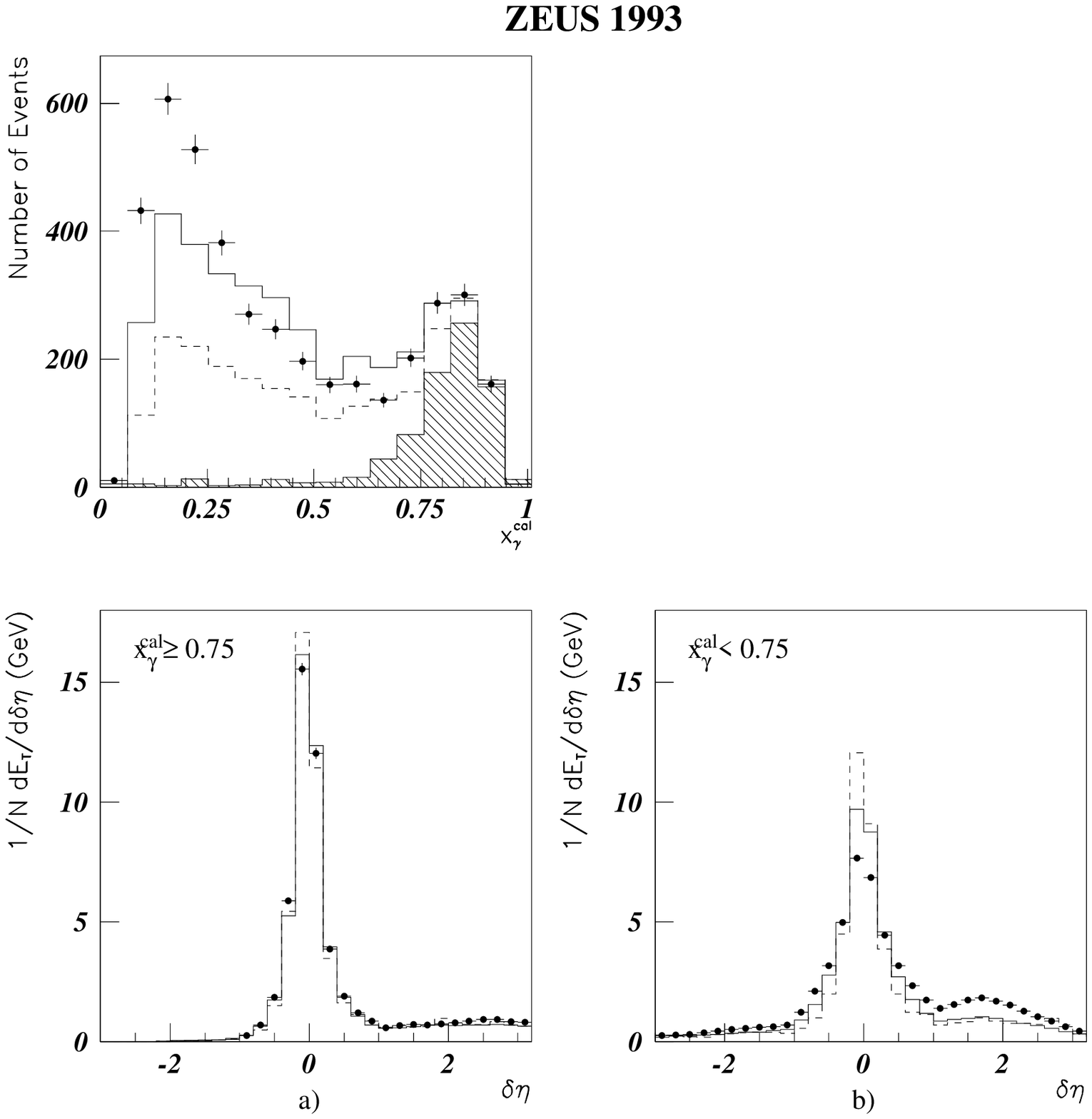,width=12cm}
\end{center}
\caption{
The uncorrected $x_\gamma^{OBS}$ distribution (indicated with $x_\gamma^{cal}$)
from events with at least two reconstructed jets measured by ZEUS. The solid
and dashed histograms represent respectively the expectations from PYTHIA and 
HERWIG Monte Carlo and have been normalized to fit the direct peak in the data.
The shaded histogram shows the direct contribution to the HERWIG distribution.}
\label{xgamma}
\end{figure}

Direct and resolved photon samples have been experimentally defined for the
first time by a cut on the $x_\gamma^{OBS}$ variable (\ref{xgobs})
in events with at least two reconstructed jets by ZEUS \cite{zeusdir1,zeusdir2}.
The observed $x_\gamma^{OBS}$ distribution after the selection cuts 
is shown in Fig.~\ref{xgamma}.
It is compared to the predictions of two Monte Carlo programs, HERWIG
\cite{marche92} and PYTHIA \cite{sjo94}, 
which include LO QCD matrix elements plus parton showers 
and hadronization models. Resolved and direct photon processes, as implemented
in the Monte Carlos, populate different $x_\gamma^{OBS}$ regions. The peak at
high values of $x_\gamma^{OBS}$ is clearly attributed to direct photon
processes. Its shift from $x_\gamma^{OBS} = 1$ is due to higher order 
QCD corrections and fragmentation effects, which are simulated in the 
Monte Carlos. Given this clear separation direct and resolved photon processes 
can be operatively defined by a cut at $x_\gamma^{OBS} = 0.75$. In the low 
$x_\gamma^{OBS}$ region the Monte Carlo curves fall considerably below the
data. The reason is the energy excess observed in the forward $\eta$ region 
in comparison with the Monte Carlo expectations, which is discussed in the next
section.

\begin{figure}
\begin{center}
\epsfig{file=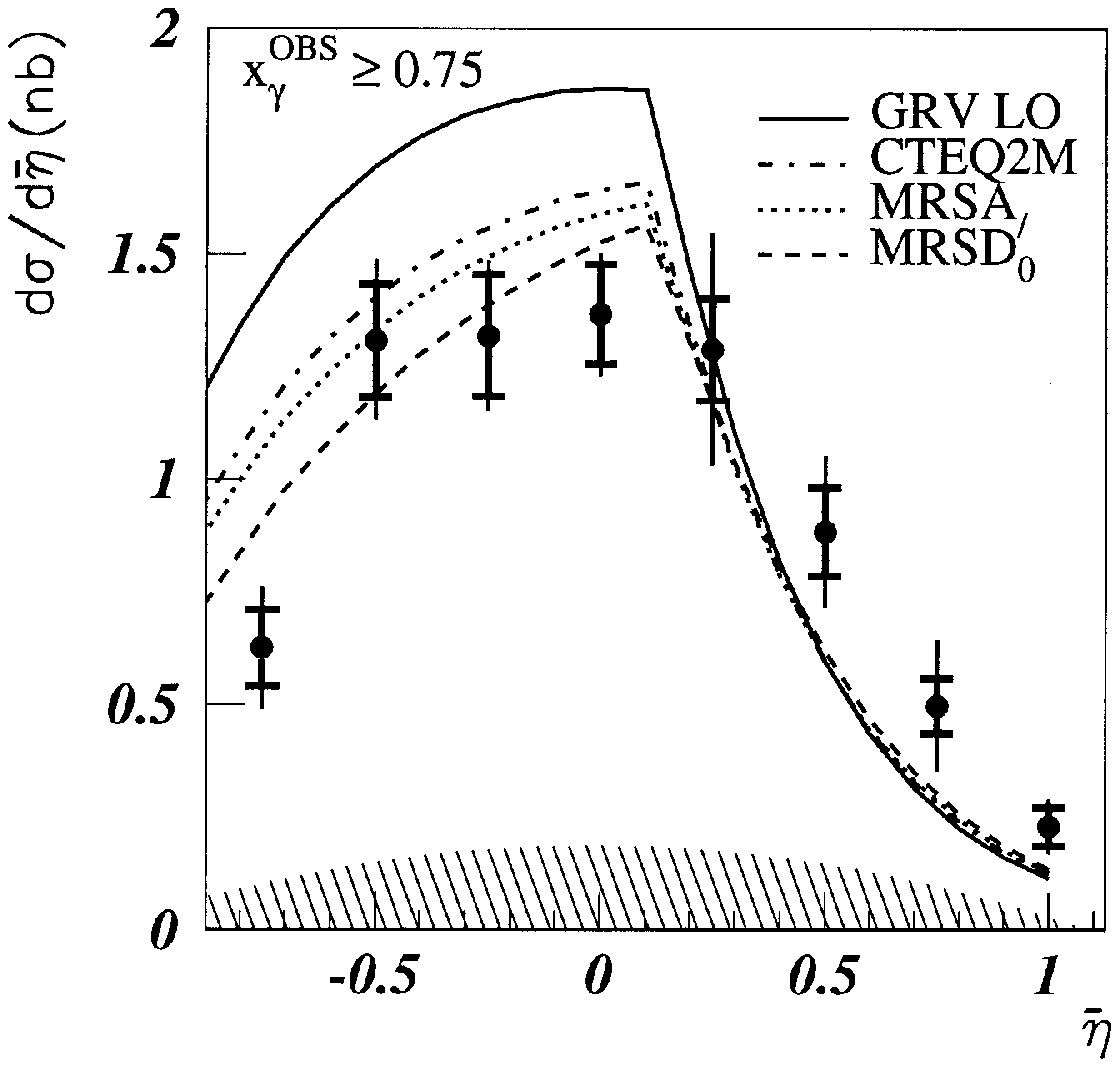}
\end{center}
\caption{Differential $e p$ cross section $d\sigma/d\bar{\eta}$ for inclusive
dijet production in the direct photon sample defined by the cut 
$x_\gamma^{OBS} > 0.75$, as measured by ZEUS.
Jets are required to have $E_T^{jet} > 6$~GeV and $|\Delta\eta| < 0.5$ and the
events to have $Q^2 < 4$~GeV$^2$ and $0.2 < y <0.8$. 
The shaded band gives the uncertainty due to the energy scale of the jets.
The data are compared to analytic LO QCD calculations \protect\cite{for93} 
using the GS2 photon 
parton distributions \protect\cite{gs} and different sets for the proton: 
GRV-LO \protect\cite{grvhadold},
CTEQ2M \protect\cite{cteq3}, MRSA \protect\cite{mrsg} and MRSD0'
\protect\cite{mrsdp}.}
\label{dircr}
\end{figure}

Dijet cross sections have been measured by ZEUS separately for 
$x_\gamma^{OBS} > 0.75$ and $x_\gamma^{OBS} < 0.75$ from $0.55$~pb$^{-1}$ of
untagged data \cite{zeusdir2}.
Jets have been required to have $E_T^{jet} > 6$~GeV. Moreover a cut on the
pseudorapidity separation between the two jets has been applied: 
$|\Delta\eta| =  \eta_1 - \eta_2 < 0.5$.
The cross section has been
measured in terms of $\bar \eta = (\eta_1 + \eta_2) / 2$, the average
pseudorapidity of the jets. Assuming two jets with the same $E_T^{jet}$ the
kinematic variables $x_\gamma^{OBS}$ and $x_p^{OBS}$ can be expressed
as:
\begin{equation}
x_\gamma^{OBS} = \frac{E_T^{jet} e^{- \bar \eta}} {E_\gamma} cosh
\frac{\Delta\eta}{2}
\end{equation}
\begin{equation}
x_p^{OBS} = \frac{E_T^{jet} e^{\bar \eta}} {E_p} cosh\frac{\Delta\eta}{2}
\end{equation}
For $\Delta \eta = 0$ the hyperbolic cosine takes its minimum value of unity and
thus for given $E_T^{jet}$ and $\bar\eta$ the minimum values of $x_\gamma^{OBS}$
and $x_p^{OBS}$ are probed \cite{for93}.
Moreover there is a strong correlation between
$\bar\eta$ and $x_p^{OBS}$ in the direct cross section and between $\bar\eta$
and $y x_\gamma^{OBS}$ in the resolved cross section. 
In Fig.~\ref{dircr} $d\sigma/d\bar\eta$ is shown for events defined as
direct ($x_\gamma^{OBS} > 0.75$), together with results of a LO QCD
calculations using GS2 parton distributions for the photon and different
parametrizations for the proton. The agreement is
reasonable and improves when comparing with a QCD Monte Carlo 
allowing for transverse momentum 
of the incoming partons, parton showers and hadronization effects. 
NLO QCD calculations have become available only recently. The first results
compared to ZEUS data had resolved diagrams calculated only at LO 
\cite{klakra96,owe96}. These calculations improved the agreement with data.
Moreover they revealed an interesting sensitivity of the dijet 
cross section at high $x_\gamma^{OBS}$ on the quark content of the photon,
with data preferring GS over GRV parametrizations.

\begin{figure}
\begin{center}
\epsfig{bbllx=15pt,bblly=160pt,bburx=525pt,bbury=647pt,clip=,%
file=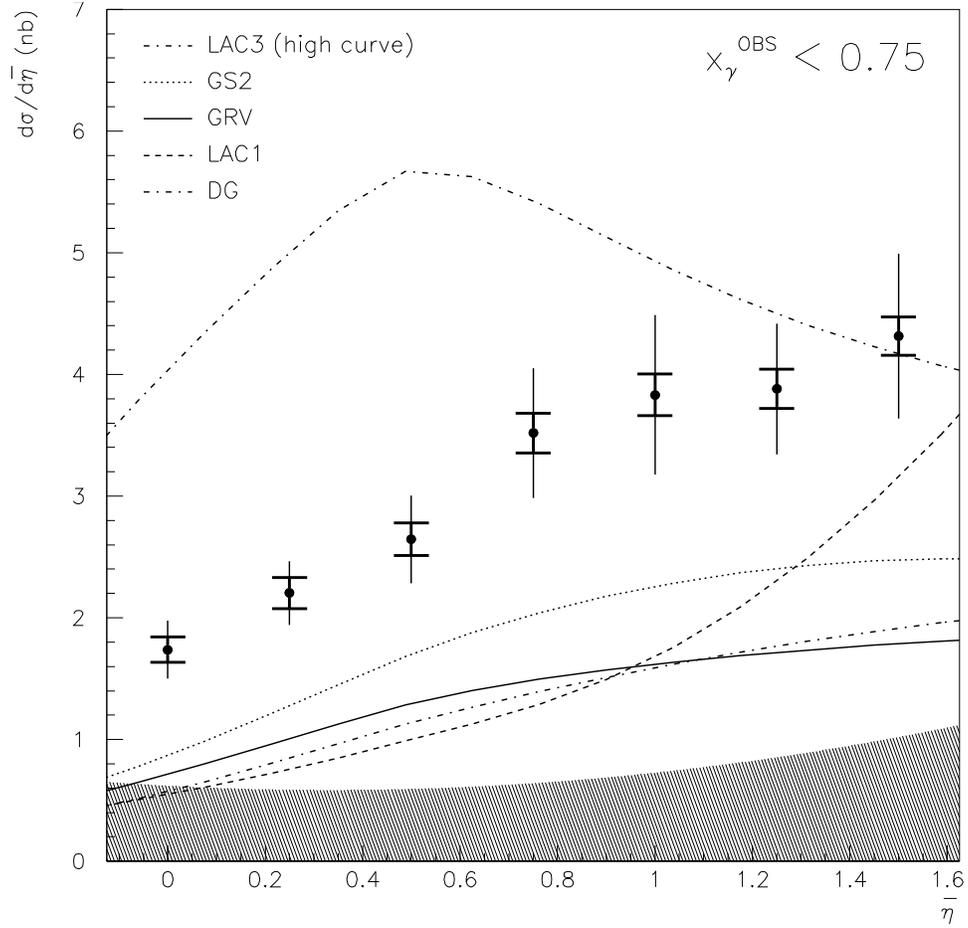,width=13cm}
\end{center}
\caption{Differential $e p$ cross section $d\sigma/d\bar{\eta}$ for inclusive
dijet production in the resolved photon sample defined by the cut 
$x_\gamma^{OBS} < 0.75$, as measured by ZEUS.
Jets are required to have $E_T^{jet} > 6$~GeV and $|\Delta\eta| < 0.5$ and the
events to have $Q^2 < 4$~GeV$^2$ and $0.2 < y <0.8$. 
The shaded band gives the uncertainty due to the energy scale of the jets.
The curves are LO QCD calculations using the MRSA \protect\cite{mrsg} 
parton distribution set for the proton and different sets for the photon:
LAC1-LAC3 \protect\cite{lac}, GS2 \protect\cite{gs}, GRV \protect\cite{grv} and DG
\protect\cite{dg85}.}
\label{rescr}
\end{figure}

The resolved cross section ($x_\gamma^{OBS} < 0.75$) is shown in
Fig.~\ref{rescr}. Data are compared to a LO calculation
using MRS(A) proton parton distributions and different parametrizations for
the photon. Here only the shape can be reproduced by the theoretical curves, 
the normalization is below the data by a factor of $1.5$ to $2$ (except for
LAC3 which is excluded by independent measurements from TRISTAN,
as we showed at the end of section \ref{introd}). 
This discrepancy can again be attributed to underlying soft or semi-hard 
interactions between the beam remnants. 

Complete NLO calculations for dijet cross sections have been performed very
recently \cite{kkslast} and compared to preliminary ZEUS data from a 
larger data sample. They obtain a good agreement for direct photon cross sections
and even for resolved photon provided a minimum $E_T^{jet} > 11$~GeV. For jets
with lower $E_T^{jet}$ the comparison is still not good and the possibility of
multiple interactions is thus confirmed on a more solid base. 

\begin{figure}[p]
\begin{center}
\epsfig{file=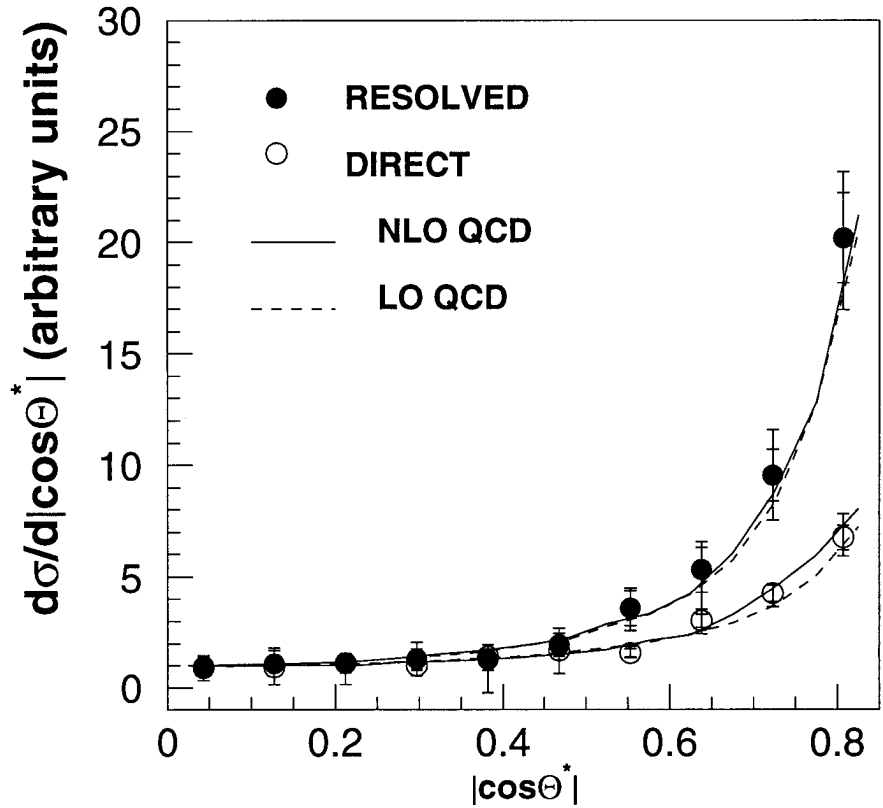}
\end{center}
\caption{
Differential $e p$ cross sections $d\sigma/d |cos \theta^\ast|$ for
dijet production in direct and resolved photoproduction as measured by ZEUS,
normalized to unity at $|cos \theta^\ast| = 0$.
Events are selected with $M_{jj} > 23$~GeV, 
$E_T^{jet} > 6$~GeV, $\bar\eta < 0.5$,$Q^2 < 4$~GeV$^2$ and $0.2 < y <0.8$.
Data are compared to LO and partial NLO QCD calculations \protect\cite{baer89}.}
\label{angdist}
\end{figure}

Besides the sensitivity to photon parton distributions, jet data can test the
QCD dynamics of the hard parton scattering. In the dijet analysis described
above, the cut $|\Delta \eta| < 0.5$ constrained the angle
$\theta^\ast$ between the jet-jet axis and the beam axis in the dijet center of
mass system to be close to $90^o$. In fact one gets:
\begin{equation}
cos\theta^\ast = tanh \left( \frac {\Delta\eta}{2} \right)
\end{equation}
Experimentally one can measure only the absolute value of $cos\theta^\ast$
because the two jets are indistinguishable. Different angular distributions are
expected for direct and resolved processes. Direct
processes, as the diagram in Fig.~\ref{dires}/a, have a quark
propagator in the $s$, $t$ or $u$ channel, with $t$ and $u$
channel processes dominating. The dominant resolved processes $g g \rightarrow
g g$, $q g \rightarrow q g$, $q q \rightarrow q q$, as the diagram in
Fig.~\ref{dires}/b, have $t$ or $u$ channel gluon exchange. 
Due to the different spin of the propagators, the angular
dependence of the cross section is thus approximately $(1 - |cos
\theta^\ast|)^{-2}$ for resolved processes and $(1- |cos\theta^\ast|)^{-1}$
for direct processes. A steeper cross section is then expected 
at small scattering angle for resolved processes. This property is expected
to be preserved even in NLO calculations \cite{baer89}.
The ZEUS experiment has measured the angular distribution in dijet events with
$E_T^{jet} > 6$~GeV and $| \bar\eta | < 0.5$ \cite{zeusangular}. 
The last cut keeps events with small boost of the dijet system in the HERA 
frame, which have the jets well inside the calorimeter acceptance:
\begin{equation}
\bar\eta \simeq \eta_{boost} = \frac{1}{2} ln \left( \frac{E_p x_p}{E_\gamma
x_\gamma} \right)
\end{equation}
A cut on the jet-jet invariant mass is applied at
$M_{jj} > 23$~GeV, to avoid the bias due to the $E_T^{jet}$ cut. The
resulting $d\sigma/d |cos \theta^\ast|$ is shown in Fig.~\ref{angdist}
for direct and resolved events, defined as before by the cut 
on $x_\gamma^{OBS}$. 
There is good agreement between data and theory. The different shapes
are not an artefact of the $x_\gamma^{OBS}$ cut. Monte Carlo studies have
confirmed that the shape is a characteristic of the spin of the propagator:
even removing the $x_\gamma^{OBS}$ cut resolved photon events generated with
Monte Carlo maintain the same shape.

\subsection{Underlying event energy}
\label{sue}

Let us come now to the problem of the observed pedestal energy. Jet
shapes are shown in Fig.~32/a-b
for direct and resolved event samples.
Here the transverse energy flow is plotted as a function of
the pseudorapidity distance from the jet axis. In the direct case 
($x_\gamma^{OBS} > 0.75$) the jet shape is well reproduced by QCD Monte Carlos
such as HERWIG or PYTHIA, including parton showers and hadronization.
Instead in the resolved case ($x_\gamma^{OBS} < 0.75$)
there is a large discrepancy on the forward side: data show a considerably
larger amount of energy. The same effect was observed by both experiments 
without a cut on $x_\gamma^{OBS}$ for jets in the 
forward hemisphere \cite{zeusincjet,h1jetold}.
The effect can be attributed consistently to resolved photon
events, since they are largely dominant for this configuration. 
A possible explanation is the occurence of additional interactions between
the proton and photon remnants, producing a soft underlying event
or a superposition of soft and semi-hard multiple interactions. These phenomena
should not happen in direct photon interactions where there is no photon 
remnant.

Multiple parton scattering has been studied in $p \bar p$ interactions
in analyses of multi-jet events \cite{hadmi}. Recently a strong signal for 
it has been reported by the CDF Collaboration \cite{cdf2p}.
The jet shapes and the pedestal energy
in $p \bar p$ interactions have also successfully been described by a
multiple interaction model \cite{sjo87}. 

Multiple interaction models are currently implemented in many Monte Carlo 
programs like PYTHIA \cite{sjo87} and  
HERWIG \cite{jimmy}, which
cover both hadron-hadron and photon-hadron interactions and in PHOJET
\cite{eng95}, which is specific to photoproduction.
The simplest model in PYTHIA generates events starting from the LO QCD 
jet cross section, where the transverse momentum cutoff $p_T^{min}$ is set
quite low, in the range of $1-2$~GeV. In this region the LO parton cross section
diverges and becomes larger than the total non diffractive cross section
$\sigma_{nd}$. The so-called  unitarization procedure consists schematically in
allowing more than one interaction per event, with mean $\langle n \rangle =
\sigma_{parton}(p_T^{min})/\sigma_{nd}$ and fluctuations calculated from
Poisson statistics. The resulting distributions are quite sensitive to 
$\langle n \rangle$,
the cutoff $p_T^{min}$ and the chosen parton densities. A similar model has
recently been implemented in HERWIG. The PHOJET generator is intended to
simulate at once all the components contributing to the total photoproduction
cross section. It is based on the two component Dual Parton Model \cite{cap94}
and includes multiple soft and hard interactions on the basis of a
unitarization scheme \cite{cap87}. Due to this scheme the dependence on the
$p_T^{min}$ separating the soft and hard regions is rather weak. The soft part
is described by soft pomeron phenomenology, with parameters tuned to $p \bar p$
and low energy $\gamma p$ interactions. 

These models have been used by the H1 Collaboration in a detailed study of the
properties of photoproduction events related to underlying multiple
interactions \cite{h1incjet}.
The average level of $E_T$ outside jets has been studied in dijet events as a
function of $x_\gamma^{OBS}$. In this study the pseudorapidity difference
$\Delta\eta$ between the two jets was required to be $\Delta\eta < 1.2$,
to avoid a possible mismatch of the photon remnant as a hard jet. 
In Fig.~\ref{etped} the average transverse energy density 
$\langle E_T \rangle/\Delta\eta
\Delta\phi$ is shown as a function of $x_\gamma^{OBS}$. It is determined from 
the summed $E_T$ in the central region of the $\gamma p$ collision 
($-1 < \eta^\ast < 1$), excluding the energy contained in cones of radius 
$1.3$ around the jet
axes. The long-dashed line is the energy density found in minimum bias events,
where of course $x_\gamma^{OBS}$ is undefined. The energy density increases
significantly going from the direct photon ($x_\gamma^{OBS} \approx 1$), 
where it is close to the level found in minimum bias events, 
to low $x_\gamma^{OBS}$ events.
The increase is much higher than what is expected from QCD Monte Carlos without
multiple interactions like the PYTHIA dotted curve, and cannot be attributed to
gluon radiation or fragmentation effects. Instead data can be described by
models with interactions between the beam remnants as PHOJET or the PYTHIA
dashed line.

This becomes more convincing when one looks at the energy-energy correlation
with respect to the central region. The pseudorapidity correlation function
$\Omega (\eta^\ast)$ is defined as:
\begin{equation}
\Omega (\eta^\ast) = \frac{1}{N_{ev}} 
\frac {\sum_{i=1}^{N_{ev}} (<E_{T,\eta^\ast=0}> - E_{T,\eta^\ast=0,i}) 
                           (<E_{T,\eta^\ast}> - E_{T,\eta^\ast,i})} 
{(E_T^2)_i}
\end{equation}
H1 has measured this quantity with respect to $\eta^\ast = 0$, where the average
transverse energy density is at the maximum, for a sample of high $E_T$ events.
The result is shown in Fig.~\ref{eec}: data show short range correlations
around $\eta^\ast = 0$ and long range anticorrelations in the photon hemisphere,
with a minimum at $\eta^\ast \approx -1.8$. From PYTHIA without multiple
interactions
one would expect stronger correlation and anticorrelation than data
show. Adding multiple
interactions the correlation is reduced in both signs and agreement with data
greatly improves. The conclusion is that the addition of uncorrelated energy to
the events results in the correct average $E_T$ level in the underlying event
(Fig.~\ref{etped}) and also gives the correct correlation strength 
(Fig.~\ref{eec}).

\subsection{Unfolding the gluon density in the photon}

The study of the underlying energy carried out by H1 has also been used to
correct their measured jet energies by subtracting the pedestal. The applied
correction is parametrized as a function of $\eta^{jet}$ and has size of 
$0.3 - 2.3$~GeV. 
From an analysis of dijet events with pedestal subtraction
H1 \cite{h1xgx} has unfolded LO parton
variables, based on the PYTHIA Monte Carlo with multiple interactions included.
This introduces some model dependence on the results. The data refer to
$290$~nb$^{-1}$ tagged photoproduction data from the 1993 run: 
the final sample is not large, 292 events with selected
two jets with $E_T^{jet} > 7$~GeV and $0 < \eta^{jet} < 2.5$. 
First it is demonstrated the need of a gluon component in the photon and then
the gluon momentum distribution is unfolded at LO. 
The distribution of the selected events in the unfolded $x_\gamma$ is shown 
in Fig.~\ref{h1x}. The full histogram represents the expected contribution of 
quarks and antiquarks in the photon, the dashed histogram the direct photon 
contribution, which is added on top
of the other. These two are simply obtained from a LO QCD calculation at the
parton level, using GRV LO parton distributions for the proton and the
photon. The proton parton distributions at the relevant momentum fraction are
well known and do not weaken the resolving power on the photon parton
distributions. Data are well above the histogram at low $x_\gamma$, 
which is right the region where the
contribution of gluons in the photon is expected to be important. The need
for a gluon component has been checked with a $\chi^2$ test on this
distribution, trying to fit data with LO QCD predictions without gluons in the
photon. Only the sum of direct-$\gamma$ and
quark-antiquarks in $\gamma$ has been considered and systematic corrections 
have been applied to both data and theoretical prediction in the sense to
decrease the discrepancy. 
The resulting probability is found to be $0.1\%$. 
The unfolded gluon momentum distribution is shown in Fig.~\ref{h1xgxfig} 
divided by $\alpha_{em}$. The average scale of the selected events is 
$\langle {p_T}^{jet} \rangle ^2 =
75$~GeV$^2$, where ${p_T}^{jet}$ is used as factorization and renormalization
scale for the QCD calculation. The resulting gluon distribution is compared to
LO parametrizations of GRV-LO, LAC1 and LAC3. This latter is clearly excluded,
in agreement with previous observations at HERA \cite{h1jetold}, TRISTAN
\cite{topajet,amyjet} and LEP \cite{lepjet}. The strong rise of LAC1 at low $x$
is disfavoured, while GRV-LO is consistent with data. This measurement
constituted the first extraction of a LO gluon density down to 
$x_\gamma = 0.04$.

\subsection{Photon remnant}

As we have seen, direct and resolved processes are physically distinct as is
apparent from the $x_\gamma^{OBS}$ distribution in dijet events, and enriched
samples can be obtained by cutting on this distribution.
Another possible separation criterion relies on the different topology of the
events. Figures 37 and 38
show respectively a direct and a
resolved event candidate as reconstructed in the ZEUS detector. In these
pictures the $z$-axis (proton beam direction) points to the left. The direct
candidate has no activity in the rear calorimeter, i.e. in the photon
direction. Instead in the resolved candidate several particles are found to go
in the photon direction. Both the events have in addition two high $E_T$ jets.
The activity in the rear direction is interpreted as coming from the
fragmentation of the photon remnant, 
which is present only for a resolved photon.

The ZEUS Collaboration has studied the properties of the photon remnant in
events with two hard jets with $E_T^{jet} > 6$~GeV and $\eta^{jet} < 1.6$
\cite{zeusphremn}. To look for a photon remnant jet, emerging with small
$p_T$ at small angles in the photon direction, the cone algorithm is not well
suited. The jet finding was instead carried out with the $k_T$-algorithm
\cite{cat93}. It is applied starting from single calorimeter cells, as if they
were particles. The prescription of the algorithm is to evaluate for each pair
of clusters (initially single cells) the quantity:
\begin{equation}
k_T^2 = 2 ~min(E_i^2,E_j^2) ~(1 - cos \theta_{ij})
\end{equation}
This is the transverse momentum of the lower energy cluster with respect to the
higher energy cluster, for small angular separation $\theta_{ij}$. The proton
remnant (escaping detection in the forward direction) is considered as a
pseudoparticle with infinite momentum along $+z$. After all the $k_T^2$ values
have been calculated, the two clusters with minimum $k_T^2$ are merged and 
their four-momenta added to give the four-momentum of the new cluster.
The procedure is iterated until for all pairs the resolution
variable $Y$ becomes larger than some threshold $Y_{cut}$. Here $Y$ is defined
as: $Y = k_T^2 / ({E_T^{tot}})^2 $, where $E_T^{tot}$ is the total
transverse energy in the event.
In the ZEUS analysis this $Y_{cut}$ value has been chosen to vary from event
to event in order to find always three clusters (apart from the proton
remnant), ordered according to their $E_T^{jet}$. 
After comparison between data and Monte Carlo events the photon
remnant has been identified in most cases as the source of the third cluster,
which appears typically in the photon fragmentation region.
The final selection required for it a minimum energy $E_3 > 2$~GeV 
and pseudorapidity $\eta_3 < -1$. 
After these cuts the $x_\gamma^{OBS}$ distribution is
found to peak at low values as expected from resolved photon events.
In Fig.~\ref{grem}/a-b the pseudorapidity and transverse momentum
distributions are shown for the third cluster, identified as the photon remnant. A
comparison with the standard QCD simulation by PYTHIA shows that 
data are shifted towards higher $p_T$ and more central $\eta$ values. 
The transverse momentum
for the photon remnant is generated in PYTHIA with gaussian distribution 
$dN/dk_T^2 \sim e^{- k_T^2/k_0^2}$ with default $k_0 = 0.44$~GeV. 
This parameter determines the
hardness of the $k_T$ spectrum. A fit to the data requires a higher $k_0$ of
$1.7 - 1.9$~GeV. This observation agrees with studies suggesting that NLO
contributions or fluctuations of the photon into $q \bar q$ pairs with high
virtuality may lead to a photon remnant with sizeable transverse momentum with
respect to the photon direction \cite{gremn}.
The properties of the jet from the photon remnant 
have been compared to those of the hard jets, 
assumed to originate from the hard parton scattering. Fig.~\ref{compa}/d
shows the mean value of the particle transverse energy with respect to the
cluster axis as a function of the cluster energy; Fig.~\ref{compa}/e shows the
average values of the scalar sums of transverse and longitudinal particle
energies with respect to the cluster axis as a function of the cluster energy.
Particle quantities are measured from calorimetric islands, i.e.
groups of adjacent
calorimeter cells, and corrected to the hadron level by a complete detector
simulation. Figure \ref{compa}/f shows the energy flow as a function of the
distance from the cluster axis, expressed as $1 - cos~\theta$. In all the
plots the photon remnant and the hard jets exhibit the same hadronization
properties. 
The average particle transverse energy with respect to the jet axis 
$\langle E_T^i \rangle$ is of the order of few hundred MeV and increases 
only weakly with the jet energy. 
The longitudinal component of the jet energy increases much faster
than the transverse energy w.r.t the jet axis.
Fig.~\ref{compa}/a-b-c compares the photon remnant distributions
with expectations from PYTHIA; they are in good agreement
after the tuning of the intrinsic $k_T$ distribution of the photon remnant.

\subsection{Open charm production}

Heavy quark photoproduction has been the subject of many theoretical studies
\cite{heavy}. 
Recent NLO QCD calculations for the HERA energies have been published for both
total and differential cross sections \cite{fri95}. 
A large amount of data exist from fixed target experiments with center of mass
energies below $20-30$~GeV \cite{fixccbar}.

In leading order the dominant production mechanism is expected to be the direct
photon-gluon fusion process $\gamma g \rightarrow Q \bar Q$, with $Q = c,b$,
which is represented by the diagram in Fig.~\ref{dires}/a, where the fermion
line is a heavy quark.
This is mostly a photoproduction process, with a tail at high $Q^2$ which is
not dominant. Charm production is much more abundant than
beauty production; the predicted cross sections at HERA 
are about $1$~$\mu$b for charm and about $10$~nb for beauty. 
Monte Carlo studies show that the dominance of charm production 
still holds after the application of experimental tagging and selection
criteria, that is why for the present analyses, with the achieved HERA
luminosity, beauty production has not yet been considered. In LO QCD charm
production can occur also in resolved photon interactions, mainly through 
the process $g g \rightarrow c \bar c$. 
The relative fraction of direct and resolved photon is
very sensitive to the photon parton densities. For parametrizations not
steeply rising at low $x_\gamma$, like GRV, the resolved contribution is of the
order of $20\%$, while for LAC1 it is more than $50\%$ of the total cross
section. Another strong source
of variations in the theoretical predictions is the value of the charm mass
$m_c$, which is generally taken in the range $1.2-1.8$~GeV. Variation of the
mass in this range changes the total cross section by more than a factor of 2.
Charm production is somewhat at the boundary of applicability of perturbative
QCD, due to the relative smallness of the charm mass, which is here the relevant
scale for the perturbative expansion.
The renormalization scale is set to $m_c$ and the factorization
scales for proton and photon parton distributions are set to $2 m_c$, according
to \cite{fri95}. Large variations occur by changing the scales. 
The large spread in the predictions justifies the interest in measuring
the charm production at HERA, aiming to fix the input parameters of the
calculation.
The most interesting topic may be the determination of the gluon density in the
proton, due to the predicted dominance of the direct photon production in the 
experimentally accessible region. In contrast to the measurements 
carried out from the scaling violations in DIS \cite{h1f2,xgdiscav}, 
the gluon density enters directly at the leading order of QCD in charm 
production. Several possibilities to extract it have been investigated 
\cite{vwoude,friglup}.

HERA experiments have measured the charm production mainly through
the reconstruction of the $D^\ast(2010)^\pm$ meson decays. The golden channel
is ${D^\ast}^+ \rightarrow D^0 \pi_s^+$, due to the very small mass difference
between ${D^\ast}^+$ and $D^0$, $\Delta M = 145.42$~MeV; the slow pion
$\pi_s^+$ has only $40$~MeV momentum in the $D^\ast$ rest frame. This special
kinematic constraint allows $\Delta M$ to be measured more accurately than the 
$D^\ast$ mass itself \cite{fel77}. The $D^0$ has been reconstructed 
in the channels $D^0 \rightarrow K^- \pi^+$ or 
$ D^0 \rightarrow K^- \pi^+ \pi^+ \pi^-$. The charge conjugate decay chain from
the ${D^\ast}^-$ is implicitly considered.

Results on the total cross section $\gamma p \rightarrow c \bar c X$
have been published by both ZEUS \cite{zeusds} and H1 \cite{h1ds}. 
H1 results come from 1994 data, amounting to $2.77$~pb$^{-1}$
tagged photoproduction data ($Q^2 < 0.01$~GeV$^2$, $159 < W < 242$~GeV) and
$1.29$~pb$^{-1}$ untagged data ($Q^2 < 4$~GeV$^2$, $95 < W < 268$~GeV). The
accepted kinematic region is $p_T(D^\ast) > 2.5$~GeV and 
$-1.5 < \rap(D^\ast) < 1$, where $\rap$ indicates the rapidity.
The $p_T$ cut depends on the minimum $p_T$ needed to measure the tracks
from the $D^\ast$ decay.
The rapidity cuts select a well understood region of the tracking
detectors.  No particle identification has been used yet;
well measured tracks are simply combined to form invariant masses. 
These measurements are
shown in Fig.~41
as a function of the average $\gamma p$ center of
mass energy $W$, together with lower energy measurements from fixed target
experiments \cite{fixccbar}. The cross
section rises by almost one order of magnitude compared to the energy
range of fixed target experiments. The curves are NLO QCD calculations
\cite{fri95} with parton density sets MRS(G) \cite{mrsg} for the proton and 
GRV-HO for the photon.
They limit the variation produced by a change of the renormalization scale
$\mu$ in the range $m_c/2 < \mu < 2 m_c$. 
The limited acceptance of the HERA experiments makes the measurement of the
total charm cross section rather model dependent, because of the need to
extrapolate from the accepted kinematic region. 
The extrapolated $\sigma_{\gamma p}$
increases by $75\%$ if LAC1 photon parton densities are used instead of GRV-HO;
it decreases by $35\%$ if the MRS(A') proton parton densities are used instead
of MRS(G). These uncertainties are included in the error bars. Data are 
not able to discriminate between different available gluon density 
parametrizations, even though they favour rising distributions at low $x$. 
The visible cross
section is instead almost insensitive to the choice of the parton distributions
and to the mixture of direct and resolved processes assumed in the Monte Carlo
to calculate the efficiency. Then cross sections within the experimental cuts
can be safely compared with theory in a model independent way.

Differential cross sections versus the transverse momentum or the
(pseudo)rapidity of the $D^\ast$ have been measured by both the experiments
\cite{h1ds,zeusds94}. ZEUS data come from $3$~pb$^{-1}$ of the 1994 run.
Fig.~\ref{dstarpt} shows the differential cross section
$d\sigma/dp_T$ for the observed $D^{\ast\pm}$. 
The theoretical curves are two very different NLO QCD calculations:
the dotted and the dashed-dotted curve stem from an exact fixed-order 
NLO calculation including the charm mass \cite{fri95}; 
the full curve is instead a NLO 
resummed calculation, where the charm mass is neglected in comparison to the
$c$-quark transverse momentum \cite{kni95,spira96} (limit of large $p_T/m_c$).
The latter approach has been pursued also by another group
\cite{cagre96,newcaccia} and is sometimes referred to as the massless approach. 
Within the limits of
applicability of this approach the charm is considered an active flavour in
both the proton and the photon structure. On the contrary in the massive
approach charm is generated from light partons and is excluded from the initial
state. 
Both the calculations take into account the  probability for a charm
quark to fragment to $D^\ast$: ${\cal P} (c \rightarrow {D^\ast}^+) = 26.0
\pm 2.1 \%$ \cite{opalds}.
The $D^\ast$ spectrum is obtained from the calculated 
$c$-quark spectrum convoluting it with the Peterson fragmentation 
function \cite{pet83}, which apart from the normalization is: 
\begin{equation}
f(z) \sim \left[ z \left( 1 - \frac{1}{z} - \frac{\epsilon_c}{1-z} \right)^2
\right]^{-1}
\label{peteff}
\end{equation} 
Here $z$ is the fraction of the charm quark momentum taken by the $D^\ast$, and
$\epsilon_c = 0.06$ is a free parameter which is fitted
from experimental data \cite{chr87}.
The massive calculation with default parameters gives the dash-dotted curve.
It has the correct shape but the normalization
is about half that of data. With an extreme choice of $m_c = 1.2$~GeV and
renormalization scale equal to half the transverse mass the dotted curve is
obtained. 
The effect of changing the charm mass in the range $1.2 - 1.8$~GeV gives only a
$\pm 20\%$ variation inside the accepted region.
Instead the resummed calculation with equal parameters gives the correct 
normalization. 
This agreement is somewhat unexpected at least in the lowest $p_T$ region
covered by the data, where the massive calculation should be 
definitely better. 
A possible explanation to this situation is suggested in \cite{newcaccia}. 
Here it is argued that the $\epsilon_c$ parameter in the
Peterson fragmentation function (\ref{peteff}) varies sensibly depending on
whether the fit to data is done using a LO or a NLO QCD calculation for the
partonic piece of the cross section. Using their NLO resummed calculation the
authors find the best fit for $\epsilon_c \simeq 0.015 - 0.02$ to be compared
to the commonly used value $0.06$, which is found with a LO calculation. This 
change would increase sizeably the $D^\ast$ cross section within the
experimental cuts, and would bring also the massive calculation closer to data.  
In Fig.~\ref{dstarrap} the differential cross section as a function of the
$D^\ast$ rapidity is shown, from the H1 analysis of untagged photoproduction
data. Data are above the theoretical prediction of the massive calculation
\cite{fri95} in the forward region. 
Different curves show that the effect of changing the proton and photon
structure functions is small. In particular using the LAC1 instead of the 
GRV(HO) set for the photon makes only a small variation.

\subsection{Inelastic $J/\psi$ production}

Another relevant process to access the gluon density in the proton is inelastic
$J/\psi$ production. Here again the leading mechanism is thought to be a
photon-gluon fusion, $\gamma g \rightarrow J/\psi ~ g$, in the so-called
Colour-Singlet Model \cite{ber81}, represented by the diagram in 
Fig.~\ref{colsin}. The name is due to the assumption that 
the $c \overline{c}$ pair giving the $J/\psi$ lies in a colour-singlet state.
Extraction of the gluon density from this process has been the aim of
analyses at fixed target experiments \cite{emc83}. There, a discrepancy was 
found in the absolute normalization, with the 
data being higher than the predictions by a
factor from 2 to 5. This was ascribed to missing higher order contributions
(K-factor). Recently NLO QCD calculations have been performed \cite{kra95,kra96}
and agreement with those data was obtained.
HERA data have been published by H1 from an integrated luminosity of 
$2.7$~pb$^{-1}$ \cite{h1psi94}. 
The selected decay mode was $J/\psi \rightarrow \mu^+ \mu^-$. 
The direct photon-gluon fusion process has to be separated by other processes 
with quite different characteristics, like diffractive production with proton
dissociation and production via resolved photon \cite{jun92}. They are best
discriminated by measuring the inelasticity variable $z$, defined as:
\begin{equation}
z = \frac{p_{\psi} \cdot P} {q \cdot P} 
\end{equation}
where $p_\psi$, $P$, $q$ are respectively the four-momenta of the $J/\psi$, the
initial proton and the photon. This expression is evaluated from measured
energy and momentum variables as:
\begin{equation}
z = \frac{(E - p_z)_{\psi}} {(E - p_z)_\gamma} = \frac{(E - p_z)_{\psi}} 
{\sum_h (E - p_z)_h }
\label{zexpr}
\end{equation}
with the sum extended to all the final state particles except the scattered
electron. In the proton rest frame $z$ would be simply the ratio of the $J/\psi$
to the photon energy. $z \simeq 1$ in the elastic process 
$\gamma p \rightarrow J/\psi ~p$, since in that 
case almost all the photon energy is transferred to the $J/\psi$. $z \approx 1$
also for events where the proton diffractively dissociates.
Experimentally if particles are detected from the proton breakup
they are at very forward angles, with $p_z \simeq E$, thus
contributing only a little to expression (\ref{zexpr}). 
On the other hand resolved photon production contributes
at low $z$, due to the reduced energy that the photon brings into the hard
subprocess. A clean sample of photon-gluon fusion events is selected in the
range $0.45 < z < 0.9$ after detailed studies and Monte Carlo simulation of all
the relevant processes. The cross section has been measured extrapolating 
to the low $z$ range and is shown in Fig.~\ref{psicros}/a. Data are
compared to a NLO calculation in the colour singlet model \cite{kra96} with
different gluon density parametrizations. All the curves assume $m_c =
1.4$~GeV, $\Lambda_{\overline{MS}} = 300$~MeV and renormalization and factorization
scales set to $\sqrt{2} m_c$. The agreement seems better for the distributions
rising more steeply at low $x$, like MRS(G). Actually the NLO calculation is 
not reliable at high $z$ and low $p_T$ of the $J/\psi$. This is apparent from 
the differential cross section $d\sigma / d p_T^2$ in Fig.~\ref{inelpsiptz}/a, 
where the theoretical curve bends over
at $p_T^2 < 1$~GeV$^2$. A safe kinematic region to avoid theoretical problems
is $z < 0.8$, $p_T^2 > 1$~GeV$^2$: this corresponds to the requirement that the
emitted gluon is hard. The cross section in this restricted region is shown in
Fig.~\ref{psicros}/b. Unfortunately the sensitivity to the input gluon
distribution is decreased and all the curves agree with data. An exponential
fit to $d\sigma /d p_T^2$ with an exponential shape $e^{-b p_T^2}$ 
gives a slope $b = 0.39
\pm 0.06 \pm 0.03$~GeV$^{-2}$, in agreement with the theoretical calculation
which gives $b = 0.3$~GeV$^{-2}$ for $p_T^2 > 1$~GeV$^2$.

Data on inelastic $J/\psi$ and $\psi '$ production at the Tevatron 
\cite{tevapsi} have shown a huge discrepancy with the predictions 
of the Colour Singlet Model. 
They can however be fitted assuming additional colour octet terms 
in the general formalism of Bodwin, Braaten and Lepage \cite{braaten}. 
This is not a new model: 
the basic idea is that the $c \bar c$ pair may be either in a colour
singlet or in a colour octet state before producing the $J/\psi$. In fact a 
colour octet state becomes a singlet simply by the emission of soft gluons. 
This non-perturbative step cannot in principle be neglected. The 
calculation needs phenomenological inputs which have to be fitted from data. 
To verify if HERA data are consistent with those observations,
calculations have recently been performed to assess the effect of possible
colour octet contributions at HERA \cite{cac96,amu96}. 
The prediction would be of a
very steep increase of the production cross section for inelastic $J/\psi$ at
high $z$. In Fig.~\ref{inelpsiptz}/b $d\sigma/dz$ as measured by H1 is
compared to the prediction of NLO colour singlet alone and to LO colour octet
fixing its parameters to the CDF data \cite{tevapsi}. It is clear that HERA
data exclude any large colour octet contribution.

\section{Hard diffraction in photoproduction}
\label{hdiff}
 
Both ZEUS and H1 have reported the
measurement of a large diffractive component in photoproduction, about $40\%$
of the total $\gamma p$ cross section \cite{h1stot,zeustot}, as
discussed in section \ref{totsgp}. This is expected in view of the
analogy of $\gamma p$ to hadron-hadron interactions. Diffractive processes are
not limited to soft interactions. The first observations of diffraction in hard
collisions at HERA have come from DIS measurements \cite{disrapgap}. 
The experimental signature for
diffractive events was the observation of a large rapidity gap
around the proton beam direction. This is customarily specified by the variable
$\eta_{max}$, 
defined as the pseudorapidity of the most forward energy deposit in the
calorimeter with energy above $400$~MeV. Diffractive event samples have been
defined having $\eta_{max} < 1.5 - 2$, which means that no particles are seen 
in the detectors within a cone of $25^o - 15^o$ degrees around the proton beam 
direction.
This cut selects mainly single photon dissociation events, where the initial
proton is not broken up. Minor components are events where the proton
dissociates to a low mass system which is not detected or non-diffractive
events with unusual energy flow. About $10\%$ of the DIS events were found
to have the large rapidity gap signature. 
The dependence of the event rate on $W$ as well
as the approximate scaling observed in the DIS data indicated a diffractive
process of leading twist. The diffractive structure function $F_2^{D(3)}$
\cite{f2diff,h1f2diff} and the cross section as a function of the mass of the 
diffractive system $M_X$ \cite{mxdiff} have then been measured.
Hard diffractive events have also been observed in photoproduction 
\cite{gprapgap}. Here events with high $E_T$ jets and the large rapidity gap
signature were found.

\subsection{Jets and heavy flavours from photon diffractive dissociation}

The picture of a hard diffractive event in photoproduction is
shown in Fig.~\ref{evtdiff}. No activity is seen in the forward calorimeter,
although two hard jets are visible in the central region. 

The interest in hard diffractive processes is motivated by the possibility of
perturbative QCD calculations. This can help to understand the nature of the
pomeron, considering it as a partonic system \cite{low75}. Ingelman and Schlein
\cite{is85} gave a particle-like description of the pomeron taking inspiration 
from ISR $p p$ data \cite{r608}. 
Their model was confirmed by the UA8 experiment \cite{ua8}, which
observed jet production in $p \bar p$ diffractive interactions with a tagged
leading proton.

Let us describe in some detail how the Ingelman-Schlein model is
implemented. It assumes that the proton emits a pomeron: 
a colourless object with vacuum quantum numbers. 
The variable $x_{\Pma}$ is the fraction of the proton momentum taken by the 
pomeron. The square of the photon-pomeron center of mass energy is
given by: $M_X^2 \simeq x_{\Pma} y s$. The flux of pomerons from the proton is
factorized in the cross section and its expression is
derived from hadron-hadron data. It is a function of $x_{\Pma}$ 
and $t$, the momentum transfer from the proton. Different forms for the pomeron
flux factor are currently used. 
The Donnachie-Landshoff form \cite{don88} is calculated in Regge theory with
parameters fitted to hadronic data:
\begin{equation}
 f_{\Pma/p}(x_{\Pma},t) = \frac{9 b_0^2}{4 \pi^2} F_1(t)^2
   x_{\Pma}^{1-2\alpha(t)}
\label{dlfluxf}
\end{equation}
where $F_1(t)$ is the elastic form factor of the proton, 
$b_0 \simeq 1.8$~GeV$^{-1}$ is the pomeron-quark coupling and 
$\alpha(t) = 1.08 + 0.25 ~t$(GeV$^2$) is the pomeron trajectory.
The systematic uncertainty 
related to this choice is about $30\%$, from the comparison of different 
procedures to extract the flux factor. 

 In the Ingelman-Schlein model the pomeron is then assumed to be a source of 
partons which may interact with
the photon in both direct and resolved processes, according to the pointlike
or hadron-like behaviour of the photon.
Parton densities of the pomeron $f_{i/\Pma}(\beta,\mu^2)$ are introduced in
perfect analogy to the proton case, where $\beta$ is the scaling variable, i.e. 
the fraction of the pomeron momentum carried by parton $i$, and $\mu$ is the
energy scale at which the pomeron is probed. The partonic cross
section is then calculated in LO QCD and factorization of these parton
densities is assumed. If this would be the case they could be extracted
independently of the experiment. Several forms for the pomeron
structure have been suggested, the most extreme ones are the following:
\begin{itemize}
\item $\Pma$ entirely made of gluons, with a hard distribution: \\
$\beta f_{g/\Pma}(\beta,\mu^2)=6\beta(1-\beta)$;
\item $\Pma$ entirely made of gluons, with a soft distribution: \\
$\beta f_{g/\Pma}(\beta,\mu^2)=6(1-\beta)^5$;
\item $\Pma$ entirely made of $q \bar q$ pairs (only two light flavours), 
with a hard distribution: \\
$\beta f_{q/\Pma}(\beta,\mu^2)= \frac{6}{4} \beta(1-\beta)$.
\end{itemize}
The momentum sum for the pomeron is written as:
\begin{equation}
\Sigma_{\Pma}(\mu^2)
\equiv \int_0^1 d\beta \sum_i \beta f_{i/\Pma}(\beta,\mu^2)
\label{msrule}
\end{equation}
and in the Ingelman-Schlein model it is assumed $\Sigma_{\Pma} = 1$.
This is however only a guess. Since the pomeron is not a real particle,
the flux factor $f_{\Pma/p}$ and the momentum sum $\Sigma_{\Pma}$ may
be not independent. In the following the dependence on $\mu^2$ is neglected.

Jet cross sections in diffractive photoproduction have been measured by ZEUS
\cite{zeusdifglu} from $0.55$~pb$^{-1}$ of untagged data.
Data are corrected at the hadron level by detector simulation. Preselected
events have a jet with $E_T^{jet} > 6$~GeV (at the calorimeter level) 
in the central region $-1 < \eta^{jet} < 1$. This corresponds to $E_T^{jet} >
8$~GeV at the hadron level. 
The invariant mass of the hadronic system is
reconstructed as: 
$M_X = \sqrt{ \sum_i E_i^2 - \sum_i \mathbf{p_i^2} }$, where the sum is
extended to all the calorimeter cells. 
Fig.~\ref{gpomdist}/a shows the correlation between $M_X$ and
$\eta_{max}$: it is apparent that there is a distinct class of events with low
$M_X$ that have low $\eta_{max}$, that is a large rapidity gap.
The distribution of $\eta_{max}$ for events with $M_X < 30$~GeV 
is shown in Fig.~\ref{gpomdist}/b. The
simulation of non-diffractive processes by the PYTHIA Monte Carlo is not able to
describe the tail at low $\eta_{max}$. Instead the POMPYT Monte Carlo
\cite{bru94}, based on
the Ingelman-Schlein model \cite{is85}, describes well the shape in this region.
Fig.~\ref{gpomdist}/c shows the $M_X$ distribution after the cut 
$\eta_{max} < 1.8$ and
Fig.~\ref{gpomdist}/d the $W$ distribution for the selected events.
The flat $W$ behaviour is in agreement with the expectations of POMPYT.

Figure \ref{setadif} shows the differential jet cross section versus 
$\eta^{jet}$ for the large rapidity gap events. Data
include a contribution from double dissociation, where the proton diffractively
dissociate to a hadronic system of mass $M_N \leq 4$~GeV which is not detected.
The dominant errors are statistical, while the largest systematic error 
($\approx 20\%$) is due to the uncertainty on the absolute energy scale 
of the calorimeter. 
LO QCD predictions obtained by Monte Carlo are superimposed. 
Large rapidity gaps may
occur in normal non-diffractive events as fluctuations in the pseudorapidity 
distribution of the final state hadrons. They are however exponentially 
suppressed with increasing gap width. The size of the non-diffractive
contribution has been estimated with PYTHIA
using MRS(D-) parton distributions for the proton \cite{mrsd} and GRV(HO) for
the photon. The result is considerably lower than data and differs in shape.
Data are instead in good agreement with the expectations of POMPYT assuming 
a hard gluon density inside the pomeron, while a pomeron made of hard quarks
gives the correct shape but too low a normalization. 
A soft gluon density gives inconsistent results on both
shape and normalization. These data are however insensitive to an additional
soft component in the pomeron, present together with the hard component.

The diffractive jet data are sensitive to both the quark and the gluon content 
in the pomeron. 
The recent HERA measurements of the diffractive structure function in DIS 
\cite{f2diff,h1f2diff} have provided information on the quark content. 
This appears to have both a hard and a soft component.
Moreover these measurements do not favour a pomeron made only of quarks and
antiquarks.
If factorization of the pomeron parton densities holds, the pomeron structure
would appear the same whether it is probed in DIS or in photoproduction.
Therefore, assuming factorization, 
the gluon content of the pomeron can be studied
combining DIS and $\gamma p$ jet data. For that purpose a pomeron 
composed of both hard gluon and quark distributions has been assumed, naming
$c_g$ and $c_q = 1 - c_g$ the relative gluon and quark fractions.
Contributions from a possible soft parton component have
been neglected, since the jet data are sensitive only to $\beta \geq 0.3$. 
The distribution of $d\sigma/d\eta^{jet}$ was corrected subtracting the 
non-diffractive and the double dissociation contributions. The first one was
estimated with PYTHIA, which gives a good description of non-diffractive jet
production in the central rapidity region. The latter was estimated from data 
to be $15\pm 10 \%$ and it has been assumed independent from
$\eta^{jet}$ according to the theoretical expectations.
Then a $\chi^2$ fit to the measured $d\sigma/d\eta^{jet}$ was carried out
varying the gluon fraction $c_g$ and leaving free the momentum sum
(\ref{msrule}) in POMPYT. 
The results are shown in Fig.~\ref{strisce} by
the solid line. The shaded band gives the $1 \sigma$ contour.

From the measurement of $F_2^{D(3)}(\beta,Q^2,x_{\Pma})$ in DIS we
can find the quark contribution to the momentum sum,
$\sum_{{\Pma} q}$, integrating over $\beta$ and $x_{\Pma}$:
\begin{equation}
  \int_{x_{{\Pma}{min}}}^{x_{{\Pma}{max}}} dx_{\Pma} \int_{0}^{1}
  d\beta \; F_2^{D(3)}(\beta,Q^2,x_{\Pma})
  = k_f \cdot  \Sigma_{{\Pma}q} \cdot I_{flux}
\end{equation}
where $I_{flux}$ is the integral of the pomeron flux factor over $t$ 
and over the
same region in $x_{\Pma}$ and $k_f$ is a number depending on the assumed number
of flavours ($5/18$ for two flavours, $2/9$ for three flavours). The integral
on the left hand side is performed using ZEUS data in the range 
$6.3 \cdot 10^{-4} < x_{\Pma} < 10^{-2}$. The flux factor
$I_{flux}$ is calculated using expression (\ref{dlfluxf}).
The result is: 
$\Sigma_{{\Pma}q} = 0.32 \pm 0.05$ ($0.40 \pm 0.07$) for two (three)
flavours. The diffractive structure function $F_2^{D(3)}$ has been
parametrized from ZEUS data in the $\beta$ region 
$0.1 < \beta < 0.8$.
Here the same form is assumed to hold in the whole range $0 < \beta < 1$.
$F_2^{D(3)}$ is approximately independent of $Q^2$ in
the range of the DIS measurements: $8$~GeV$^2 < Q^2 < 100$~GeV$^2$. Hence the
dependence of $\Sigma_{{\Pma}q}$ on $Q^2$ is neglected.
Moreover the $Q^2$ values are comparable to the values of $(E_T^{jet})^2$ 
of the selected jet events. Thus the DIS results
are expressed as a constraint in the $\Sigma_{\Pma} - c_g$ plane of figure
\ref{strisce} (dot-dashed lines):
\begin{equation}
\Sigma_{\Pma} \cdot (1-c_g) = \Sigma_{{\Pma}q} = 
0.32 ~(0.40)  ~~~~~ for ~2 ~(3) ~flavours
\end{equation}

Combining the two results and considering all the systematic uncertainties one
obtains: $0.4 < \Sigma_{\Pma} < 1.6$ and $0.3 < c_g < 0.8$. 
From this nothing can be safely concluded on the validity of a
momentum sum rule for the pomeron. However, assuming the same pomeron
flux factor in DIS and photoproduction, this determination of the gluon 
content of the pomeron is independent on the normalization and even on the 
validity of a momentum sum rule. 
From this analysis one concludes that between $30\%$ and
$80\%$ of the momentum of the pomeron carried by partons is due to hard gluons.

Further information on the partonic structure of the pomeron can be obtained
from the diffractive production of heavy flavours.
The diffractive production of $J/\psi$ has been discussed in section
\ref{vmsection} because of the similarity with the light vector meson analyses.
It illuminates the realm of hard diffraction, exhibiting a strong 
difference from light vector mesons, particularly in the steep dependence of
the cross section on the center of mass energy.
Clean data samples \cite{h1psi94,zeuspsi94}
with good statistics have already been studied at HERA and
succesfully compared to calculations based on perturbative QCD \cite{rysk}. 
However for $J/\psi$ elastic production the theory retains some weakness 
related to long distance contributions in the cross section.
Instead the diffractive production of open charm is in principle free from
these uncertainties and recently
has attracted much theoretical interest \cite{diffch}.
This process has been searched for by H1 \cite{h1ds}. They observed evidence of
diffractive charm production by inspecting the $\eta_{max}$ distribution, 
shown in Fig.~\ref{etamaxds} for selected $D^\ast$ events. Here statistical
subtraction of the background under the mass peak has been applied. The QCD
Monte Carlo representing non-diffractive production is unable to reproduce the
distribution, while the hard diffraction model implemented in the RAPGAP 
generator \cite{jun95} reproduces well
the shape when added to PYTHIA. H1 sets a lower limit on the diffractive cross
section for $p_T(D^\ast) > 2.5$~GeV, $-1.5 < \rap(D^\ast) < 1$ and 
$\eta_{max} < 2$ in untagged events ($Q^2 < 4$~GeV$^2$ and $0.1 < y < 0.8$) to:
$\sigma(e p \rightarrow D^{\ast \pm} X) > 145$~pb   at  $90\%$ confidence 
level. Comparing this limit to predictions of RAPGAP, a quark dominated pomeron
is disfavoured. In fact, the
predicted cross section for a pomeron made of hard gluons is $\approx 800$~pb, 
while for a quark dominated pomeron it is only $\approx 30$~pb.

\subsection{Hard colour singlet exchange}

Another kind of hard diffractive process leads to the appearance of large
rapidity gaps between two high $E_T$ jets. These events would result from the
exchange of a colour singlet propagator in the $t$-channel, as in the diagram
of Fig.~\ref{diacolsi}/a. The propagator could be an electroweak gauge boson or a
strongly interacting object. 
The normal QCD jet production is instead described as a strong process with the
exchange of a quark or a gluon propagator, as in Fig.~\ref{diacolsi}/b.
The colour charge of the propagator generates the colour connections 
between the hard jets. For this
reason the pattern of soft gluon radiation is distributed over all the phase
space, particularly in the central rapidity region between the two jets. 
Instead, for colour singlet exchange, the hard jets would be colour connected 
only to the beam remnants. In this case the central region would be depleted
from particle production and could present the rapidity gap signature.
This was first suggested by Bjorken \cite{bjorken92}, who also calculated
the expected rate for this kind of events in $p \bar p$ interactions
at the Tevatron ($\sqrt{s} = 1.8$~TeV).
He estimated the contribution from the exchange of two gluons in a colour 
singlet state to be $10\%$ of the cross section for single gluon exchange. Other
calculations valid for high $t$ have been published based on perturbative QCD
\cite{mutang92,ddtan}.

The topology of the
relevant events is shown in Fig.~\ref{diacolsi}/c. Two jets are shown in
the $\eta$-$\phi$ plane, back-to-back in azimuth and separated by an interval
$\Delta\eta$ in pseudorapidity. This is the pseudorapidity difference between
the axes of the two jets. No particles are found between the jets in rapidity
gap events.

The gap-fraction $f(\Delta\eta)$ is defined 
as the ratio of the number of dijet events having a gap $\Delta\eta$ to the
total number of dijet events with jet-jet separation equal to $\Delta\eta$. It 
is expected to fall exponentially with increasing $\Delta\eta$ for
non-diffractive processes. This follows
assuming a constant pseudorapidity distribution of particles between the jets
and is supported by analytic QCD calculations \cite{ddtan} 
and Monte Carlo results. Instead for colour singlet exchange the gap
fraction is expected to be almost flat with $\Delta\eta$
\cite{bjorken92,ddtan}. At large enough $\Delta\eta$ the colour-singlet
exchange would then be dominating, as is shown in Fig.~\ref{diacolsi}/d. 

Allowing secondary
interactions between the photon and proton remnants, the gap could be sometimes
filled. A survival probability $\cal{P}$ has been defined in
\cite{bjorken92}, such that the observed gap-fraction is the product of
$\cal{P}$ and the gap probability at the parton level, which is
a calculable quantity. Estimates of the survival probability for $p \bar p$
collisions at the Tevatron are in the range $5 - 30 \%$ \cite{bjorken92,got93}.
The same quantity at HERA could be higher due to the substitution of the $\bar
p$ with the $\gamma$ and the lower center of mass energy. 

Both the Tevatron experiments D0 \cite{d0cs} and CDF \cite{cdfcs} 
reported the observation of an excess of events
with two hard jets separated by a rapidity gap, 
with respect to the expectations from normal QCD processes with colour exchange.
The measured excess is about $1 \%$ for D0, $0.8 \%$ for CDF. 

ZEUS published results based
on $2.6$~pb$^{-1}$ of 1994 data \cite{zeuscolsin}, 
requiring two jets with $E_T^{jet} > 6$~GeV,
$\eta^{jet} < 2.5$ and small boost $|\bar \eta | = |\eta_1 +
\eta_2 | / 2 < 0.75$. A cone algorithm with $R=1$ was used and the distance
between the two jets was required to be $|\Delta \eta | > 2$ to have
non-overlapping cones in $\eta$. Gap events were found by inspecting the
calorimeter pattern of energy, identifying particles by ``islands'' of
neighbouring calorimeter cells with summed $E_T$ above a threshold of $250$~MeV. 
A gap event was defined as an
event having no particles between the two jets with $E_T > 300$~MeV 
(after correction of the calorimeter energies).
The result is shown in Fig.~\ref{csfinal}, where the excess of gap
events is evident in the last bin. The largest systematic
uncertainty in data comes from the cut on the $E_T$ of the island and is
included in the error bars. Data have been fitted with a
function $f(\Delta\eta) = k e^{\alpha \Delta\eta} + \beta$, with $\alpha$ and
$\beta$ free parameters and $k$ a normalization fixed by the obvious condition
$f(\Delta\eta) = 1$ at $\Delta\eta = 2$. The result is shown as the
solid curve, and $\chi^2 = 1.2$ for two degrees of freedom. The fitted 
parameters are: $\alpha = -2.7 \pm 0.3 (stat.) \pm 0.1 (syst.)$, 
$\beta = 0.07 \pm 0.02 (stat.) ^{+ 0.01} _{-0.02} (syst.)$.
The size of the excess has been also checked comparing data with the
expectations from PYTHIA, simulating non-diffractive jet production.
From this comparison the
excess is found to be $7\%$, in agreement with the result of the fit.
The actual number of events due to colour singlet exchange could be even higher
than $7 \%$, accounting for a survival probability for the gaps.
The larger excess of gap events found at HERA in comparison to the Tevatron
results could come from the same underlying dynamics, given different survival
probabilities for gaps in the two environments.  
At HERA the contribution
to the gap events from $\gamma/Z^0$ or $W^\pm$ exchange cannot account for the
observed excess, since $\sigma^{EW}/\sigma^{QCD} < 7 \cdot 10^{-4}$ in this
kinematic range. Instead data can be explained by a hard diffractive 
process based on the exchange of a colour singlet combination of gluons.

\section{Outlook}
\label{conclu}

In this paper we have reviewed the published HERA results on photoproduction. 
The understanding of the hadronic structure of the photon as well as the
understanding of photon induced reactions has greatly improved in the last 
few years. HERA data have played a central role. 
This progress will continue in the next years. 
We will not summarize here the covered contents, but rather prefer 
to have a glimpse on the future prospects for selected topics.

Most of the HERA measurements will have significant improvements in
the next years with the expected luminosity increase.
The final HERA results presented here include data up to the 1994 run, with
integrated luminosity of about $3$~pb$^{-1}$. 
The achieved luminosity during 1995-96 runs is about $20$~pb$^{-1}$ and almost 
the same should be achieved during 1997. 
Future upgrade plans aim at an integrated luminosity exceeding
$100$~pb$^{-1}$ per year usable by the experiments.

Hard photoproduction processes at HERA are complementary to the $e \gamma$ 
deep inelastic scattering, being directly sensitive to both the quark and the 
gluon component of the photon. 
Data on jet photoproduction are presently consistent with the expectations 
of perturbative QCD. Ongoing analyses are studying extended domains in 
$E_T^{jet}$ and $\eta^{jet}$, with different jet algorithms. 
The most severe experimental limitation on jet measurements
is the energy scale uncertainty of the calorimeters, which can however
be improved to about $2 - 3 \%$ from the $5 \%$ quoted in the results shown
here. With such improvement large statistic samples with high $E_T$ jets 
will allow more precise QCD tests.
The measurement of high $p_T$ inclusive charged particle spectra has been
recently proposed as a valid alternative to have access to the parton 
distributions in the photon \cite{binproc}. 
It has the advantage of being insensitive to the energy scale of the 
calorimeter, at the price of an uncertainty due to the fragmentation functions,
needed to describe the transition from partons to hadrons. 
The estimated accuracy on the extraction of a NLO gluon density by this method
is about $10 \%$. In parallel to HERA, the knowledge of the parton distributions
in the photon (particularly the quark component) will be improved by high 
luminosity data from LEP2.
From it high precision measurements of $F_2^\gamma$ are expected up to
$Q^2$ of few hundred GeV$^2$ and minimum $x_\gamma$ about $10^{-3}$.

The heavy flavour studies at HERA are presently all limited by statistics.
The comparison of open charm ($D^\ast$) data with the existing NLO QCD 
calculations has proved to be an interesting and intriguing benchmark.
In particular the discrepancy of
the exact massive NLO calculation with HERA data has to be clarified. 
It could be due to uncertainties of the charm fragmentation function or, 
if nothing else, it might reveal the importance of higher order contributions 
(NNLO), which are far from being calculated.
Apart from the normalization of the cross section, it will be interesting to see
if the reported excess of $D^\ast$ data in the forward pseudorapidity region
will be confirmed.
At reasonably high $p_T$ values a safe comparison of the
NLO massive calculation with the calculation resumming the large logarithms of
$p_T / m_c$ (the ``massless'' approach) should also help to understand the 
situation.
Open charm data could also be used for a direct extraction of a NLO gluon 
density in the proton when a large data sample ($\approx 100$~pb$^{-1}$) will be 
available, as it has been proposed in \cite{friglup}.
Charm tagging capabilities will benefit from detector upgrades of the
interaction region with the introduction of a silicon microvertex detector.
This has already been installed in H1 and is currently being 
developed in ZEUS. 

Elastic vector meson production at HERA is being studied with larger statistics
while varying the involved energy scales ($Q^2$, $t$).
The early results suggested the picture of a hard production mechanism, when a
hard scale is present in the interaction, be it a high mass (as the 
$J/\psi$ mass) or high $Q^2$ or $t$. Recent fixed target data on $\rho$
production at high $Q^2$ make difficult to draw safely this conclusion for
light vector mesons.
Detailed studies varying the scales will possibly shed some light on the 
transition from a non-perturbative description of the pomeron 
(the soft pomeron), to the hard perturbative pomeron.
Hard diffraction data from HERA (in which the final state has jets or heavy 
flavour particles) are presently indicating that the gluon content of the 
diffractive exchange is substantial. Many different models are appearing on
the market and are being tested, trying to unveil the nature of the hard 
pomeron.

In conclusion, the study of $\gamma p$ interactions at HERA has yielded 
a lot of results. Several detector upgrades have already been performed
(or are being done), which will result in improved purities of the selected 
samples as well as a wider kinematic coverage. Adding to this the expected 
much larger statistics, there is a bright future for photoproduction physics 
at HERA.

\newpage
\begin{large}
\bf{Acknowledgements}
\end{large}
\vspace{0.5cm}
\\
This work has been supported by an European Community fellowship.
It is a pleasure to thank M.~Arneodo, R.~Brugnera, J.~Dainton and Y.~Yamazaki
for useful discussions. I am specially grateful to  M.~Cacciari, A.~Caldwell 
and S.~Bhadra for a critical reading of the manuscript and many useful comments.


\clearpage
\begin{table}[t]
  \begin{center}
    \begin{tabular}{||c||rcrcrcr|crc||} \hline\hline
      {\em process} &
      \multicolumn{7}{c|}{{\em cross section} ($\mu$b)} &
      \multicolumn{3}{c||}{{\em full error} ($\mu$b)} \\
      \hline\hline
      $\sigma(\gamma{p}\rightarrow\,XY)$,\,  {\em DD} &
       \multicolumn{7}{c|}{20 $\pm$ 20 (assumed)} & $\quad$
        & --- &  \\
      $\sigma(\gamma{p}\rightarrow\,Xp)$,\, {\em GD} &
       23.4 & $\pm$ & 2.6 & $\pm$ & 4.3  & $\pm$ & 10.2 && 11.3 & \\
      $\sigma(\gamma{p}\rightarrow\,VY)$,\,  {\em PD} &
       8.7 & $\pm$ & 1.5 & $\pm$ & 1.5  & $\pm$ &  3.0 &&  3.6 & \\
      $\sigma(\gamma{p}\rightarrow\,Vp)$,\, {\em EL} &
      17.1 & $\pm$ & 1.6 & $\pm$ & 3.7  & $\pm$ &  1.4 &&  4.3 & \\
      \hline
      {\em EL + GD + PD + DD } &
      69.2 & $\pm$ & 3.4 & $\pm$ & 8.8 & $\pm$ &  9.3 && 13.2 & \\
             {\em ND} &
      96.1 & $\pm$ & 3.5 & $\pm$ & 14.7 & $\pm$ &  9.6 && 17.9 & \\
      \hline
      {\em Total} &
                165.3 & $\pm$ & 2.3 & $\pm$ & 10.9 & $\pm$ &  1.3 && 11.2 & \\
      \hline\hline
    \end{tabular}
  \end{center}
\label{photos}
\caption{\small
  Total and partial $\gamma p$ cross sections as measured by H1
  \protect\cite{h1stot}. The contribution of the double dissociation is 
  assumed in the range: $0 < \sigma_{DD} < 40 ~\mu b$. 
  The first error is statistical, the second one is systematic, 
  and the third error is the systematic uncertainty due to the assumption on 
  $\sigma_{DD}$. The full error is obtained as the quadratic sum of all of
  these.}
\vspace*{-5.cm}
\end{table}
\begin{table}[h]
\vspace*{6cm}
  \label{h1sparz}
  \begin{center}
    \begin{tabular}[h]{|c||c|c|c|c|}
      \hline
      &\multicolumn{4}{c|}{{\em Cross Sections}~($\mu$b)} \\ \cline{2-5}
      \raisebox{1.4ex}[2.ex][0ex]{\em Reaction} &
      Data & CKMT & SaS & GLM \\
      \hline\hline
      $\sigma(\gamma{p}\rightarrow\,Vp)$,\, {\em EL} &
      $17\pm\,4$ & 17 & 16 & 17 \\
      \hline
      $\sigma(\gamma{p}\rightarrow\,Xp)$,\, {\em GD} &
      $26\pm\,5$ & 25 & 13 & 18 \\
      \hline
      $\sigma(\gamma{p}\rightarrow\,VY)$,\,  {\em PD} &
      $9\pm\,2$ & 7 & 10 & 15 \\
      \hline
      $\sigma(\gamma{p}\rightarrow\,XY)$,\,  {\em DD} &
      $ 15 $ & 15 & 13 & 15 \\
      \hline
    \end{tabular}
  \end{center}
\caption {\small
    Comparison of diffractive cross section
    calculations \protect\cite{ckmt95,sas93,glm95}
    with the H1 measurements \protect\cite{h1stot}. The data are
    given for a fixed value $\sigma_{DD}=15\mu$b.}  
\vspace*{-6cm}
\end{table}
\begin{table}[b]
\vspace*{1cm}
$$
\begin{tabular}{|c|c|c|c|c|}
\hline
reaction & reference & $\langle W \rangle$~[GeV] & $b$~[GeV$^{-2}$] &
$|t|$-range [GeV$^2$]
\\ \hline
$ \gamma p \to \rho^0 p$ & \cite{zeusrho} & 70 & 10.4 $\pm$ 0.6 $\pm$ 1.1 & 
$<$~0.15 \\
 \hline
$ \gamma p \to \rho^0 p$ & \cite{zeuslps} & 70 & 9.8 $\pm$ 0.8 $\pm$ 1.1 &
0.073 -- 0.4 \\
 \hline
$ \gamma p \to \rho^0 p$ & \cite{h1rho} & 55 & 10.9 $\pm$ 2.4 $\pm$ 1.1 & 
0.025 -- 0.25 \\
 \hline
$ \gamma p \to \omega p$ & \cite{zeusomega} & 80 & 10.0 $\pm$ 1.2 $\pm$ 1.3 &
$<$~0.6 \\
 \hline
$ \gamma p \to \phi p$ & \cite{zeusphi} & 70 & 7.3 $\pm$ 1.0 $\pm$ 0.8 & 
0.1 -- 0.5 \\
 \hline
\end{tabular}
$$
\caption{\small
Summary of the determinations of the $b$ slope for elastic 
light vector meson photoproduction at HERA. 
The values of $b$ have been obtained from a fit of
the differential cross section of the form $ \frac{d\sigma}{d|t|} = 
A \cdot~{e}^{-b|t|}$ in the given $t$~range.}
\label{slopes}
\end{table}
\clearpage
\clearpage
\clearpage
\clearpage
\clearpage
\newpage
\clearpage
\clearpage
\clearpage
\clearpage
\clearpage
\begin{figure} [p]
\begin{center}
\epsfig{file=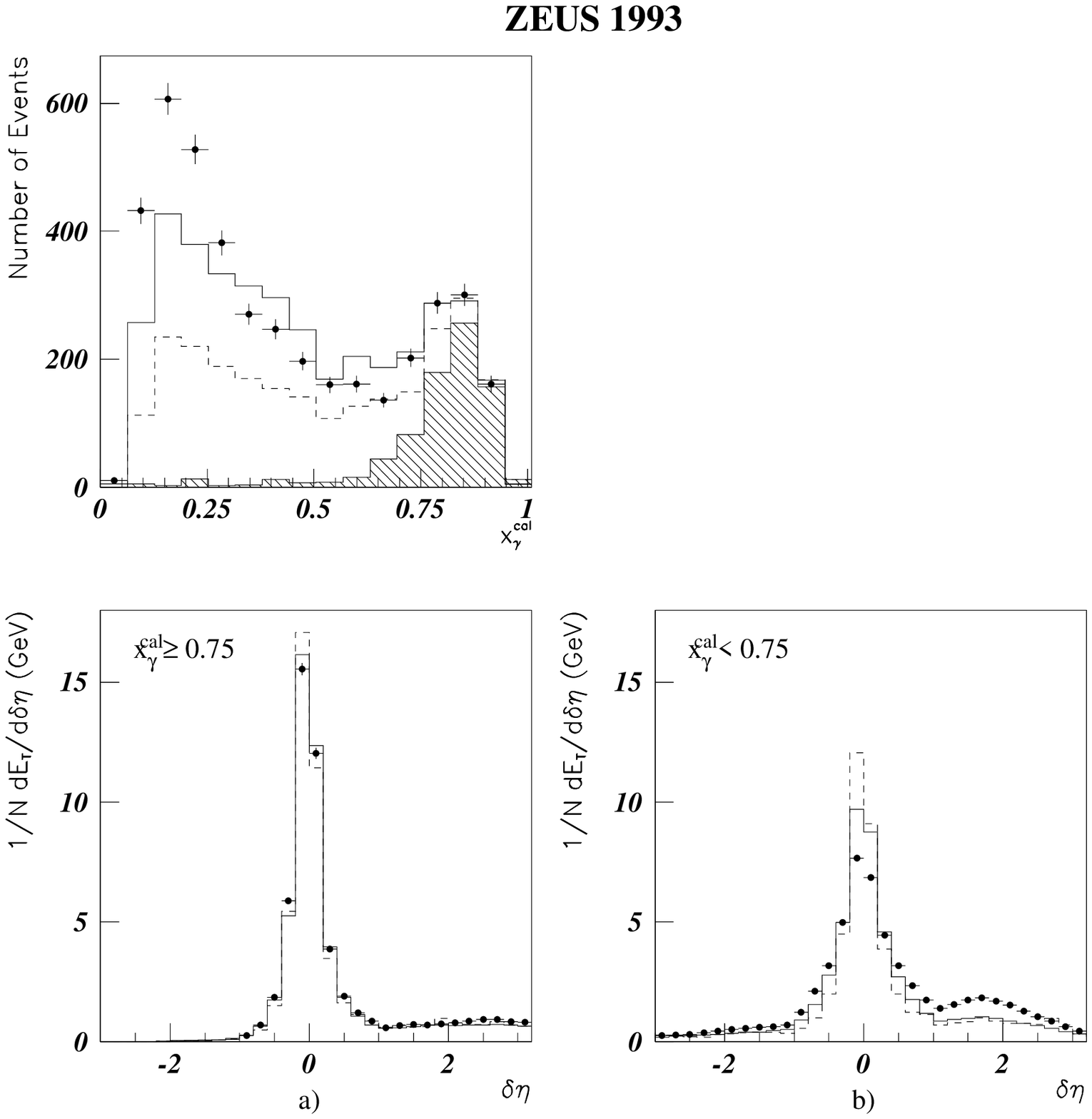,width=16cm,%
bbllx=30pt,bblly=160pt,bburx=530pt,bbury=400pt,clip=}
\caption{
Uncorrected transverse energy flow $1/N ~dE_T/d\delta\eta$ around the
jet axis, for cells within one radian in $\phi$ of the jet axis for (a) direct
and (b) resolved photon events, from the ZEUS analysis in
\protect\cite{zeusdir2}. The solid and dashed lines represent respectively the
distribution from PYTHIA and HERWIG.}
\end{center}
\end{figure}
\clearpage
\begin{figure} [p]
\begin{center}
\epsfig{file=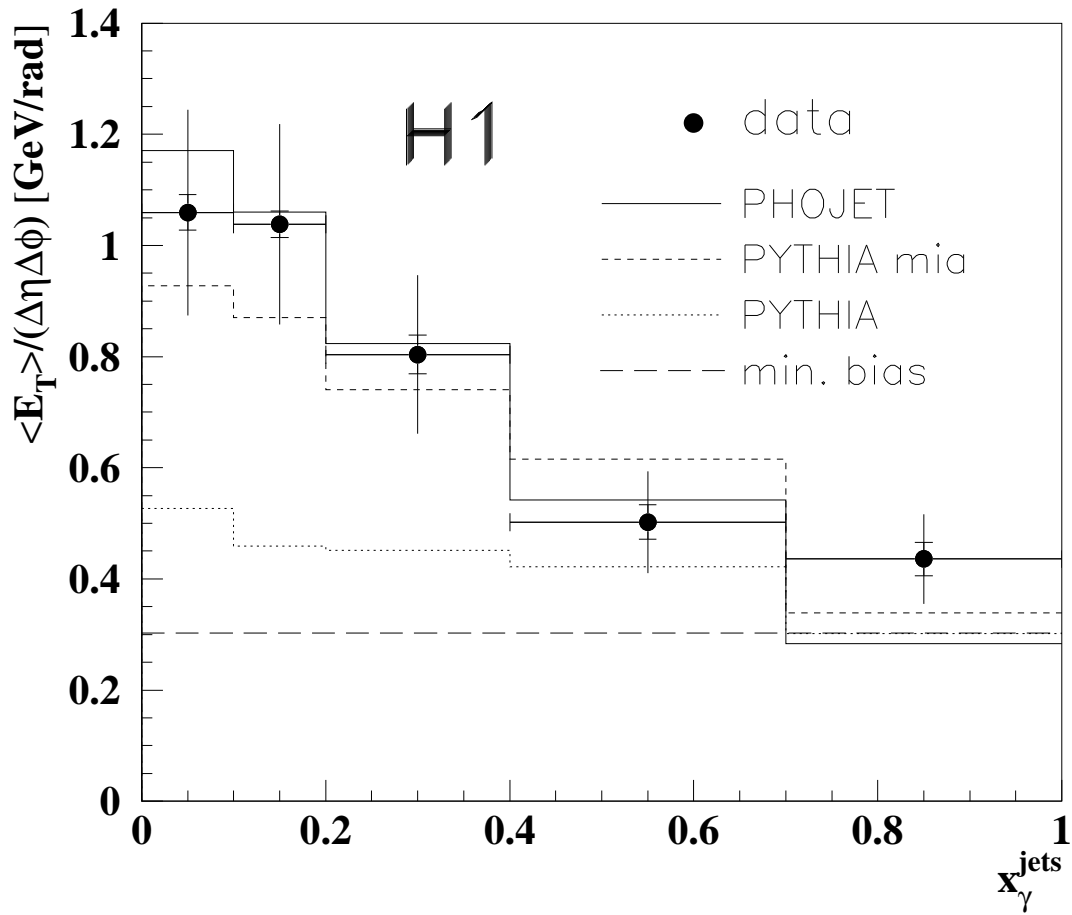,width=16cm}
\end{center}
\caption{
Average transverse energy density (per unit area in the $\eta - \phi$ space)
in the central region $|\eta^\ast| < 1$ outside
the two jets with the highest $E_T^{jet}$ as a function of $x_\gamma$.
The inner error bars are statistical, the outer ones result from the 
quadratic sum of statistical and systematic errors.
The long-dashed line indicates the energy density measured in minimum bias 
events.
The histograms show the expectations of different QCD generators
with interactions of the beam remnants (full=PHOJET, short-dashed=PYTHIA), 
and without them (dotted=PYTHIA).}
\label{etped}
\end  {figure}
\begin{figure} [p]
\epsfig{file=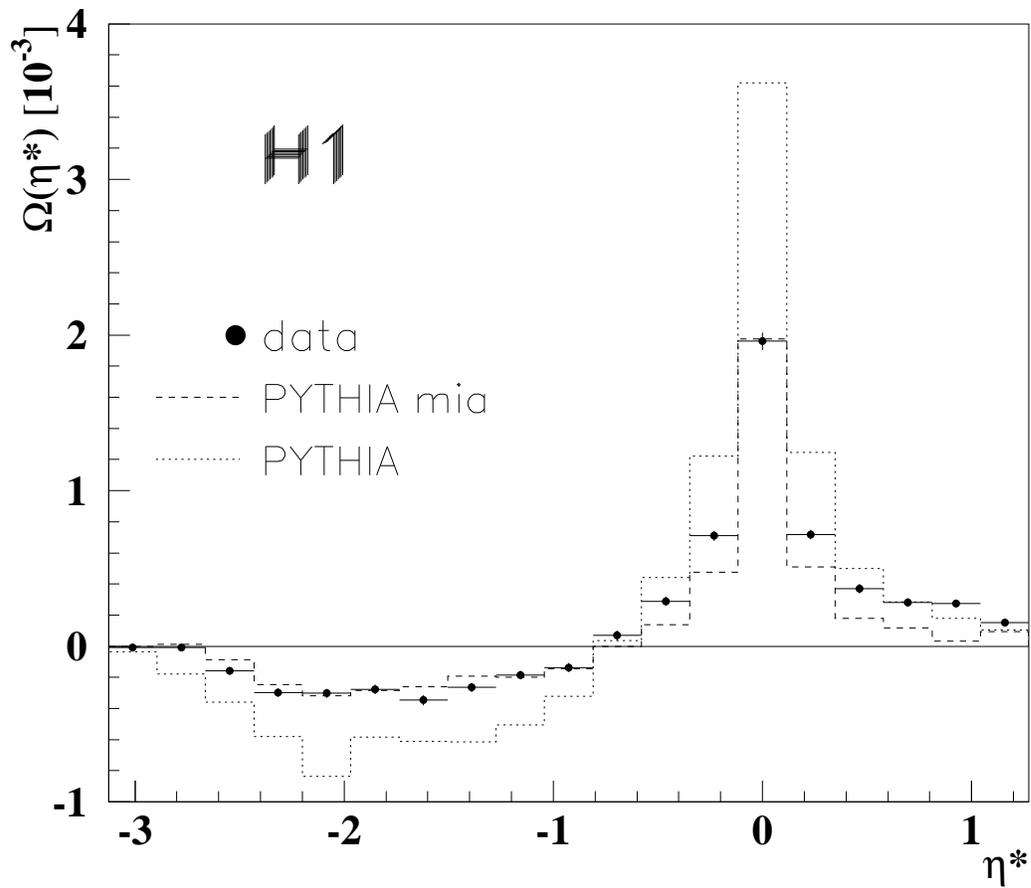,width=16cm}
\caption{Energy-energy correlation with respect to $\eta^\ast = 0$ as a
function of $\eta^\ast$ for a high $E_T$ event sample \protect\cite{h1incjet}. 
The dashed (dotted) histogram represents the prediction of the QCD
generator PYTHIA with (without) interactions of the beam remnants.}
\label{eec}
\end  {figure}
\begin{figure}[p]
\epsfig{file=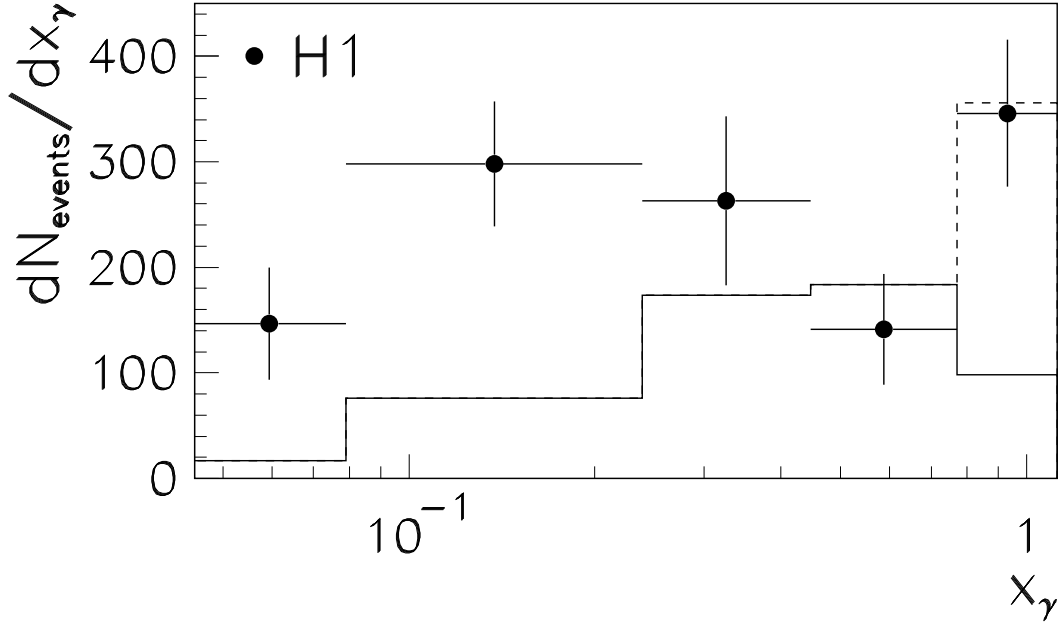,width=15cm}
\vspace{-0.3cm}
\caption {
Distribution of the unfolded $x_\gamma$ from a two-jet event sample 
\protect\cite{h1xgx}.
Only the statistical errors are shown.
The histograms are PYTHIA Monte Carlo calculations showing the contribution of
quarks from the resolved photon (solid) and of direct photon events (dashed).}
\label{h1x}
\end  {figure}
\begin{figure}[p]
\vspace{0.5cm}
\epsfig{file=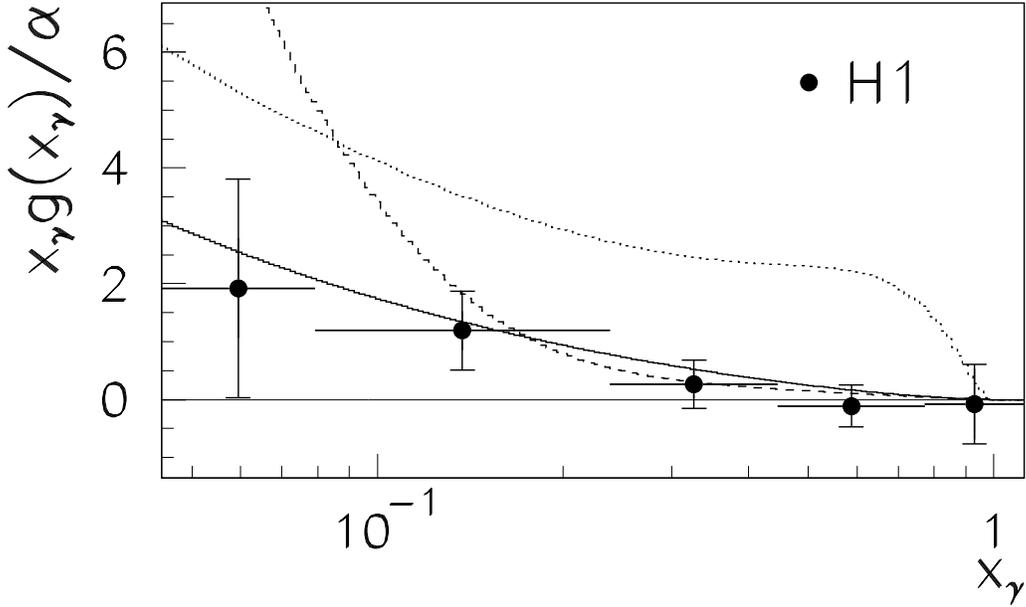,width=15cm}
\vspace{-0.3cm}
\caption{
The gluon momentum distribution of the photon divided by the fine structure 
constant $\alpha$ as obtained in the LO analysis of H1 \protect\cite{h1xgx} 
at a scale $\langle p_T^{jet} \rangle ^2 = 75$~GeV$^2$, compared to LO 
parametrizations: GRV (full) \protect\cite{grv}, 
LAC1 (dashed) and LAC3 (dotted) \protect\cite{lac}.}
\label{h1xgxfig}
\end  {figure}
\begin{figure}[p]
\begin{center}
\epsfig{file=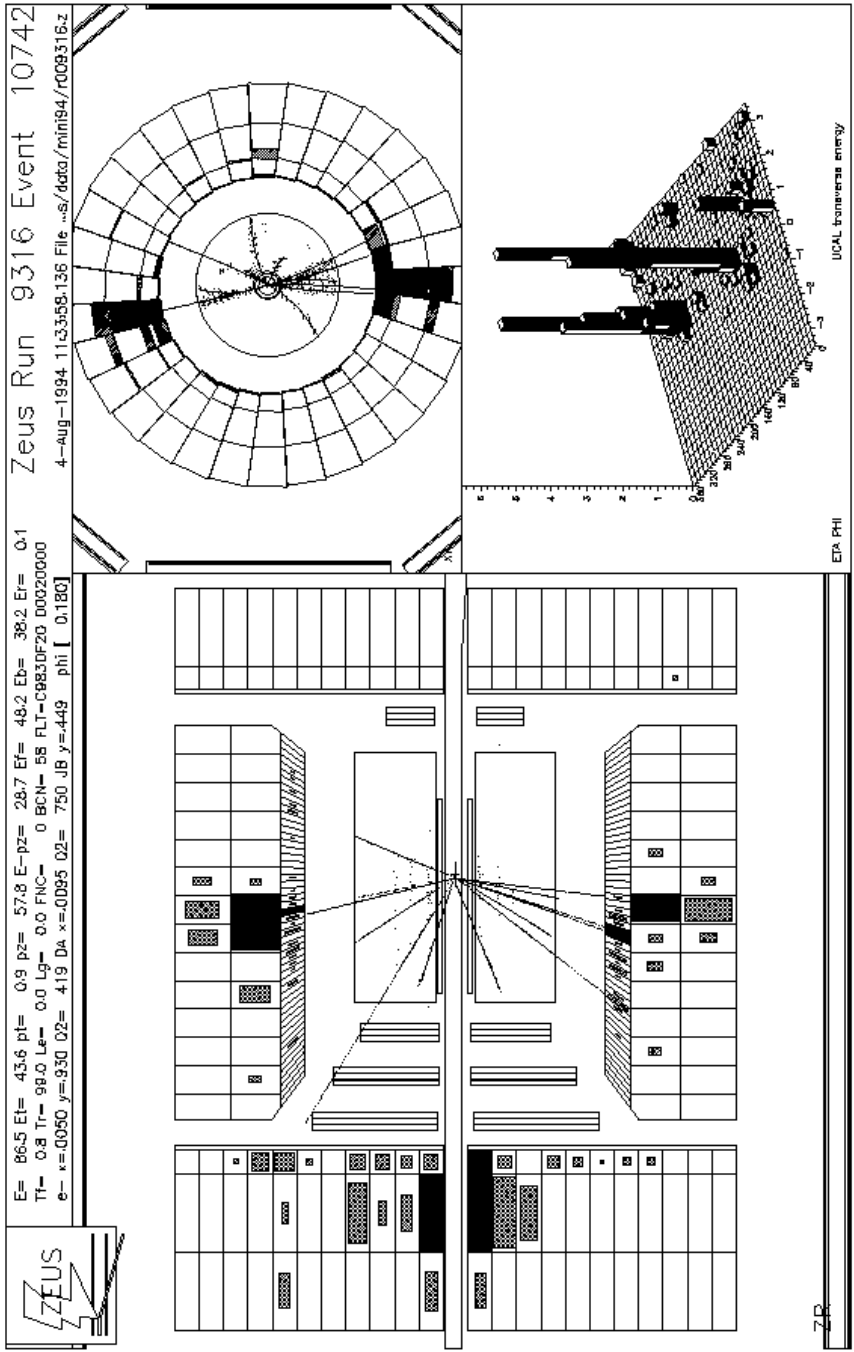,height=18.5cm}
\vspace{1cm}
\caption{Direct photoproduction of jets as seen in the ZEUS detector: the
figure on the left is a longitudinal section of the detector, in which the 
electron beam comes from the left and the proton beam from the right. Two
central jets are visible, balancing each other in $p_T$. The upper right
picture is a transverse view of the same event and the lower right figure is a
lego-plot showing the transverse energy deposited in the calorimeter in bins of
$\eta-\phi$.}
\end{center}
\end  {figure}
\begin{figure}[p]
\begin{center}
\epsfig{file=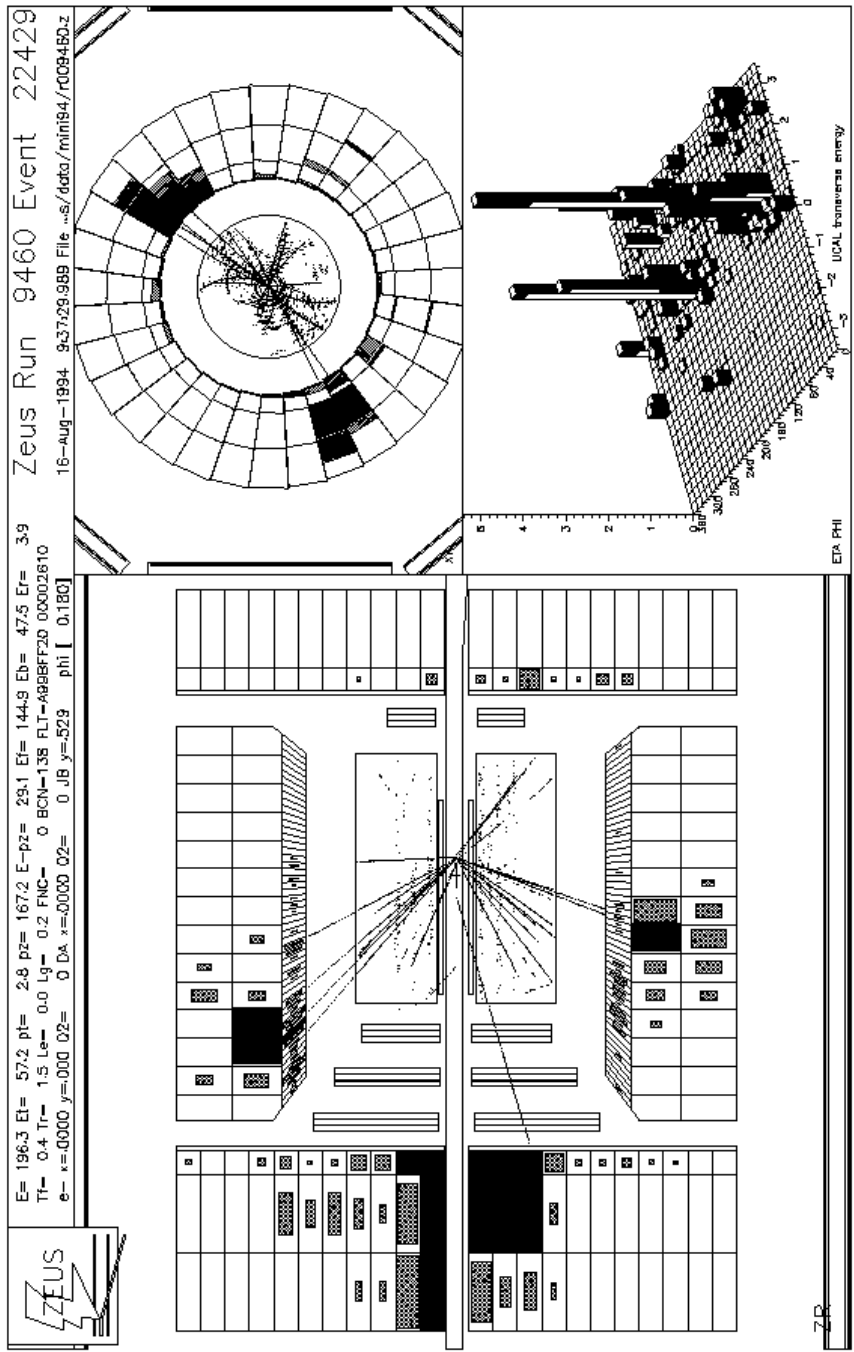,height=18.5cm}
\vspace{1cm}
\caption{
Resolved photoproduction of jets as seen in the ZEUS detector, with the same
meaning as in the previous figure. On the left plot the photon remnant 
(going to the right) is visible.}
\end{center}
\end  {figure}
\begin{figure}[p]
\begin{center} 
\epsfig{file=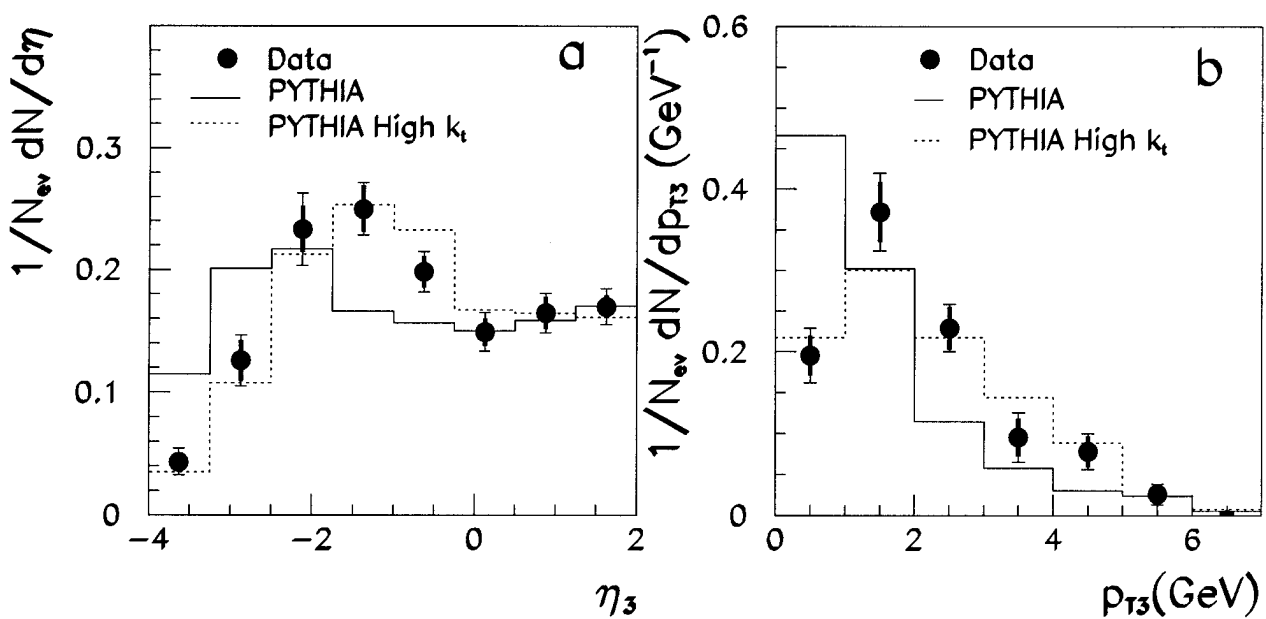}
\end{center}
\caption{
Corrected distributions of the third cluster in photoproduction events with 
two other jets with 
$E_T^{jet} > 6$~GeV and $\eta^{jet} < 1.6$ as measured by ZEUS 
\protect\cite{zeusphremn}: 
(a) pseudorapidity; (b) transverse momentum.
The histograms are expectations from PYTHIA Monte Carlo: the full line is the
default version, the dotted line is obtained with a higher intrinsic 
transverse momentum ($k_t$) for the partons in the photon, as explained in the
text.}
\label{grem}
\end{figure}
\begin{figure}[p]
\begin{center}
\epsfig{file=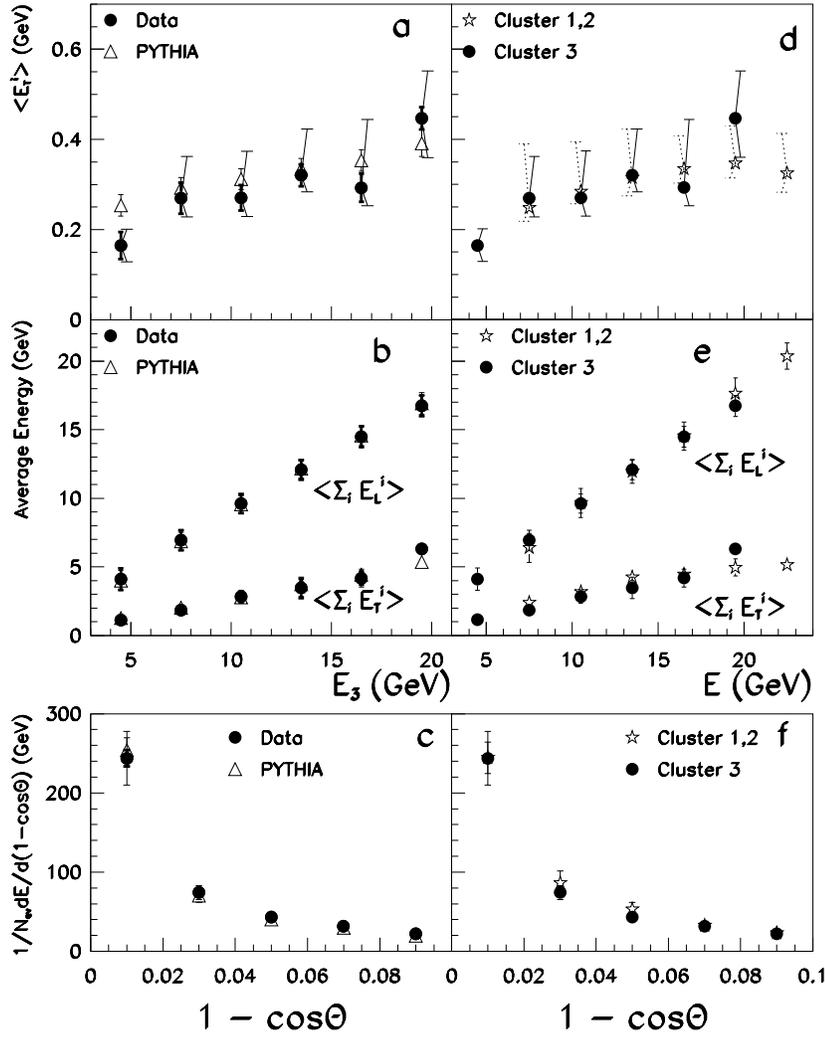,width=12cm}
\end{center}
\caption{Fragmentation properties of the photon remnant compared to those
of the hard jets in selected resolved photon events measured by ZEUS 
\protect\cite{zeusphremn} and to the expectations of PYTHIA Monte Carlo:
(a) and (d) show the mean value of $\langle E_T^i \rangle$, 
the average energy transverse to the cluster axis per particle, 
as a function of the cluster energy;
(b) and (e) show the average values of the total transverse 
($\Sigma_i E_T^i$) and total longitudinal 
($\Sigma_i E_L^i$) energy;
(c) and (f) show the energy flow around the cluster axis.}
\label{compa}
\end{figure}
\begin{figure}[p] 
\begin{center}
\epsfig{file=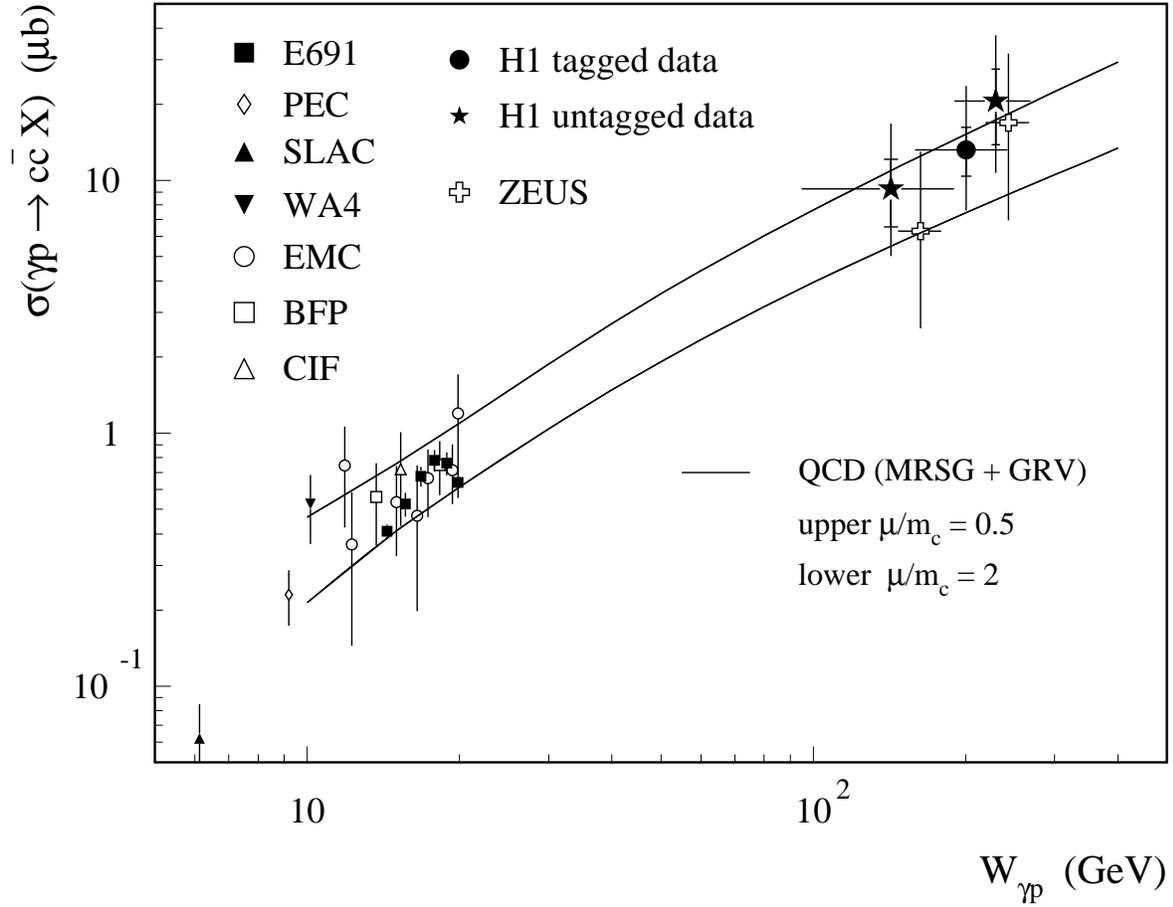,width=17cm}
\caption {Total charm photoproduction cross section as a function 
of $W_{\gamma p}$: measurements from ZEUS, H1 and previous fixed target 
experiments are shown. The error bars give the statistical and systematic
errors added in quadrature, including the uncertainty in the extrapolation.
The lines are predictions of a NLO QCD calculation \protect\cite{fri95}
showing the effect of
changing the renormalization scale from $\mu = 0.5 ~m_c$ (upper curve) to 
$\mu = 2 ~m_c$ (lower curve). The parton densities used in the calculation are
MRSG for the proton and GRV-HO for the photon.}
\end{center}
\end{figure}
\clearpage
\begin{figure}[p]
\begin{center}
\epsfig{file=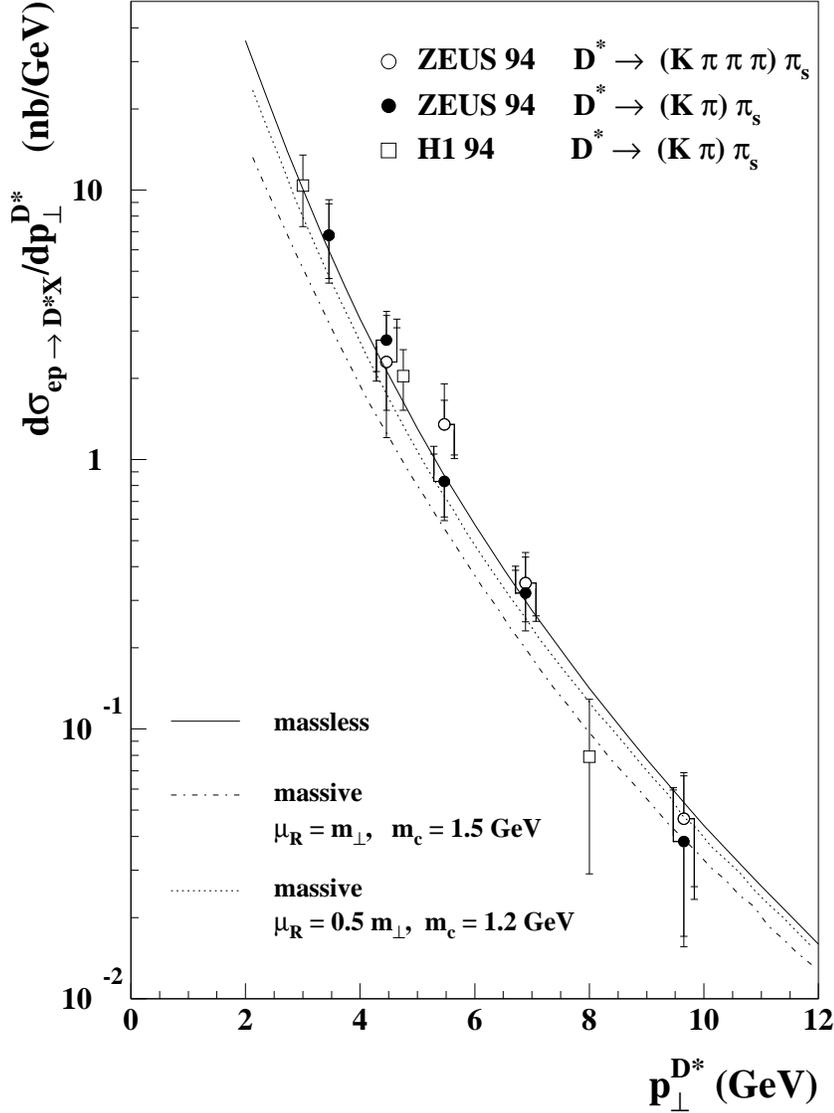,height=16.5cm}
\end{center}
\caption{The differential $e p$ cross section $d\sigma/dp_T^{D^\ast}$ for
$D^\ast$ photoproduction in the kinematic range $Q^2 < 4$~GeV$^2$ 
and $115 < W <280$~GeV integrated over the $D^\ast$ pseudorapidity range 
$-1.5 <\eta^{D^\ast} < 1.0$. NLO QCD predictions are shown: 
the dot-dashed curve is the result of the massive calculation of
\protect\cite{fri95} using MRSG and GRV-HO parton densities for the proton 
and photon respectively, $m_c = 1.5$~GeV, renormalization scale 
$\mu_R = m_T$ and 
fragmentation parameter $\epsilon_c = 0.06$; the dotted curve is the same but
with $\mu_R = 0.5~m_T$ and $m_c = 1.2$~GeV; the full curve comes from the
resummed calculation in \protect\cite{spira96} with the same parameters as the
dot-dashed curve except the proton parton densities CTEQ4M 
\protect\cite{cteq4}.} 
\label{dstarpt}
\end{figure}
\clearpage
\begin{figure}[p] 
\epsfig{file=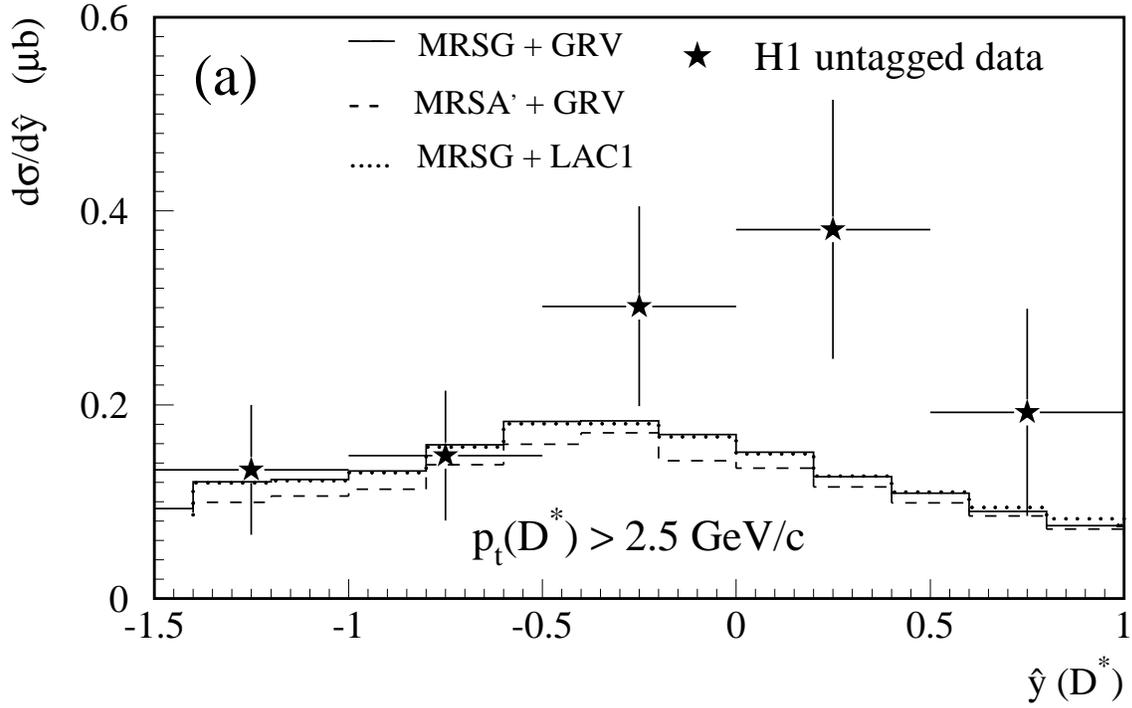,width=17.cm}
\caption{
Differential $\gamma p$ cross section for $D^\ast$ photoproduction 
with respect to the $D^\ast$ rapidity $\hat y$ integrated over $p_T(D^\ast) >
2.5$~GeV. The histograms are predictions of a NLO QCD calculation
\protect\cite{fri95} with $m_c = 1.5$~GeV and different parton densities:
MRSG + GRV-HO (solid), MRSA' + GRV-HO (dashed), MRSG + LAC1 (dotted) for the
proton and the photon respectively.}
\label{dstarrap}
\end{figure}
\begin{figure}[p]
\begin{center}
\epsfig{bbllx=300pt,bblly=315pt,bburx=430pt,bbury=415pt,clip=,%
file=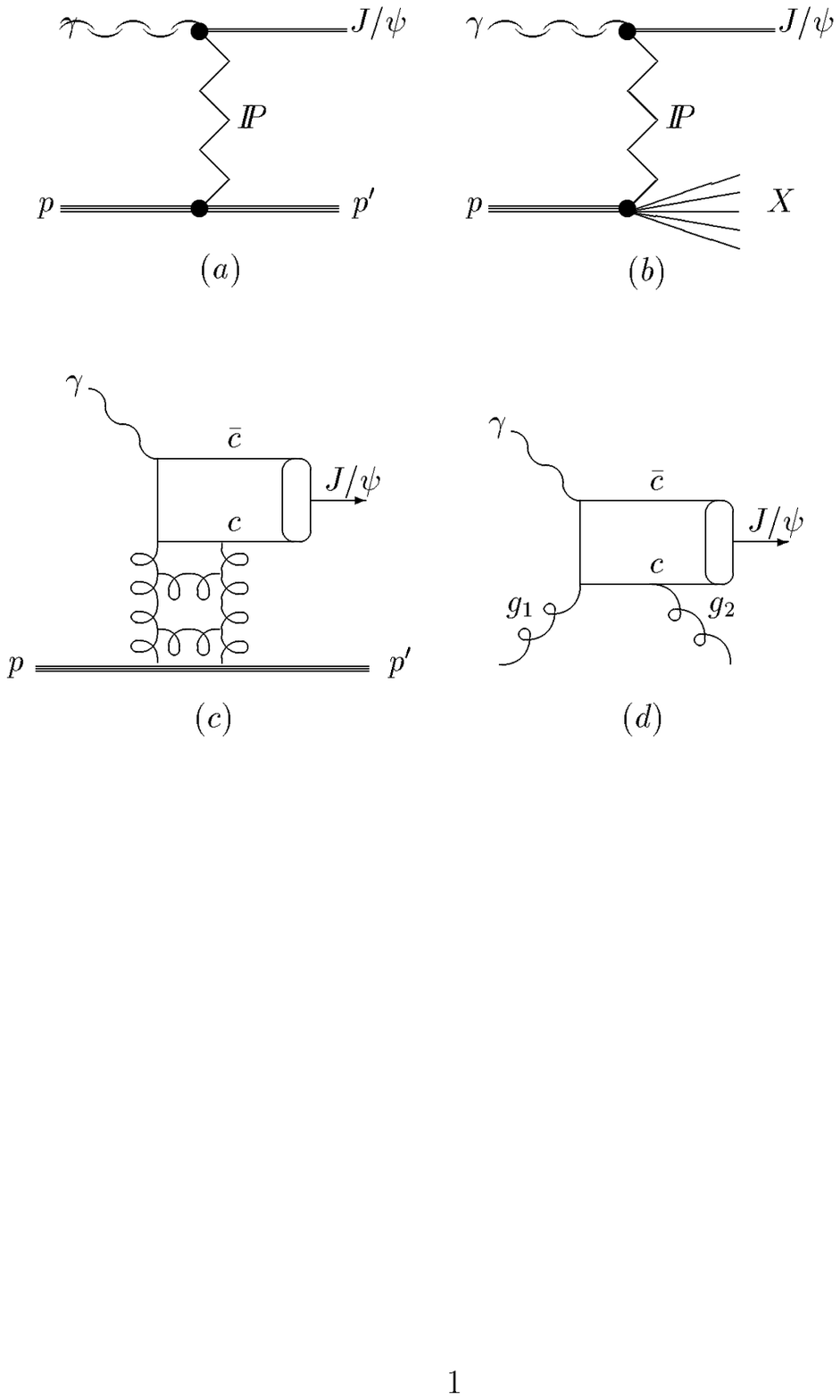,width=8cm}
\end{center}
\caption{Photon-gluon fusion diagram for inelastic $J/\psi$ production.}
\label{colsin}
\end{figure}
\begin{figure}[p]
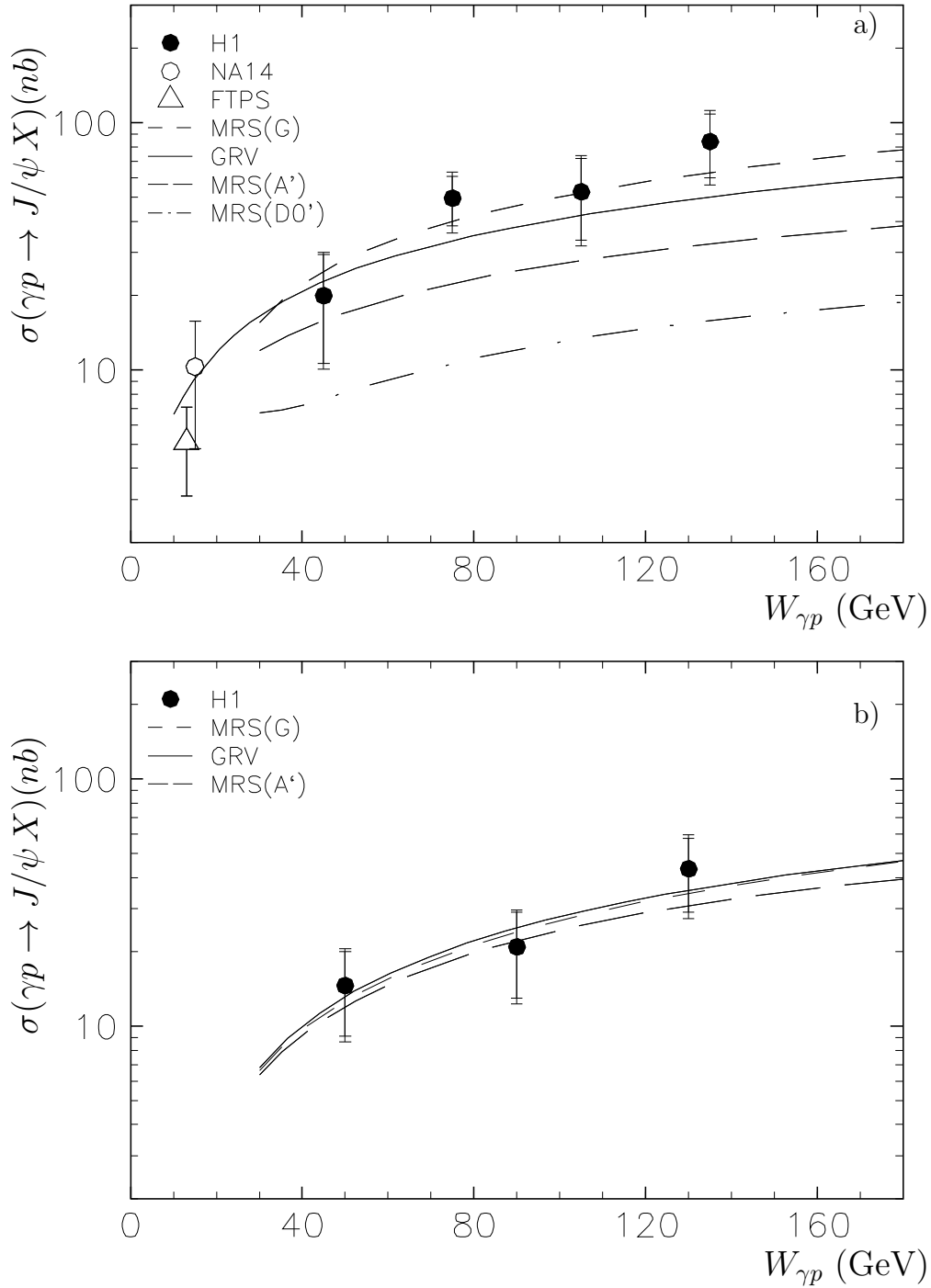

\vspace*{13pt}
\setlength{\unitlength}{1cm}
\begin{picture}(14.0,16.0)
\put(-0.75,7.5){\epsfig{file=psicros.eps,height=24cm,angle=90}}
\put(1.5,13.0){\begin{rotate}{-90}\large $\sigma(\gamma p \rightarrow J/\psi \,
X)(nb)$\end{rotate}}
\put(12.,9.){\large $W_{\gamma p}$~(GeV)}
\put(-0.75,-2.0){\epsfig{file=psicroscut.eps,height=24cm,angle=90}}
\put(1.5,3.0){\begin{rotate}{-90}\large $\sigma(\gamma p \rightarrow J/\psi \, 
X)(nb)$\end{rotate}}
\put(12.,-0.5){\large $W_{\gamma p}$~(GeV)}
\put(13.3,17.5){a)}
\put(13.3,7.5){b)}
\end{picture}
\vspace*{13pt}
\caption{Total cross section for inelastic $J/\psi$ photoproduction 
as a function of the $\gamma p$ center of mass energy with: 
(a) $z<0.9$; (b) $z<0.8$ and $p_T^2>1$~GeV$^2$.
The curves are NLO QCD calculations \protect\cite{kra96} with different input
gluon distributions and contain a $15\%$ correction accounting for 
$\psi '$ background.}
\label{psicros}
\end{figure}
\clearpage
\begin{figure}[p]
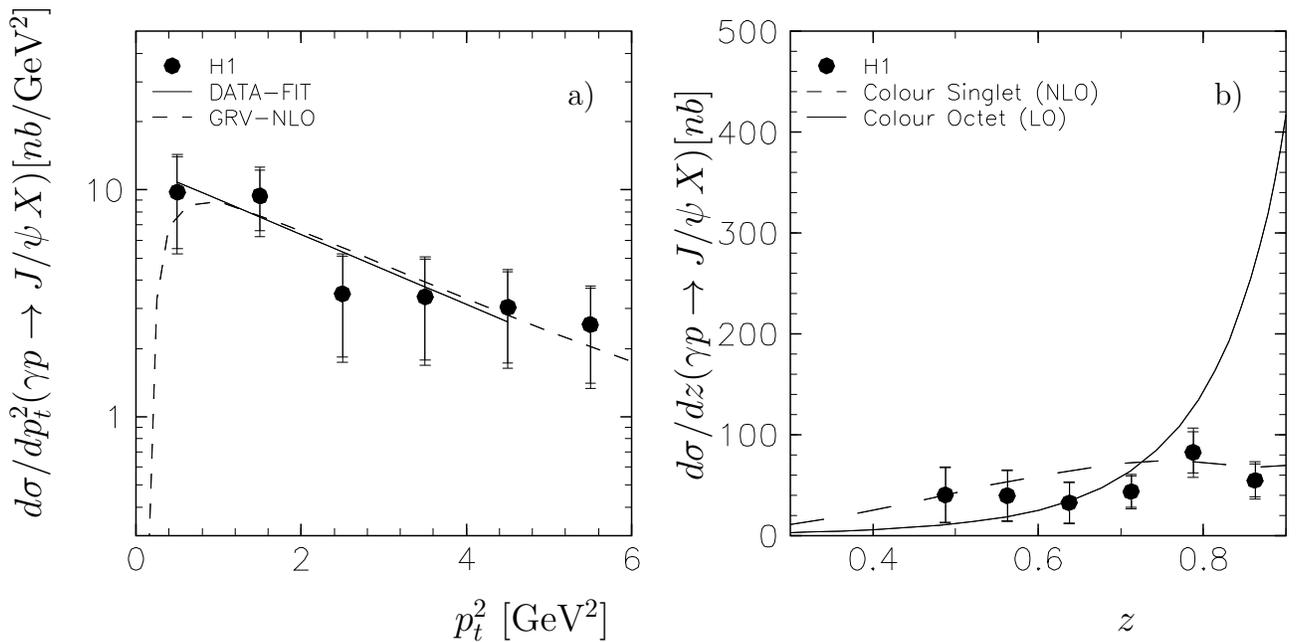

\vspace*{13pt}
\setlength{\unitlength}{1cm}
\begin{picture}(14.0,7.5)
\put(-1.6,-1.6){\epsfig{file=inelpsipt.eps,height=23cm,angle=90}}
\put(.2,1.0){\begin{rotate}{-90}\large $d\sigma/dp_t^2(\gamma p \rightarrow
J/\psi \,X )[nb/$GeV$^2$]\end{rotate}}
\put(5.7,-0.5){\large $p_t^2$ [GeV$^2$]}
\put(7.2,6.5){a)}
\put(15.8,6.5){b)}
\put(7.1,-1.6){\epsfig{file=inelpsiz.eps,height=23cm,angle=90}}
\put(9.0,1.5){\begin{rotate}{-90}\large $d\sigma/dz(\gamma p \rightarrow J/\psi
\, X)[nb]$\end{rotate}}
\put(14.5,-0.5){{\large $z$}}
\end{picture}
\vspace*{13pt}
\caption{Differential $\gamma p$ cross sections for inelastic $J/\psi$
production as measured by H1 in the energy range $30 < W < 150$~GeV: 
(a) with respect to $p_T^2 (J/\psi)$, integrated over $z<0.9$; 
(b) with respect to $z$, integrated over $p_T^2 > 1$~GeV$^2$.
The dashed curves are results of a NLO QCD calculation \protect\cite{kra96}
in the Colour-Singlet Model, with GRV parton densities.
The full line in (a) is a fit to the data at $p_T^2\leq 5 $~GeV$^2$.
The solid curve in (b) is the expected contribution of colour octet terms 
in a LO calculation \protect\cite{cac96}.}
\label{inelpsiptz}
\end{figure}
\begin{figure}[p]
\begin{center}
\epsfig{file=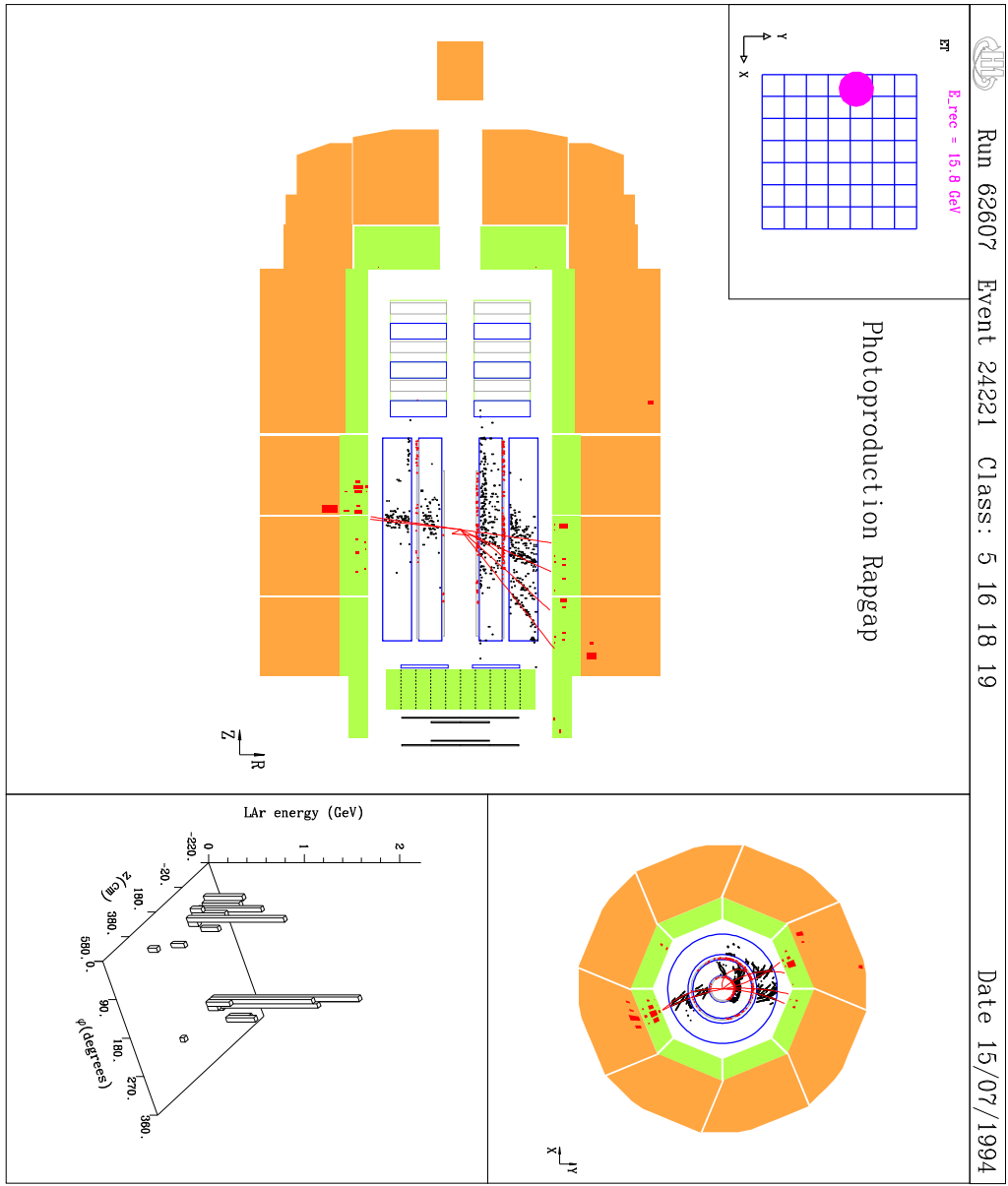,height=17cm,%
bbllx=5pt,bblly=80pt,bburx=600pt,bbury=790pt,clip=}
\end{center}
\vspace{1cm}
\caption{
Dijet photoproduction with a large rapidity gap around the proton direction
as seen in the H1 detector. The left picture is a longitudinal section
in which the proton beam comes from the right, the
upper right picture is a transverse view and the lower right one is a lego-plot 
of the calorimeter energy pattern.
This event has the scattered electron detected in the low-angle tagger,
represented in the upper left corner.}
\label{evtdiff}
\end  {figure}
\begin{figure}[p] \centering
\epsfig{file=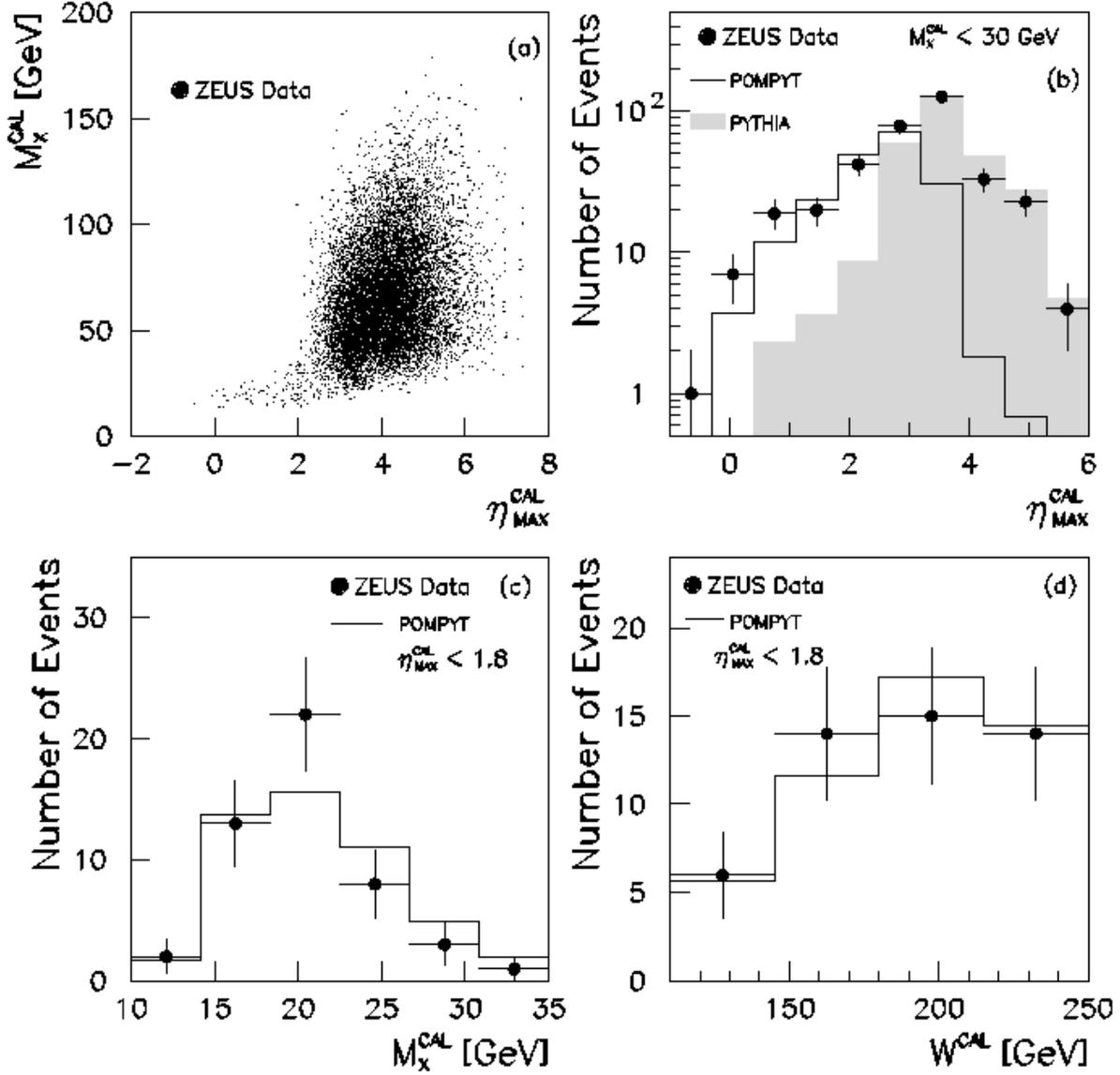,width=16cm}
\caption{
(a) The scatter plot of $M^{cal}_X$ versus $\eta^{cal}_{max}$ for a jet event
sample with $E^{jet}_{T,cal} > 6$~GeV and $-1<\eta^{jet}_{cal}<1$;
(b) the distribution of $\eta^{cal}_{max}$
  for the events with $M^{cal}_X < 30$~GeV along with the predictions
  of PYTHIA (shaded area) and POMPYT with a hard gluon density in the
  pomeron (solid line). The predictions are normalised to the number of
  data events above and below $\eta^{cal}_{max} = 2.5$, respectively;
(c) the distribution in $M^{cal}_X$ for the events with
  $\eta^{cal}_{max}<1.8$ together with the prediction of POMPYT with a
  hard gluon density in the pomeron (solid line) normalised to the
  number of data events; (d) the distribution in $W^{cal}$ for the
  events with $\eta^{cal}_{max} < 1.8$ and the
  prediction of POMPYT as in (c).}
\label{gpomdist}
\end{figure}
\begin{figure}[p] \centering
\epsfig{file=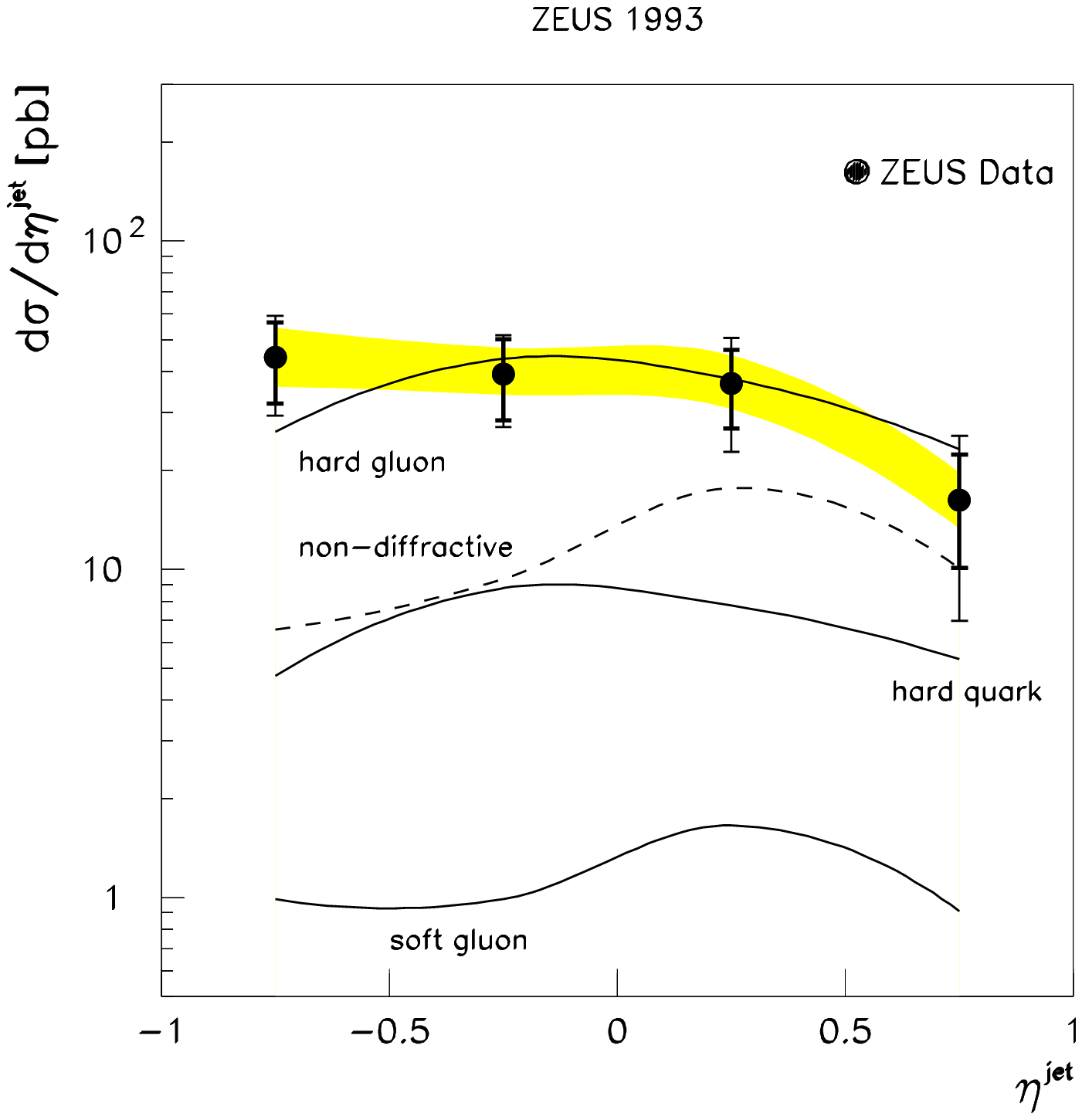,width=16cm}
\caption{Differential $ep$ cross section 
$d\sigma/d\eta^{jet}(\eta^{had}_{max}<1.8)$ for inclusive jet
  production with $E_T^{jet} >$~8~GeV in the kinematic region
  $Q^2 \leq$~4~GeV$^2$ and $ 0.2 < y < 0.85$. The measurements
  are not corrected for the contributions from non-diffractive
  processes and double dissociation. The shaded
  band displays the uncertainty due to the energy scale of the jets.
  The full curves are predictions of POMPYT for single diffractive jet
  production using the DL flux factor with different parametrizations of the
  pomeron parton densities (hard gluon, hard quark, soft gluon) and the GS-HO
  photon parton distributions. 
  The expectation from non-diffractive processes is shown as the dashed line,
  which is obtained by PYTHIA with MRSD- and GRV-HO parton distributions 
  for the proton and the photon respectively.}
\label{setadif}
\end{figure}
\begin{figure}[p] \centering
\epsfig{file=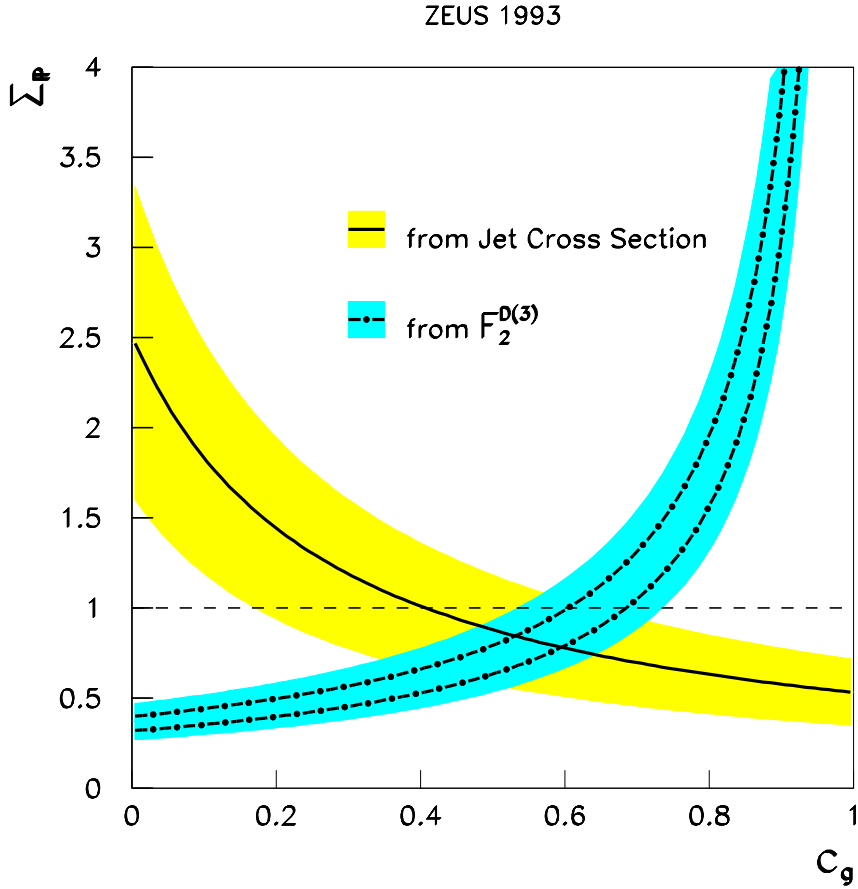,width=16cm}
\caption{Constraints on the $\Sigma_{\Pma} - c_g$ plane, the momentum sum and 
the fraction of hard gluons in the pomeron. The solid line is the result of the
$\chi^2$ fit of the measured $d\sigma/d\eta^{jet}$ for the diffractive event
sample with the predictions of POMPYT. 
The two dot-dashed curves are obtained from the measurement of the diffractive
structure function in DIS \protect\cite{f2diff},
 assuming two (lower) or three (upper) quark flavours.
The shaded bands represent $1~\sigma$ contours.}
\label{strisce}
\end{figure}
\begin{figure}[p] \centering
\epsfig{file=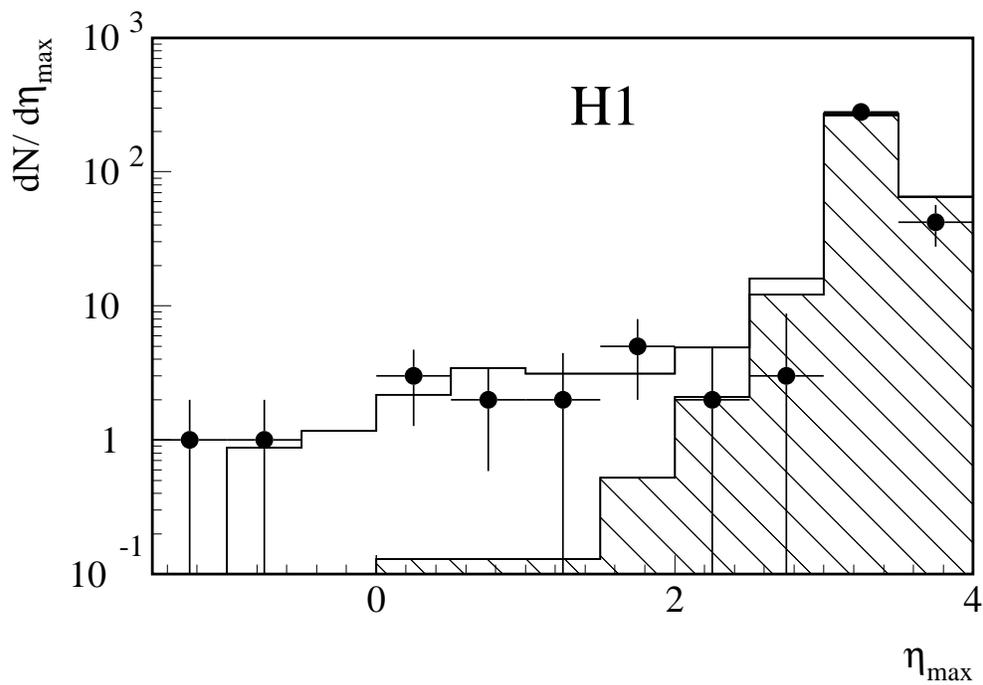,width=16cm}
\caption{$\eta_{max}$ distribution of $D^{*}$ candidate events.
The hatched histogram represents the contribution of non-diffractive processes,
from PYTHIA, the solid histogram is the sum of the previous one with the 
contribution of diffractive processes from RAPGAP. The two components are 
respectively normalized to the number of events with $\eta_{max} > 3$ and 
$\eta_{max} < 2$.}
\label{etamaxds}
\end{figure}
\begin{figure}[p] \centering
\epsfig{file=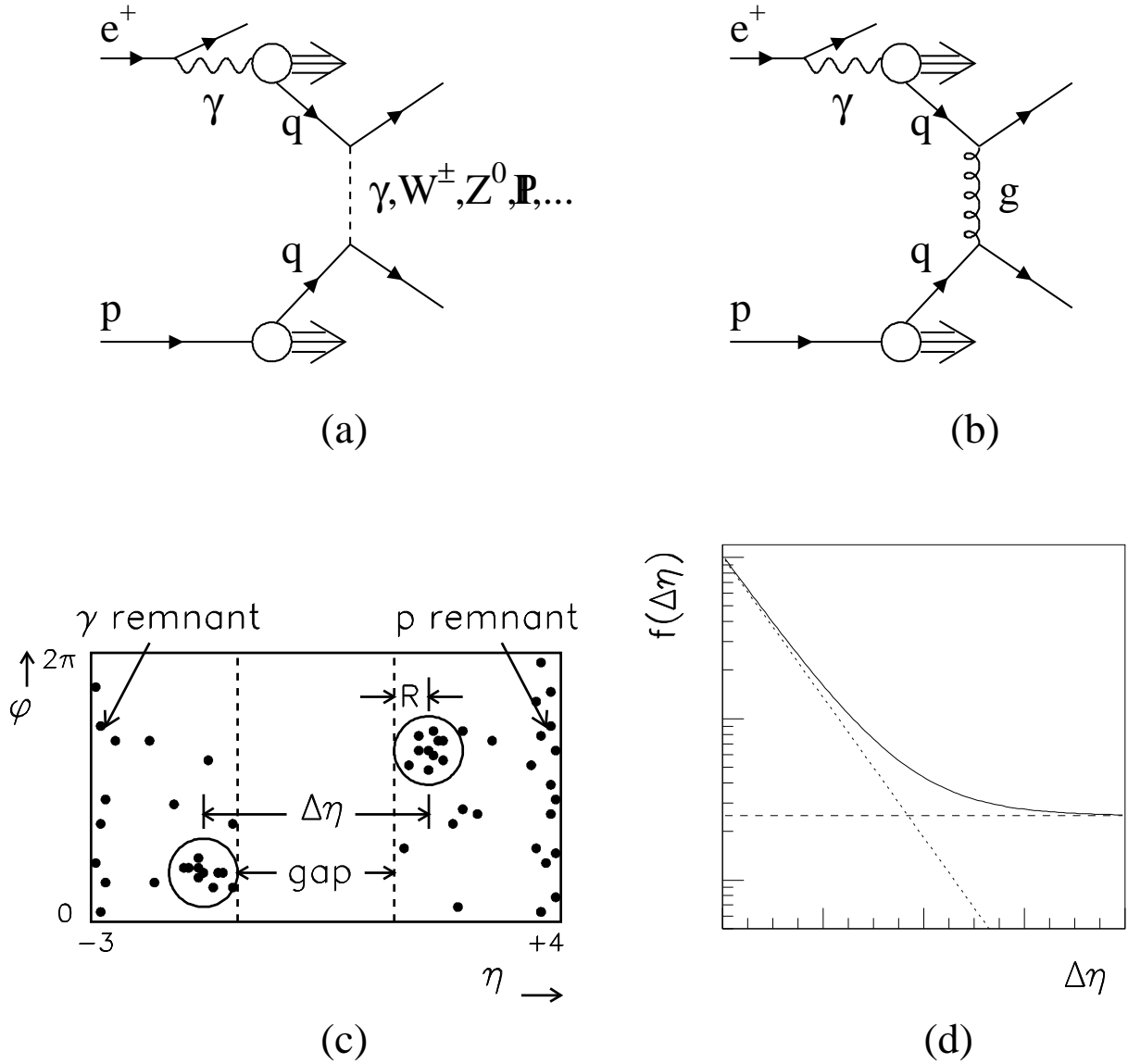,width=16cm}
\caption{
Resolved photoproduction via (a) colour singlet exchange and (b)
colour non-singlet exchange.
The rapidity gap event morphology is shown in (c) where black dots
represent final state hadrons and the boundary illustrates the limit
of the ZEUS acceptance.  Two jets of radius $R$ are shown, which are back
to back in azimuth and separated by a pseudorapidity interval
$\Delta\eta$.
An expectation for the behaviour of the gap fraction is shown
in (d)(solid line).
The non-singlet contribution is shown as the dotted line and
the colour singlet contribution as the dashed line.}
\label{diacolsi}
\end{figure}
\begin{figure}
\begin{center}
\epsfig{file=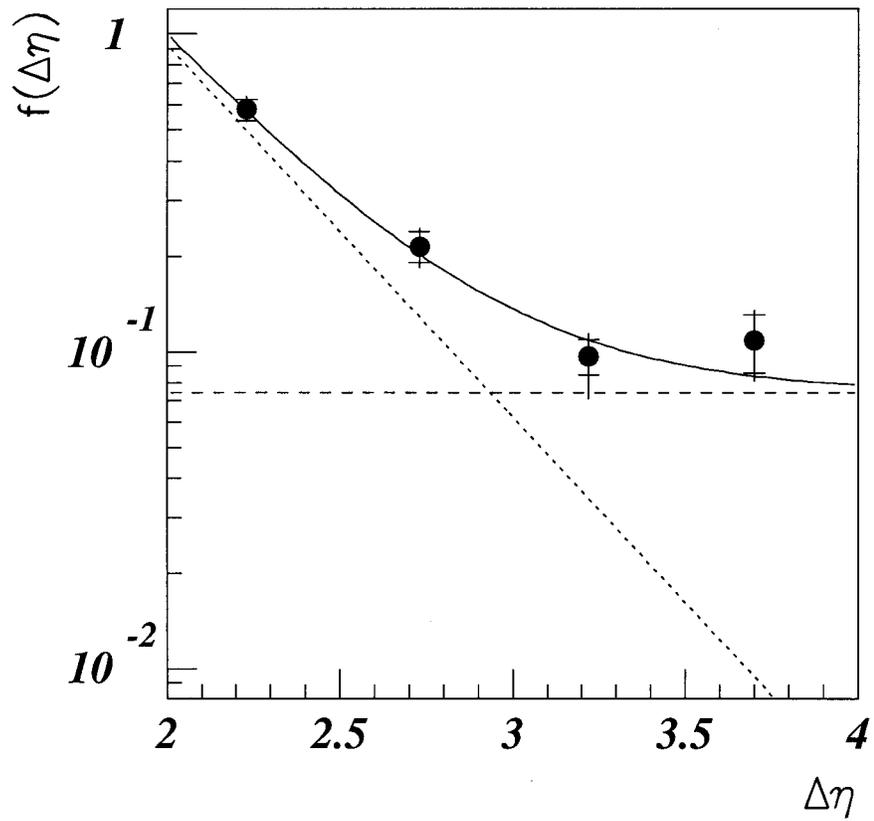}
\end{center}
\caption{The gap-fraction as a function of the gap width from the ZEUS
analysis of dijet events \protect\cite{zeuscolsin}. 
Data are compared to a fit to an exponential plus a constant.}
\label{csfinal}
\end{figure}
\end{document}